\documentclass[prb,nofootinbib,twocolumn,superscriptaddress,showpacs,floatfix]{revtex4}
\usepackage{bbding}
\usepackage{mathrsfs}
\usepackage{amsmath,amsfonts,amssymb}
\usepackage{epsfig}
\usepackage{graphicx}
\usepackage{wasysym}
\usepackage{bbm}
\usepackage{psfrag}
\usepackage{color}
\usepackage{pstool}
\usepackage{braket}
\usepackage{lmodern}
\usepackage[dvips, bookmarks, colorlinks=true, plainpages = false, citecolor = blue, linkcolor = blue, urlcolor = blue, filecolor = blue]{hyperref}
\def\be{\begin{equation}}
\def\ee{\end{equation}}
\def\bea{\begin{eqnarray}}
\def\eea{\end{eqnarray}}

\begin{document}
\title{Topological phases of polyacenes}
\author{Rakesh Kumar Malakar}\email{rkmalakar75@gmail.com}
\author{Asim Kumar Ghosh}
 \email{asimkumar96@yahoo.com}
\affiliation {Department of Physics, Jadavpur University, 
188 Raja Subodh Chandra Mallik Road, Kolkata 700032, India}
\begin{abstract}
The introduction of Su-Schrieffer-Heeger model 
has led to a major breakthrough in the area of one-dimensional
topological insulators, even though this model was primarily formulated
on an organic polymer called $trans$-polyacetylene
in order to explain its anomalous conductivity. 
In this study, a group of seven tight-binding models has
been introduced which are formulated on 
another organic polymer called polyacene,
where a variety of topological phases have
been observed in six models.
Topological properties of the most common geometric isomers
known as $cis$-polyacene, and $trans$-polyacene
have been investigated along with five additional 
modified polyacene structures. 
Although their geometric structures differ by
several symmetries, tight-binding band structures of 
$cis$-polyacene and $trans$-polyacene are found the same, and again   
their topological characters are found totally opposite. 
The $trans$-polyacene is nontrivial as it exhibits 
topological phase with nonzero winding number, while
the $cis$-polyacene is topologically trivial, although
both the structures adhere to the same set of symmetries
required for the topological character of BDI class. However,
$cis$-polyacene possesses additional mirror symmetry
in the real space but lacks inversion symmetry. 
Additional two modified structures of $trans$-polyacene and three 
 modified structures of $cis$-polyacene have been considered in order to
induce the nontrivial topology. 
Pair of flat bands is found both in modified $trans$-polyacene and $cis$-polyacene
structures such that they are topologically nontrivial for $trans$-polyacene
while trivial for $cis$-polyacene. Topological character of
all the models is different. 
  \vskip 1 cm
  Corresponding author: Asim Kumar Ghosh,
  \footnote{Corresponding author: Asim Kumar Ghosh}

  Email: asimkumar96@yahoo.com 
\end{abstract}
\maketitle
%%%%%%%%%%%%%%  Section I %%%%%%%%%%%%%%%%%%%%%%%%%%%%%%%%%%%%%%%%%%%%%%%%%
\section{INTRODUCTION}
Topological properties of matter have attracted 
immense research interest in the field of condensed matter
physics where basic feature of topological insulator (TI)
can be explained by the Su-Schrieffer-Heeger 
(SSH) model \cite{Qi,SSH1,SSH2}. The SSH model has been formulated on the
hydrocarbon chain with alternate single and double bonds
known as polyacetylene,  (C$_2$H$_2$)$_n$ \cite{SSH3,Rakesh1}.
This one-dimensional (1D) tight-binding (TB) model
is a staggered model with dimer unit cell as the
hopping strength for single and double bonds
has been assigned by different values. 
The nontrivial topology has been induced as soon as the
intercell hopping strength becomes stronger than 
intracell hopping strength. This phase has been characterized by 
the topological invariant called winding number ($\nu$) 
with $\nu=1$, in contrast to the
trivial phase where $\nu=0$, when intercell hopping strength
is weaker than that of intracell.
In the nontrivial phase a pair of zero-energy edge states is
found to emerge in accordance to the bulk-boundary correspondence rule,
which is absent in the trivial phase, 
although both the phases posses nonzero band gap. 
Experimental demonstration of these edge states
has been accomplished in photonic \cite{Vicencio}
and acoustic waveguides, \cite{Chen}
mechanical oscillator \cite{Merlo}, water-wave channel \cite{Pagneux}, 
system with Rydberg atoms
emulating SSH model \cite{Browaeys}, 
Bose-Einstein condensate of ${}^{87}$Rb atom into
a 1D optical superlattice potential \cite{Atala}, 
through quench dynamics of the system of ultracold atoms \cite{Xie},
and topolectrical circuits \cite{Thomale},
where the structure of SSH model has been fabricated 
in those hybrid material platforms. 

In this study, TB models on another
hydrocarbon chains known as polyacene, (C$_4$H$_2$)$_n$
have been formulated which exhibit a variety of 
topological phases \cite{Kitao}.
Polyacenes are polycyclic aromatic hydrocarbons,
formed by linearly fused benzene rings and regarded as the
1D analogue of two-dimensional (2D) graphene \cite{Clar-John,Bettinger}.
Being a member of organic optoelectronic materials
polyacene has drawn immense
interest for their huge light-harvesting efficiency,
high chemical tunability which are useful for the
fabrications of photovoltaics, solid-state lighting, 
sensors and organic solar cells \cite{Chin,Romanin}.
These hydrocarbons
are basically low-band gap organic polymers, indispensable for
enhanced optical absorption and efficient charge transport.
Apart from the astonishing optoelectronic properties, 
ployacene is found capable to exhibit nontrivial topological character. 
Multiple topological phase transitions have been induced
in polyacene under both compressive and tensile strains \cite{Chin}, and 
the double-bridged polyacene polymers are found to exhibit
different types of topological phases \cite{Romanin}. 
%which is famous for potential optoelectronic applications 
Depending on the orientation of their repeating units around
double bonds, two geometric isomers are primarily available for polyacene,
which are known as  $cis$-polyacene ($c$-pol),
and $trans$-polyacene ($t$-pol) \cite{Melo,Tanaka}. 
%In $cis$-polyacene ($c$-pol), the repeating units are on the same side of the
%double bond, while in $trans$-polyacene ($t$-pol),
%they are on opposite sides \cite{Melo,Tanaka}.
The basic structural difference can be explained in
terms of a mirror symmetry. 
However, this seemingly minor structural difference  leads to
significant difference not only in their physical and chemical properties,
but also in their topological character. 
%It will be shown that TB Hamiltonian on the 
%$t$-pol possesses the chiral symmetry, where $c$-pol does not,
%which subsequently leads to the fact that
%$t$-pol is topologically nontrivial while $c$-pol is trivial. 
Polyacetylene also has two geometric isomers, say,
$trans$- and $cis$-polyacetylene, however, for this case both the
configurations yield the same TB Hamiltonian.  

In this study, topological behavior of seven TB models
based on $t$-pol, and $c$-pol will be investigated.
Topological triviality of $c$-pol
attributes to the fact that under a transformation,
this structure can be mapped on to a pair of identical SSH chains
with the presence of staggered fields. As a result,
it exhibits topological triviality,
eventhough it holds the zero- and nonzero-edge states. 
Both $t$-pol and $c$-pol possess the 
chiral symmetry (CS), particle-hole symmetry (PHS) and
time-reversal symmetry (TRS) in terms of four-site unit cell,
which have been demonstrated extensively in the
real space as well as in the reciprocal space. 
However, in comparison to  $t$-pol,
$c$-pol lacks the inversion symmetry (IS)
but holds the mirror symmetry (MS) in addition.

In order to induce topological nontriviality in the 
$c$-pol structure, an additional intracell hopping term has been considered, 
which introduces IS in the system. This model is referred as $cb$-pol,
where this new bond bridges the open vertices of the benzene ring.
The topological behaviour of this model has been explained
by mapping this system on to a pair of identical
SSH chains with uniform fields of opposite signs for the
two different chains. As a result topological behavior
of this model is identical to that of standard SSH model. 
In the same way, effect of this bridging bond
on the $t$-pol structure
has been investigated here in 
$tb$-pol model by introducing the additional intercell hopping term
like the $cb$-pol structure. The resulting system yields a pair of
flat bands and it always remain
topologically nontrivial through out the parameter regime,
in contrast to the $t$-pol which is found topological in a limited region. 
Similarly, topological feature of $c$-pol and  $t$-pol in
the presence of cross-bridging bonds, referred as  $ccb$-pol and
$tcb$-pol have been studied which demonstrate nontrivial topology
of different types. 

In order to induce topological phases with higher winding numbers
in $c$-pol, additional hopping terms in between nearest neighbour (NN) and
next nearest neighbour (NNN) unit cells for the $c$-pol structure
in a different manner have been introduced.
Three different values of hopping amplitude  
have been considered where the resulting model has been termed as
 $cn$-pol. 
The system exhibits three different topologically
nontrivial phases characterized by $\nu=-2,2,4$. 
So, altogether seven different TB model on the polyacene
structure have been formulated which are referred as 
$t$-pol, $tb$-pol, $tcb$-pol, $c$-pol, $cb$-pol $ccb$-pol, and $cn$-pol
in this study whose topological properties have been
investigated in great details. Two different 
hopping strengths for the single and double bonds
are taken into account in the six 
models, while strength for NNN hopping in the $cn$-pol structure
is assumed different from that of single and double bonds. 

The article has been arranged in the following order.
The $t$-pol, $tb$-pol, $tcb$-pol, $c$-pol, $cb$-pol, $ccb$-pol and $cn$-pol 
models have been introduced in the Sections \ref{trans},
\ref{trans-bridged}, \ref{trans-cross-bridged},
\ref{cis}, \ref{cis-bridged}, \ref{cis-cross-bridged} and \ref{cis-nontrivial},
respectively. Topological properties of these models
have been investigated in the respective sections.
A table describing the symmetries for the models has been
presented. Finally, 
an extensive discussion based on the results has been made available 
in Sec \ref{Discussion}.
%%%%%%%%%%%%%%  Section II %%%%%%%%%%%%%%%%%%%%%%%%%%%%%%%%%%%%%%%%%%%%%%%%%
\section{THE {\em trans}-POLYACENE MODEL}
\label{trans}
Polyacene is the simplest member of the aromatic polymer group
which is composed of  benzene rings
stacked linearly one after another. This hydrocarbon
had become a subject of immense theoretical studies 
many decades ago although this polymer has been synthesized
very recently \cite{Clar-John,Bettinger}.
In 2023, polyacene was synthesized 
after successful synthesis of precursor polymer within 
metal-organic frameworks followed by its conversion \cite{Kitao}. 
The primary motivation behind the theoretical studies on
polyacene was the understanding of the electronic properties
of its possible structural isomers, $t$-pol and $c$-pol,
and etc \cite{Melo,Tanaka}.
These studies indicate that $t$-pol is 
more stable than $c$-pol \cite{Melo,Bettinger,Whangbo}. In addition, 
this hydrocarbon has been regarded as the 1D analogue of graphene, 
a 2D material made of pure carbon, which is famous for its
higher values of electrical and thermal conductivities. 
\begin{figure}[h]
\psfrag{A}{A}
\psfrag{B}{B}
\psfrag{C}{C}
\psfrag{D}{D}
\psfrag{v}{$v$}
\psfrag{w}{$w$}
\includegraphics[width=230pt]{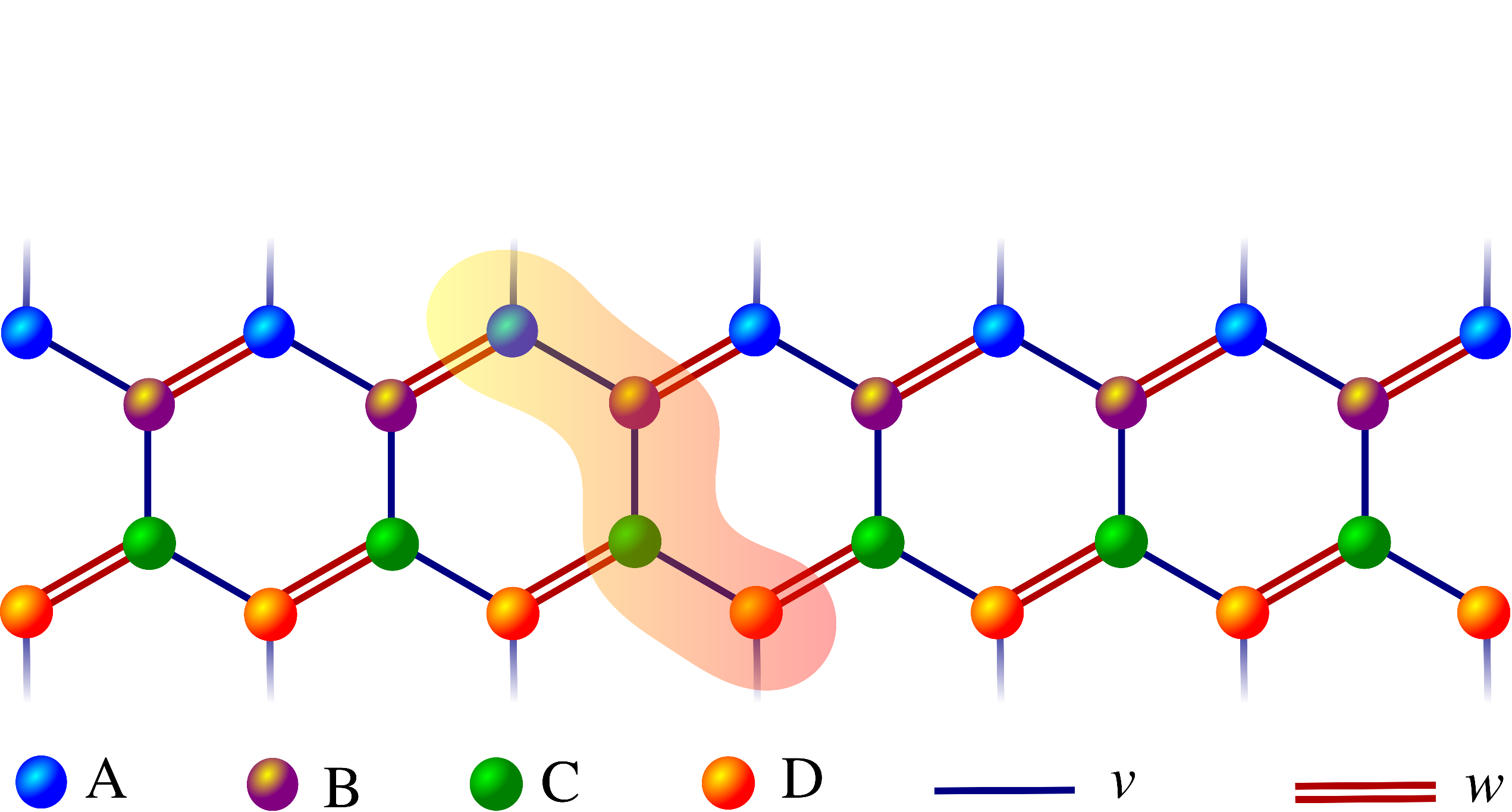}
\caption{Structure of {\em trans}-polyacene.
  Carbon atoms marked by A, B, C
  and D constitute a unit cell. One such unit cell is shown
  within shaded region.}
 \label{trans-polyacene}
\end{figure}
Structure of $t$-pol has been depicted in Fig. \ref{trans-polyacene},
whose tetramer unit cell has been shown by shaded region. 
Within a unit cell, four carbon atoms as labelled by the
sites, A, B, C and D, 
have been connected only by the single bonds,
while adjacent unit cells are connected
solely by a pair of double bonds, where B  and D have 
been shared with A and C, respectively. This structure can be
viewed as two coupled $trans$-polyacetylene chains,
one is made of sites A and B, and another is made of
sites C and D, giving rise to the tetramer unit cell 
accommodating four nonequivalent sites.

The TB Hamiltonian for the $t$-pol structure is
given by,
\bea H_{\rm t}&=&v\sum_{j=1}^N \left(a_j^\dag b_j+b_j^\dag c_j+c_j^\dag d_j \right)\nonumber \\
&\;\;+&w\sum_{j=1}^{N-1}
 \left(b_j^\dag a_{j+1}+d_j^\dag c_{j+1}\right)+h.c.,
 \label{trans-polyacene-ham}
 \eea
where $a_j$, $b_j$, $c_j$, and $d_j$ are the fermionic annihilation
operators corresponding to the A, B, C and D sites
respectively of the $j$ th cell,
while $N$ is the total number of sites.
The hopping parameters over single and double bonds are
specified by $v$ and $w$, respectively.
%Basically, it is a chain of tetramers.
In order to diagonalize the Hamiltonian
(Eq. \ref{trans-polyacene-ham}), following 
Fourier transformation of the operators is considered: 
\[\alpha_j=\frac{1}{\sqrt N}\sum_{k\in {\rm BZ}} e^{ikj}\alpha_k,\]
where $\alpha=a,b,c,d$, and $k$ is the Bloch wave vectors  
within the Brillouin zone (BZ). 
Hamiltonian in the $k$-space 
can be written in terms of a $4\times 4$ matrix
for the tetramer unit cell as:  
\[H_{\rm t}=\sum_{k\in{\rm {BZ}}}\Psi_k^\dag H_{\rm t}(k)\Psi_k.\]
The state vector is $\Psi_k=[a_k\;b_k\;c_k\;d_k]^{\rm T}$, 
where T stands for transpose,  
and
\be
H_{\rm t}(k)=\left(
\begin{array}{cccc}0&g(k)&0&0\\[0.4em]
  g^*(k) &0&v&0\\[0.4em]
  0&v&0&g(k)\\[0.4em]
  0&0&g^*(k) &0
\end{array} \right),
 \label{Htk}
 \ee
 where $g(k)=v+we^{-ik}$, and lattice spacing is assumed unity. 
Now, it is easy to check that $H_{\rm t}(k)$ satisfies the
 following symmetry relations under the three
different operators:
\[\left\{\begin{array}{l}
\mathcal K^{-1} H_{\rm t}( k) \mathcal K=H_{\rm t}(-k),\\ [0.4em]
 \mathcal T_z^{-1} H_{\rm t}( k) \mathcal T_z=-H_{\rm t}(k),\\ [0.4em]
 \mathcal P^{-1} H_{\rm t}( k) \mathcal P=-H_{\rm t}( -k),\end{array}\right.
 \]
 where $\mathcal K$ is the complex conjugation operator,
 $\mathcal T_z=I \otimes \sigma_z$, 
 $\mathcal P = \mathcal K \mathcal T_z$, $I$ is the ($2\times 2$) 
 identity matrix, and
$\sigma_z$ is the $z$-component of the Pauli matrix set.
 These relations respectively correspond to the
 conservation of TRS, CS and 
 PHS. However, it is true that any two relations
 of the above triplet set lead to the third one.
 For example, chiral operator $\mathcal T_z$ can be expressed 
 in terms of $\mathcal K$ and $\mathcal P$ as
 $\mathcal T_z=\mathcal K \cdot \mathcal P$. 
 The system belongs to the BDI class according to the
 tenfold classification scheme \cite{Ryu},
 since $\mathcal K^2=1$, $\mathcal P^2=1$, and $\mathcal T^2_z=1$.
 In addition, the system preserves the IS, as it satisfies the relation,  
 \[\mathcal T_{xx}^{-1} H_{\rm t}( k) \mathcal T_{ xx}=H_{\rm t}(-k),\]
 where $ \mathcal T_{xx}=\sigma_x\otimes \sigma_x$,
 and $\sigma_x$ is the $x$-component of the Pauli matrix set.

 These symmetries can be realized in the real space also.
 For example, CS can be understood by the operator,
 $\mathcal C=I_{N/4}\otimes \mathcal T_z$, which is applicable on
 ($N\times N$) Hamiltonian matrix in the real space,
 where $I_{N/4}$ is an identity matrix of size
 ($N/4 \times N/4$),
 since the unit cell comprises of four non-equivalent
 sites. %$\left(\dfrac{N}{4} \times \dfrac{N}{4}\right)$).
 PHS can be realized by the operator, $\mathscr P=\mathcal{ K} \mathcal C$,
 while IS by the operator, $\mathscr{ I}=I'_{N/4}\otimes
 \mathcal T_{xx}$, where $I'_{N/4}$ is an anti-diagonal
 identity matrix of size ($N/4 \times N/4$).
 The inversion implemented by the operator $\mathscr{ I}$ is accomplished by the
 transformations:
 \be
\left\{
\begin{array}{l}a_j \rightarrow d_{N+1-j},\\[0.4em]
b_j \rightarrow c_{N+1-j}, \\[0.4em]
 c_j \rightarrow b_{N+1-j}, \\[0.4em]
d_j \rightarrow a_{N+1-j}. 
\end{array} \right.
\label{transform1}
 \ee
 
 In order to define the winding number
 for this four-band systems, 
 new Hamiltonian is obtained under the unitary transformation as, 
 $H'_{\rm t}(k)=U^{-1}H_{\rm t}(k)U$, where \cite{Liu}
 \be
U=\left(
\begin{array}{cccc}0&0&0&1\\[0.4em]
  0 &1&0&0\\[0.4em]
  0&0&1&0\\[0.4em]
  1&0&0 &0
\end{array} \right). 
 \label{U}
 \ee
 Under this transformation, $H'_{\rm t}(k)$
 assumes $2 \times 2$ block off-diagonal form, where
 diagonal blocks are null, while off-diagonal blocks are
 nonzero, as shown below. 
 \be
H'_{\rm t}(k)\!=\!\left(
\begin{array}{cc}0&h_{\rm t}(k)\\[0.4em]
  h_{\rm t}^\dag(k)&0
\end{array} \right)\!,\;{\rm with}\;\;
h_{\rm t}(k)\!=\!\left(
\begin{array}{cc}g^*(k)&0\\[0.4em]
  v&g^*(k)
\end{array} \right)\!. \nonumber
 %\label{Htkprime}
 \ee
 
\begin{figure}[h]
    \psfrag{a}{(a)}
  \psfrag{b}{(b)}
  \psfrag{Energy}{ \hskip -0.5 cm Energy}
  \psfrag{I}{ \hskip -0.2 cm $I_{pr}$}
   \psfrag{win}{ \hskip -0. cm $\nu$}
  \psfrag{00}{\hskip 0.05 cm $0.0$}
  \psfrag{0.25}{$0.25$}
  \psfrag{0.5}{\hskip -0.00 cm $0.5$}
  \psfrag{1.0}{\hskip -0. cm $1.0$}
\psfrag{4}{$4$}
\psfrag{2}{$2$}
\psfrag{1}{$1$}
\psfrag{0}{$0$}
\psfrag{-1}{\hskip -0. cm$-1$}
\psfrag{-2}{\hskip -0.05 cm $-2$}
\psfrag{-4}{\hskip -0. cm $-4$}
\psfrag{v}{\hskip 0.1 cm $v/w$}
\psfrag{w}{$w$}
\hskip 0.24 cm
\includegraphics[width=240pt]{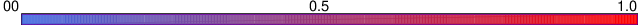}
\vskip -0.0 cm
\includegraphics[width=250pt]{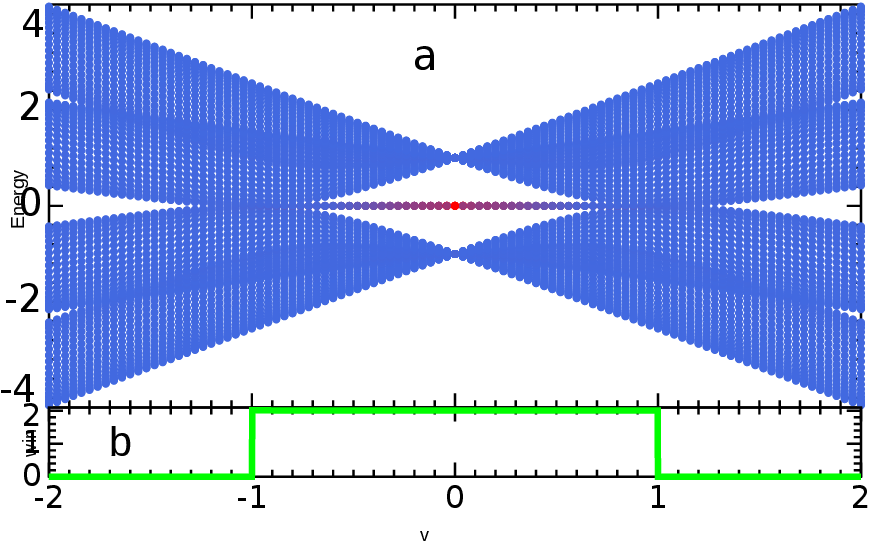}
\caption{Variation of energies (a) and winding number (b)
  with $v/w$ of {\em trans}-polyacene.
  Top colorbar indicates the variation of $I_{pr}$.}
 \label{trans-polyacene-energies-obc}
\end{figure}

 Winding number is now defined by
 \be
 \nu=\frac{1}{2\pi i}\oint dk \; {\rm Tr}\left[ \frac{1}{h_{\rm t}(k)}\frac{dh_{\rm t}(k)}{dk}\right],\ee
 which can be employed for the
 characterization of topological property of this four-band system.
 After simplification, 
 \be \nu =\frac{2}{2\pi i}\oint_{|z|=1} d\log{f_t(z)}, \nonumber \ee
 where $z=e^{ik}$ and $f_t(z)=v+wz$, which has no pole.
 Now using Cauchy principle argument,
 \[\frac{1}{2\pi i}\oint_{|z|=1} d\log{f_{\rm t}(z)}=\mathcal N_z, \nonumber\]
 where $\mathcal N_z$ is the number of zero of $f_{\rm t}(z)$ 
 within the unit circle, $|z|=1$, about the origin.
 Since,
  \be \mathcal N_z =\left\{
\begin{array}{cc}0,& {\rm if}\; v>w,\nonumber\\[0.4em]
  1,&{\rm if}\; v<w,\nonumber
\end{array} \right.\nonumber \ee
winding number is given by
  \be \nu =\left\{
\begin{array}{cc}0,& {\rm if}\; v>w,\nonumber\\[0.4em]
  2,&{\rm if}\; v<w.\nonumber
\end{array} \right.\nonumber \ee
  % \nu &=&\frac{1}{\pi}\int_0^{2\pi} dk \frac{w^2+vwe^{ik}}{v^2+w^2+2vw\cos{(k)}}%,\nonumber\\[0.4em]
 % &=&\frac{1}{\pi}\int_0^{2\pi} dk\; \frac{we^{ik}}{v+we^{ik}},\nonumber\\[0.4em]
 %&=&\frac{2w^2}{|v^2-w^2|}-\left\{
%\begin{array}{cc}\frac{2w^2}{v^2-w^2},& {\rm if}\; v>w,\nonumber\\[0.4em]
%  \frac{2v^2}{w^2-v^2},&{\rm if}\; v<w,\nonumber\\[0.4em]
%\end{array} \right.,\nonumber\\[0.4em]
%&=&\left\{\begin{array}{cc}0,& {\rm if}\; v>w,\nonumber\\[0.4em]
%  2,&{\rm if}\; v<w.
%\end{array} \right.
% \eea
 Variation of winding number with respect to $v/w$ has been shown in
 Fig. \ref{trans-polyacene-energies-obc} (b). It indicates that
 the {\em t}-pol hosts a nontrivial topological phase with
 $\nu=2$, when $-1< v/w < +1$, and otherwise trivial. 
 
\begin{figure}[h]
\psfrag{a}{(a)}
\psfrag{b}{(b)}
\psfrag{p}{$-\pi$}
\psfrag{q}{$-\frac{\pi}{2}$}
\psfrag{n}{$\frac{\pi}{2}$}
\psfrag{m}{$\pi$}
\psfrag{1}{\hskip -0.05 cm$1$}
\psfrag{0}{\hskip -0.05 cm$0$}
\psfrag{-1}{\hskip -0.05 cm$-1$}
\psfrag{k}{$k$}
\psfrag{E}{$E_{\rm t}(k)$}
\includegraphics[width=245pt]{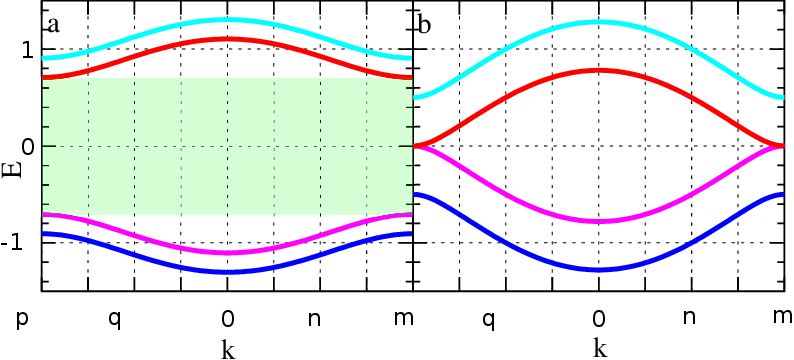}
\caption{Dispersion relations  of {\em t}-pol model
  for (a) $v=0.2$, $w=1$;  (b)  $v=0.5$, $w=0.5$. Energy gap in (a) is
shown by green shade. }
 \label{trans-polyacene-dispersions}
\end{figure}
Four eigenenergies, $E_{\rm t}(k)$ of $H_{\rm t}(k)$
are given by
 \be E_{\rm t}(k)=\pm \sqrt{\frac{p_{\rm t}\pm\sqrt{q_{\rm t}}}{2}},
   \label{Etk}  \ee
  where
  $p_{\rm t}= 3v^2+2w^2+4vw\cos{k}$, and
  $q_{\rm t}=5v^4+4v^2w^2+8v^3w\cos{k}$. 
The corresponding eigenvectors, 
$\phi_{\rm t}(k)$ of $H_{\rm t}(k)$, 
are given by, 
\[\phi_{\rm t}(k)=\left[\frac{vg}{E_{\rm t}^2-|g|^2}\;\;
  \frac{vE_{\rm t}}{E_{\rm t}^2-|g|^2}\;\;1\;\;
  \frac{g^*}{E_{\rm t}} \right]^{\rm T},\]
where four different eigenvectors are obtained from this
expression by substituting four different values of
$E_{\rm t}$, as noted in Eq. \ref{Etk}. 

Dispersion relations (Eq. \ref{Etk}) indicate that
four bands collapse onto two flat bands when $v=0$,
where $E_{\rm t}=\pm w$. It means each flat band is doubly degenerate
at this point. On the other limit, when $w=0$,
system exhibits four different flat bands. 
The four-band dispersion relations, $E_{\rm t}(k)$,
have been plotted in Fig. \ref{trans-polyacene-dispersions}, 
 for (a) $v=0.2$, $w=1$;  (b)  $v=0.5$, $w=0.5$.
 Fig. \ref{trans-polyacene-dispersions} (a) corresponds to
 the nontrivial phase where four bands are separated by
 distinct energy gaps, which actually survives in the
 regions, $-1< v/w < +1$, while at the transition points, $v/w=\pm 1$,
 energy gap between the middle bands vanishes as shown in
 Fig. \ref{trans-polyacene-dispersions} (b).
 Energy gap is shown by green shade in
 Fig. \ref{trans-polyacene-dispersions} (a).
 Minimum of the gap ($\Delta E_t$) is found at $k=\pm \pi$, 
 whose value can be obtained by
 $\Delta E_t=\sqrt 2\sqrt{p-\sqrt q}$, where
 $p=v^2+2(v-w)^2$, and $q=v^2(2p-v^2)$. 
The extended-Huckel-theory crystalline-orbital (EHCO) method
has predicted the gap, $\Delta E_t=0.45$ eV \cite{Whangbo}, where 
the minimum of band gap is noted at $k=\pm \pi$. 

In order to study the feature of edge states,
eigenenergies of the system under open boundary condition   
(OBC) have been plotted with respect to $v/w$ as shown in
Fig. \ref{trans-polyacene-energies-obc} (a),
as long as $|v/w|<2$. The diagram indicates that
energy spectra satisfy the relation, $E_{\rm t}(-v)=E_{\rm t}(v)$. 
It is observed that two pairs of zero-energy edge states exist 
within,  $-1< v/w < +1$, while they vanish beyond that,
or when $|v/w|>1$. 
Property of eigenvectors has been further examined by
estimating the value of their inverse participation ratio, 
$I_{pr}$. $I_{pr}$ is defined by the formula:
$I_{pr}=\sum_j|\psi_j|^4$, where $\psi_j$ is the
probability amplitude of a particular normalized
eigenstate $\psi$ at the lattice site $j$, for the
eigenenergy $E$, where  $\sum_j|\psi_j|^2=1$.
For an ideal localized state
$I_{pr}=1$, while for an extremely delocalized state
both $|\psi_j|^2$ and $I_{pr}$ decay linearly with the length of
the chain, leading to $I_{pr}\rightarrow 0$ in the thermodynamic limit. 
As a result, a higher value of $I_{pr}$ suggests greater localization,
while a lower value of that indicates greater delocalization. 
$I_{pr}$ of every quasienergy state has been enumerated
and the value has been color marked according to
the colorbar above the Fig. \ref{trans-polyacene-energies-obc} (a). 
Colorbar indicates that value of 
$I_{pr}$ for the zero-energy edge states 
is much higher than that for the other non-zero energy states. 
It means zero-energy sates are highly localized.
To obtain the energy diagram under OBC,
the real space Hamiltonian
in Eq. \ref{trans-polyacene-ham}, has been
spanned for $N=80$ sites, which is equivalent to
20 unit cells. After exact diagonalization of $N \times N$
Hamiltonian matrix, eigenenergy and corresponding $\psi_j$
have been enumerated. $I_{pr}$ for a particular state, $\psi$ is then 
determined simply by summing $|\psi_j|^4$ over the every site, $j$. 

\begin{figure}[h]
  \psfrag{a}{(a)}
  \psfrag{b}{(b)}
    \psfrag{p}{\scriptsize {$|\psi|^2$}}  
%  \psfrag{p}{$|\psi|$}  
  \psfrag{Energy}{ \hskip -0.15 cm Energy}
  \psfrag{I}{ \hskip -0.2 cm $I_{pr}$}
  \psfrag{00}{\hskip 0.05 cm $0.0$}
  \psfrag{0.25}{$0.25$}
    \psfrag{0.50}{\hskip -0.00 cm $0.5$}  
\psfrag{0}{$0$}
\psfrag{20}{$20$}
\psfrag{40}{$40$}
\psfrag{60}{$60$}
\psfrag{80}{$80$}
\psfrag{0}{$0$}
\psfrag{0.00}{\hskip -0. cm $0.00$}
 \psfrag{0.51}{\hskip -0.00 cm $0.50$}  
\psfrag{1.0}{\hskip -0. cm $1.0$}
\psfrag{0.5}{\hskip -0. cm $0.5$}
\psfrag{0.0}{\hskip -0.0 cm $0.0$}
\psfrag{-1.4}{\hskip -0.05 cm$-1.4$}
\psfrag{-0.7}{\hskip -0.05 cm $-0.7$}
\psfrag{0.7}{\hskip -0.0 cm $0.7$}
\psfrag{1.4}{\hskip -0.0 cm $1.4$}
\psfrag{v}{\hskip -0.15 cm $v=0.2,\;w=1$}
\psfrag{site}{\hskip 0.11 cm site}
\hskip 0.45 cm
\includegraphics[width=240pt]{top-color-box-polyacene.eps}
\vskip -0.0 cm
\includegraphics[width=250pt]{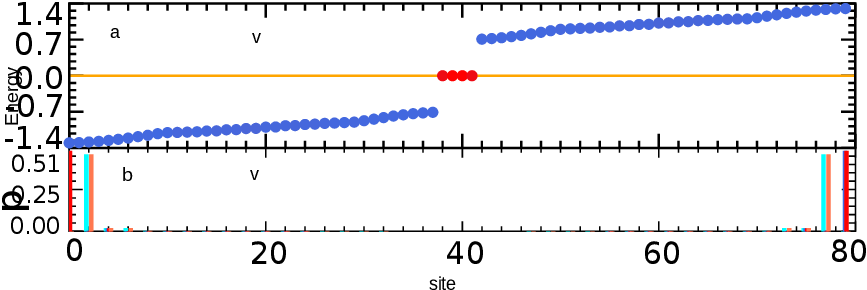}
\caption{Variation of energies for
  {\em trans}-polyacene when $v/w=0.2$, is shown in ascending order
  (a). The presence of four edge states at zero energy is found.  
  Top colorbar indicates the variation of $I_{pr}$.
  Probability density of four edge states are
  shown in the histogram by different colors (b). }
 \label{trans-polyacene-edge-energies}
\end{figure}
Energies of {\em t}-pol when $v=0.2,\;w=1$, 
have been plotted in Fig.\ref{trans-polyacene-edge-energies} (a),
in ascending order.
It reveals the presence of two pairs of zero-energy edge
states within the band gap which corresponds to 
the topological regime with $\nu=2$. 
Value of $I_{pr}$ for those states is very high. 
Probability density per site, $|\psi_j|^2$, 
of four zero-energy states has been 
  shown in the histogram by different colors
  in Fig. \ref{trans-polyacene-edge-energies} (b),
  which clearly indicates the localization of 
  four zero-energy states towards the opposite edges.
  
\begin{figure}[h]
  \psfrag{a}{(a)}
  \psfrag{b}{(b)}  
\psfrag{A}{A}
\psfrag{B}{B}
\psfrag{C}{C}
\psfrag{D}{D}
\psfrag{v}{$v$}
\psfrag{w}{$w$}
\includegraphics[width=230pt]{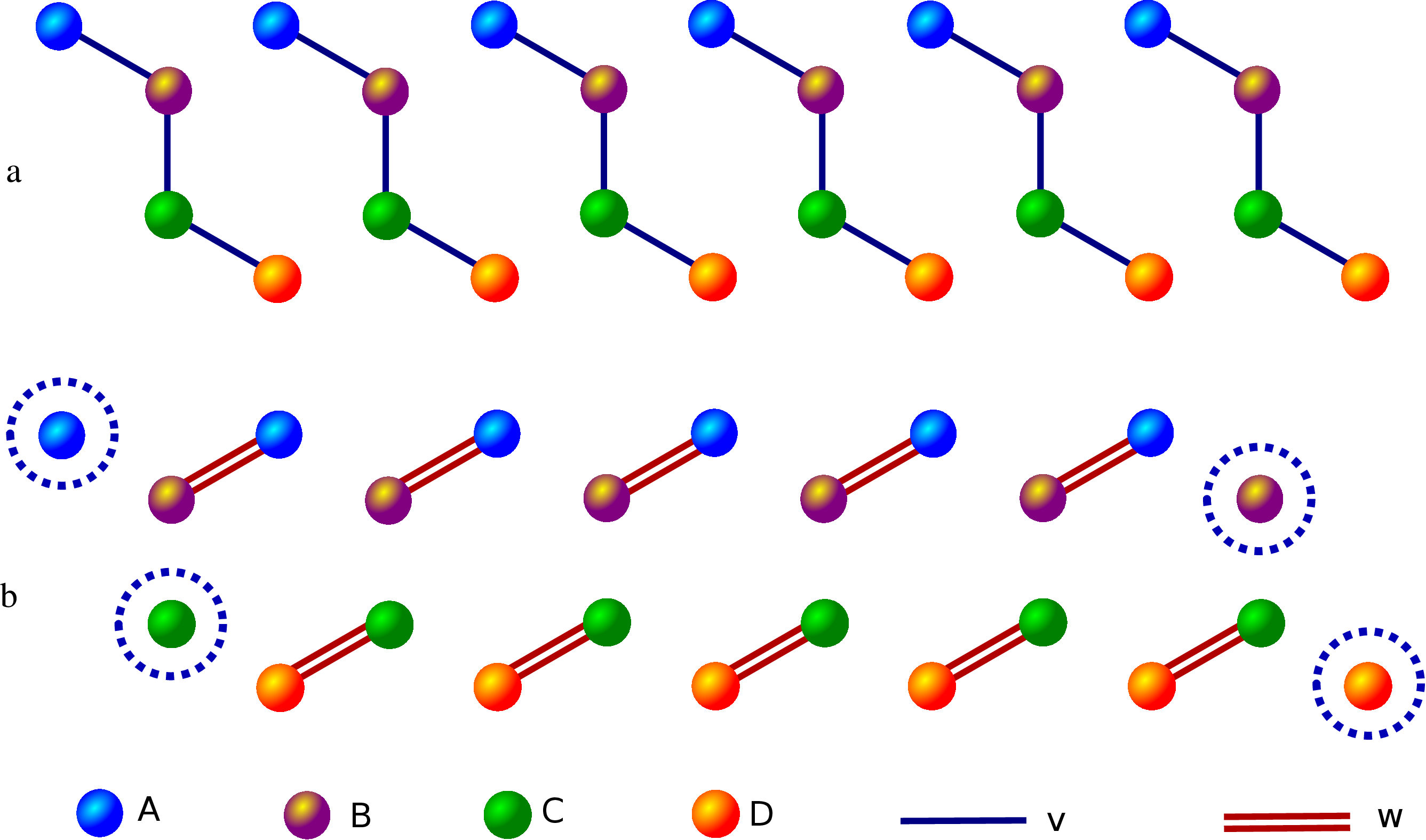}
\caption{Real space configuration of atoms and
  bonds within {\em trans}-polyacene for two extreme cases,
  (a) when $v\ne 0$ but $w=0$,
  and (b) when $w\ne 0$ but $v=0$. Free atoms
  are shown within dotted circles.} 
 \label{trans-polyacene-edge-modes}
\end{figure}

Existence of four zero-energy modes may be explained
by comparing two extreme limits, when 
one out of the two parameters, $v$ and $w$, is made zero. 
For example, when $v\ne 0$ and $w = 0$, all the atoms are connected
by the single bonds, as depicted in
Fig. \ref{trans-polyacene-edge-modes} (a). In this case,
six tetramer unit cells are shown where 
no atoms are left free, which corresponds to the
trivial phase. Whereas, when $w\ne 0$
and $v = 0$, atoms are connected only by double bonds as shown in  
Fig. \ref{trans-polyacene-edge-modes} (b).
As a result, 10 dimers are formed leaving four atoms free.
Out of these four atoms, two atoms lie in each of two extreme ends,
where one from each sublattice (A, B, C, D) 
are left unpaired, which are shown within the dotted circles.   
This configuration corresponds to the four zero-energy modes, which
are found to survive upto the limits $|v/w|<1$, or within the
nontrivial regime, $-1<v/w<1$. Estimated value of $I_{pr}$ for each of these
four zero-energy states is exactly unity, ($I_{pr}=1$). 
It is noted before that a pair of doubly degenerate flat bands, $E_{t}=\pm w$,
will appear in place of four bands when $v=0$. Therefore,
the system exhibits topological flat bands at this particular
point with $\nu=2$. 
%%%%%%%%%%%%%%  Section III %%%%%%%%%%%%%%%%%%%%%%%%%%%%%%%%%%%%%%%%%%%%%%%%%
\section{THE {\em trans}-POLYACENE MODEL with bridging bonds}
\label{trans-bridged}
\begin{figure}[h]
\psfrag{A}{A}
\psfrag{B}{B}
\psfrag{C}{C}
\psfrag{D}{D}
\psfrag{v}{$v$}
\psfrag{w}{$w$}
\includegraphics[width=240pt]{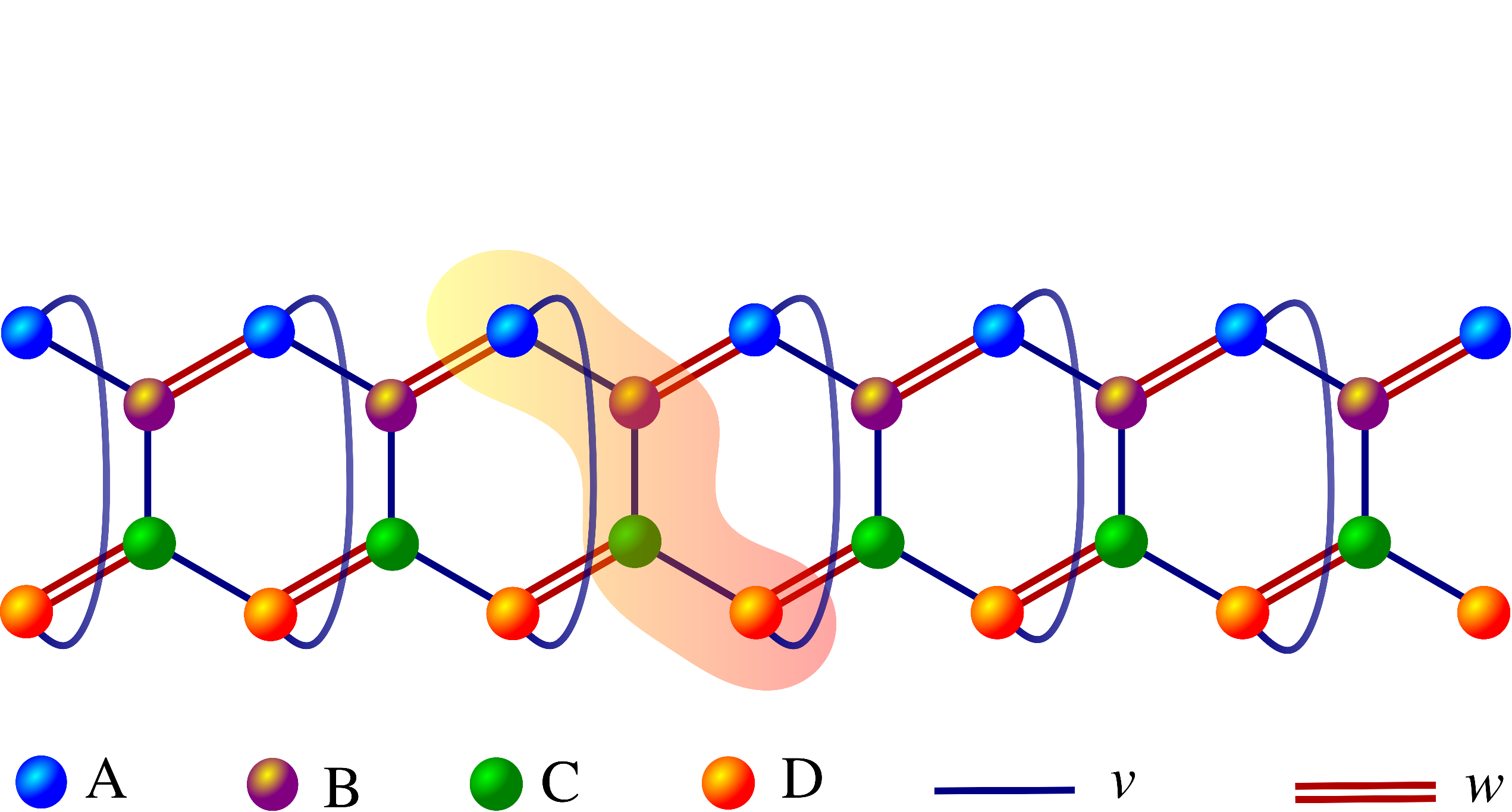}
\caption{Structure of {\em trans}-polyacene with bridged bonds ($tb$-pol). 
  Carbon atoms marked by A, B, C
  and D form a unit cell. One such unit cell is highlighted.}
 \label{trans-bridged-polyacene}
\end{figure}

The Hamiltonian for the $t$-pol structure along with the
additional bridged bond ($tb$-pol) is given by,
\be H_{\rm {tb}}=H_{\rm {t}} + v\sum_{j=1}^N \left(d_j^\dag a_{j+1}+h.c.\right).
 \label{trans-bridged-polyacene-ham}
 \ee
 Structure of $tb$-pol has been depicted in Fig. \ref{trans-bridged-polyacene},
 in which the additional hopping term acts between A and D sites of NN cells.
 This structure includes an extra covalent bonding between the carbon atoms
 sitting in opposite para positions of benzene ring. 
Existence of this kind of bonds bridging opposite
vertices of the benzene ring
as shown in Fig. \ref{trans-bridged-polyacene} has been reported before
\cite{Kruger,Jancarik}.
In the $k$-space, the system can be defined by the Hamiltonian, 
\be
H_{\rm {tb}}(k)=\left(
\begin{array}{cccc}0&g(k)&0&ve^{-ik}\\[0.4em]
  g^*(k) &0&v&0\\[0.4em]
  0&v&0&g(k)\\[0.4em]
  ve^{ik}&0&g^*(k) &0
\end{array} \right).
 \label{Htbk}
 \ee
%  where $g(k)=v+we^{-ik}$. 
  This system preserves the same set of symmetry like the $t$-pol,
 or conserves TRS, PHS, CS and IS, thus belongs to BDI class. 
 The dispersion relations of four bands are given by
 \[E_{\rm {tb}}=\pm\sqrt{2v^2+w^2+2vw\cos{k}\pm 2v(v+w)\cos{\frac{k}{2}}}.\]
 The corresponding eigenvectors, 
 $\phi_{\rm tb}(k)$ are given by
 \[\phi_{\rm tb}(k)=\left[\frac{E_{\rm tb}-vx}{g^*}\;\;
  1\;\;x\;\;\frac{E_{\rm tb}x-v}{g} \right]^{\rm T},\]
where
\[x=\frac{E_{\rm tb}^2g-|g|^2g+v^2\lambda}{vE_{\rm tb}(g+\lambda)},\;
{\rm with}\;\lambda=g^*e^{-ik}.\]
 \begin{figure}[h]
\psfrag{a}{(a)}
\psfrag{b}{(b)}
\psfrag{c}{(c)}
\psfrag{d}{(d)}
\psfrag{p}{$-\pi$}
\psfrag{q}{$-\frac{\pi}{2}$}
\psfrag{n}{$\frac{\pi}{2}$}
\psfrag{m}{$\pi$}
\psfrag{1}{$1$}
\psfrag{0}{$0$}
\psfrag{2}{$2$}
\psfrag{3}{$3$}
\psfrag{-1}{\hskip -0.02 cm$-1$}
\psfrag{-2}{\hskip -0.02 cm$-2$}
\psfrag{-3}{\hskip -0.02 cm$-3$}
\psfrag{k}{$k$}
\psfrag{E}{$E_{\rm tb}(k)$}
\includegraphics[width=240pt]{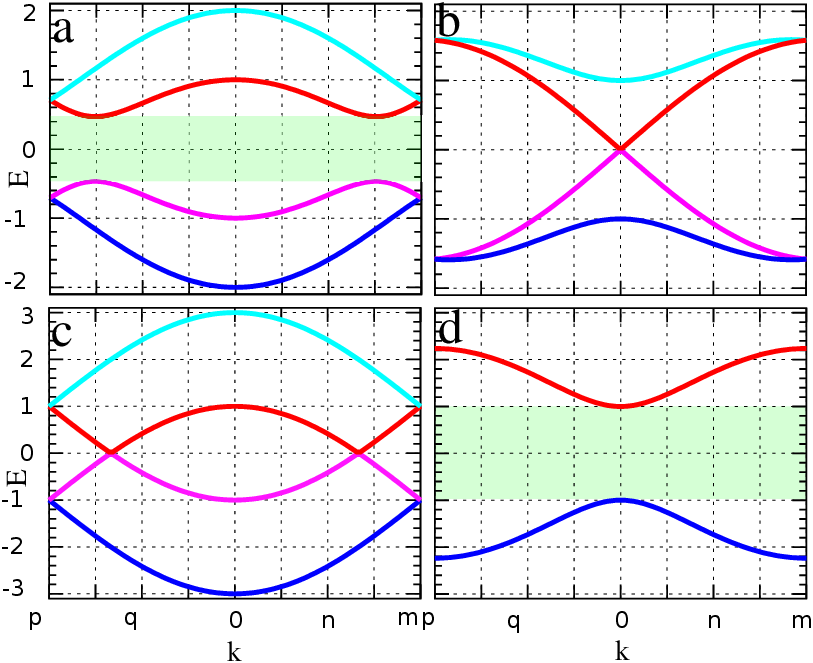}
\caption{Dispersion relations of $tb$-pol for (a) $v=1/2$, 
  (b)  $v=-1/2$,  (c)  $v=1$, and (d)  $v=-1$, when $w=1$.
  Energy gaps in (a) and (d) are shown
by shades.}
 \label{trans-bridged-polyacene-dispersions}
 \end{figure}

Dispersion relation, $E_{\rm tb}(k)$ has been plotted in 
Fig. \ref{trans-bridged-polyacene-dispersions},
for four different values, $v=0.5,\,-0.5,\,1$, and $-1$, 
in (a), (b), (c) and (d), respectively, when $w=1$.
 Energy gaps in (a) and (d) are shown
 by shaded regions.
 
Interesting features of the dispersion relations are noted below:
\begin{itemize}
  \item Two upper and two lower bands always touch
 each other separately at the Brillouin zone boundaries,
 $k=\pm \pi$, irrespective of the values of the
 parameters, $v$ and $w$. This feature is evident from all the
 diagrams, (a), (b), (c) and (d). 
\item The four bands of the system collapse into two
  different bands when $v=-w$. The two upper and two lower
  bands become degenerate throughout the BZ. This particular feature
  has been shown in Fig. \ref{trans-bridged-polyacene-dispersions} (d). 
\item The four bands of the system yield a pair of non-dispersive flat bands,
  $E_{\rm {tb}}=\pm w$, if $v=0$, which is similar to $t$-pol. 
\end{itemize}
Band diagrams shown in Fig. \ref{trans-bridged-polyacene-dispersions}
(b) and (c)
have been drawn for two different phase transition points,
where no band gap survives.
On the other hand, band gap is found to exist in 
Fig. \ref{trans-bridged-polyacene-dispersions} (a) and (d),
when the system lies in two different topological phases,
which will be discussed later. 

\begin{figure}[h]
\psfrag{a}{(a)}
\psfrag{b}{(b)}
\psfrag{z}{\hskip 0.1 cm $Z_{2,3}$}
  \psfrag{I}{ \hskip -0.2 cm $I_{pr}$}
  \psfrag{00}{\hskip 0.05 cm $0.0$}
  \psfrag{0.25}{$0.25$}
  \psfrag{0.5}{\hskip -0.00 cm $0.5$}
 \psfrag{1.0}{\hskip -0. cm $1.0$} 
\psfrag{E}{ \hskip -0.5 cm Energies}
\psfrag{4}{$4$}
\psfrag{2}{$2$}
\psfrag{1}{$1$}
\psfrag{0}{$0$}
\psfrag{-1}{\hskip -0.14 cm$-1$}
\psfrag{-2}{\hskip -0.14 cm $-2$}
\psfrag{-4}{\hskip -0.14 cm $-4$}
\psfrag{v}{\hskip -0.15 cm $v/w$}
\psfrag{w}{$\nu$}
\hskip 0.16 cm
\includegraphics[width=240pt]{top-color-box-polyacene.eps}
\vskip 0.02 cm
\includegraphics[width=250pt]{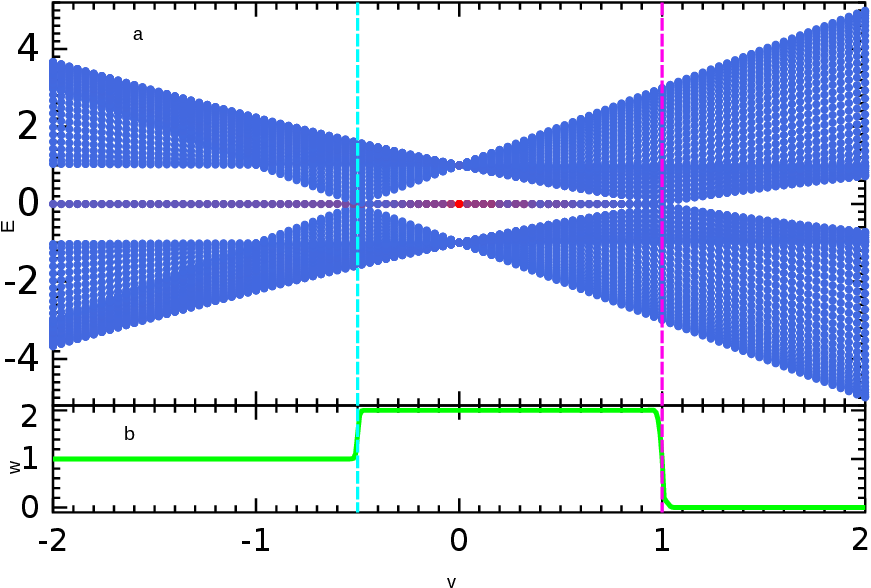}
\caption{Variation of energies for trans-bridged polyacane open chain (a), and 
   winding number (b) with respect to $v/w$. 
  Top colorbar indicates the variation of $I_{pr}$.
Vertical dashed lines are drawn at the phase transition points.}
 \label{trans-bridged-polyacene-energies-obc}
\end{figure}

Under the unitary transformation $H'_{\rm tb}(k)=U^{-1}H_{\rm tb}(k)U$,
  $H'_{\rm tb}(k)$ has become 
 block off-diagonal,  
 \be
H'_{\rm tb}(k)\!=\!\!\left(
\begin{array}{cc}0&h_{\rm tb}(k)\\[0.4em]
  h_{\rm tb}^\dag(k)&0
\end{array}\!\! \right)\!,\,{\rm with}\;
h_{\rm tb}(k)\!=\!\left(
\begin{array}{cc}g^*(k)&ve^{ik}\\[0.4em]
  v&g^*(k)
\end{array}\!\! \right)\!.\nonumber
 \label{Htbkprime}
 \ee
Winding number in this case can be evaluated as
\bea
\nu&=&\frac{1}{2\pi i}\oint dk \;
    {\rm Tr}\left[ \frac{1}{h_{\rm tb}(k)}\frac{dh_{\rm tb}(k)}{dk}\right],\nonumber\\[0.4em]
    &=&  \frac{1}{2\pi i}\oint_{|z|=1} d\log{f_{tb}(z)}, \nonumber
\eea
where $f_{tb}(z)=w^2z^2+v^2(1-z)+2vwz$, which has no pole.
The roots of $f_{tb}(z)$ are 
\[z_\pm=\frac{v^2-2vw\pm v\sqrt{v^2-4vw}}{2w^2},\;{\rm with},\,
z_+z_-=\frac{v^2}{w^2}.\]
The number of zero for $f_{tb}(z)$, within $|z|=1$, has been given as,
\be
\mathcal N_z=\left\{
\begin{array}{ll}0,& {\rm if}\; v>w,\nonumber\\[0.4em]
  2,& {\rm if}\; -w/2<v<w,\nonumber\\[0.4em]
  1,&{\rm if}\; v<-w/2.\nonumber
\end{array} \right.\nonumber\ee
As a result, using the Cauchy principle argument, 
topological phases in this case have been characterized by the 
winding number as,
\be \nu =\left\{\begin{array}{ll}0,& {\rm if}\; v>w,\nonumber\\[0.4em]
  2,& {\rm if}\; -w/2<v<w,\nonumber\\[0.4em]
  1,&{\rm if}\; v<-w/2.\nonumber
\end{array} \right.\nonumber\ee
Variation of winding number with $v/w$ has been shown in
Fig. \ref{trans-bridged-polyacene-energies-obc} (b) in green line.
It reveals that system exhibits topologically nontrivial phase
at the point $v/w=0$, since $\nu=2$. So, the system demonstrates
topological flat bands like the $t$-pol. 
The configuration of atomic bonds in 
real space for this particular is similar to the Fig.
\ref{trans-polyacene-edge-modes} (b). However, the
system exhibits an additional topological phase with $\nu=1$,
in comparison with the  $t$-pol as an effect of the
presence of vertical bridging bond. 

Variation of energy spectrum of the model for open
chains with respect to $v/w$ have been
shown in Fig. \ref{trans-bridged-polyacene-energies-obc} (a),
in order to search the zero-energy edge modes in the
topological regimes.
The diagram shows that energy spectra satisfy the
relation, $E_{\rm tb}(-v) \ne E_{\rm tb}(v)$, unlike the $t$-pol. 
Value of  $I_{pr}$ for the individual
mode has been demonstrated according to the top colorbar placed 
in Fig. \ref{trans-bridged-polyacene-energies-obc}.
Redness of the zero-energy modes indicates their higher localization. 
Boundaries of different topological phases have been 
marked by dashed vertical lines. 
\begin{figure}[h]
  \psfrag{a}{(a)}
  \psfrag{b}{(b)}
   \psfrag{c}{(c)}
  \psfrag{d}{(d)}
   \psfrag{p}{\scriptsize {$|\psi|^2$}}  
  \psfrag{Energy}{ \hskip -0.15 cm Energy}
  \psfrag{I}{ \hskip -0.2 cm $I_{pr}$}
  \psfrag{00}{\hskip 0.05 cm $0.0$}
  \psfrag{0.25}{$\hskip -0.05 cm 0.25$}
  \psfrag{0.00}{\hskip -0.05 cm $0.00$}
    \psfrag{0.50}{\hskip -0.1 cm $0.5$}  
\psfrag{0}{$0$}
\psfrag{20}{$20$}
\psfrag{40}{$40$}
\psfrag{60}{$60$}
\psfrag{80}{$80$}
\psfrag{0}{$0$}
\psfrag{1.0}{\hskip -0. cm $1.0$}
\psfrag{0.2}{\hskip -0. cm $0.2$}
\psfrag{-2.4}{\hskip -0.05 cm $-2.4$}
\psfrag{0.5}{\hskip -0. cm $0.5$}
\psfrag{0.0}{\hskip -0.0 cm $0.0$}
\psfrag{-1.2}{\hskip -0.05 cm$-1.2$}
\psfrag{1.2}{\hskip -0.05 cm$1.2$}
\psfrag{0.6}{\hskip -0. cm $0.6$}
\psfrag{-0.6}{\hskip -0. cm $-0.6$}
\psfrag{v}{\hskip -0.15 cm $v=0,\;w=1$}
\psfrag{v1}{\hskip -0.15 cm $v=-1,\;w=1$}
\psfrag{site}{\hskip 0.1 cm site}
\hskip 0.22 cm
\includegraphics[width=230pt]{top-color-box-polyacene.eps}
\vskip -0.0 cm
\includegraphics[width=250pt]{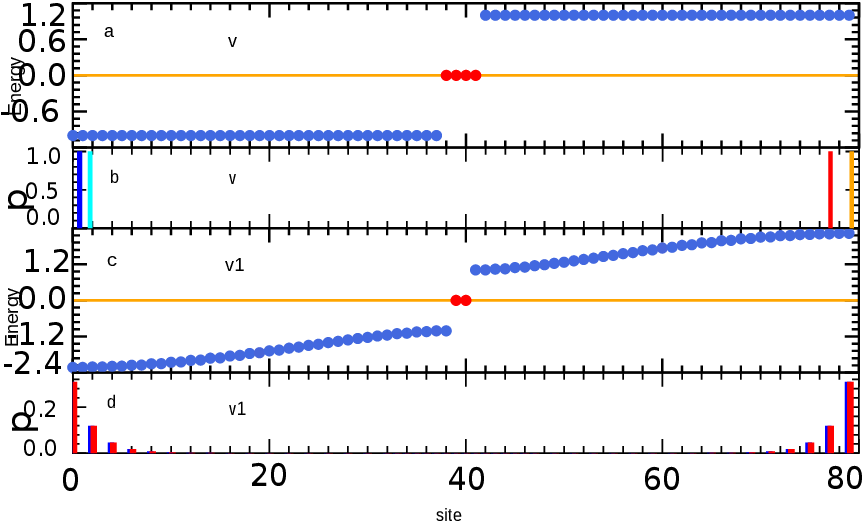}
\caption{Variation of energies in ascending order for $tb$-pol open
  chain when $v=0.0$ (a), and
   $v=-1.0$ (c), when $w=1$.
  Top colorbar indicates the variation of $I_{pr}$.
  Probability density, $|\psi_j|^2$ of edge states are
  shown in the histograms by different colors when $v/w=0.0$ (b), and
   $v/w=-1.0$. (d)}
 \label{trans-bridged-polyacene-edge-energies-set-I-II}
\end{figure}

In order to examine the existence of zero-energy modes and
their nature of localization in the topological phases,
energy spectra of the system at the points $v/w=0$, and $v/w=-1$,
have been demonstrated in
Fig. \ref{trans-bridged-polyacene-edge-energies-set-I-II} (a) and (c),
respectively. Four and two zero-energy modes are found to
exist in the respective topological phases when $\nu=2$, and $\nu=1$,
which is consistent with the bulk-boundary correspondence rule for
chiral systems.
Fig. \ref{trans-bridged-polyacene-edge-energies-set-I-II} (a)
shows the existence of two flat bands at $E_{\rm tb}=\pm w$,
by the points drawn on
two parallel lines for energies $w=\pm 1$,
those are topologically nontrivial. 
Probability density of the zero-energy modes has been
depicted in the histograms for the respective phases in the 
Fig. \ref{trans-bridged-polyacene-edge-energies-set-I-II} (b) and (d),
respectively. Localization of these modes towards the opposite edges
confirm that these states truly represent the edge modes. 
Therefore, $tb$-pol exhibits one additional
topological phase with $\nu=1$, compared to the $t$-pol. 
%%%%%%%%%%%%%%  Section IV %%%%%%%%%%%%%%%%%%%%%%%%%%%%%%%%%%%%%%%%%%%%%%%%%
\section{THE {\em trans}-POLYACENE MODEL with cross-bridging bonds}
\label{trans-cross-bridged}
The Hamiltonian for the $t$-pol structure along with the
additional cross-bridged bond ($tcb$-pol) is given by,
\be H_{\rm {tcb}}=H_{\rm {t}} + v\sum_{j=1}^N \left(a_j^\dag d_j+h.c.\right). 
 \label{trans-cross-bridged-polyacene-ham}
 \ee
\begin{figure}[h]
%\psfrag{A}{A}
%\psfrag{B}{B}
%\psfrag{C}{C}
%\psfrag{D}{D}
%\psfrag{v}{$v$}
%\psfrag{w}{$w$}
\includegraphics[width=240pt]{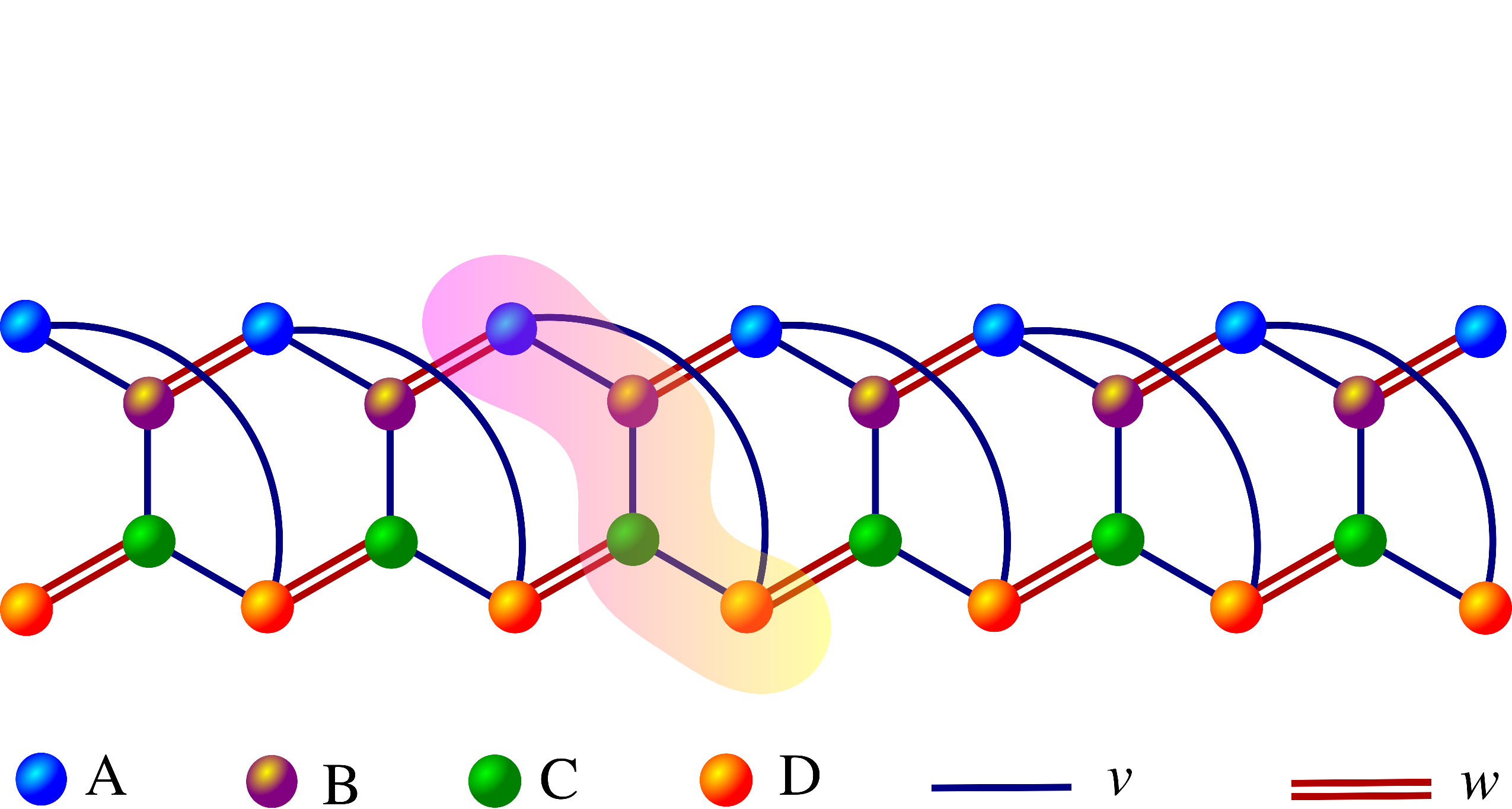}
\caption{Structure of {\em trans}-polyacene with cross-bridged bonds. 
  Carbon atoms marked by A, B, C
  and D form a unit cell. One such unit cell is highlighted.}
 \label{trans-cross-bridged-polyacene}
\end{figure}
The structure of $tcb$-pol has been shown in Fig. \ref{trans-cross-bridged-polyacene},
where the additional bond represents the hopping of strength $v$,
between A and D sites of the same unit cell. 
The system has been expressed in the $k$-space by the Hamiltonian, 
\be
H_{\rm {tcb}}(k)=\left(
\begin{array}{cccc}0&g(k)&0&v\\[0.4em]
  g^*(k) &0&v&0\\[0.4em]
  0&v&0&g(k)\\[0.4em]
  v&0&g^*(k) &0
\end{array} \right).
 \label{Htcbk}
 \ee
%  where $g(k)=v+we^{-ik}$. 
  This system preserves the same set of symmetry like the $t$-pol,
  thus belongs to BDI class. 
 The dispersion relations are given by
 \[E_{\rm {tcb}}=\pm w, \pm\sqrt{4v^2+w^2+4vw\cos{k}},\]
 which always yield
 a pair of flat bands with $E_{\rm {tcb}}=\pm w$.
In addition, these bands become doubly degenerate when $v=0$. 
However, the system yields three flat bands when $w=0$,
with energies $E_{\rm {tcb}}=\pm 2v$,
and $E_{\rm {tcb}}=0$, which is doubly degenerate.  

\begin{figure}[h]
\psfrag{a}{(a)}
\psfrag{b}{(b)}
\psfrag{c}{(c)}
\psfrag{d}{(d)}
\psfrag{p}{$-\pi$}
\psfrag{q}{$-\frac{\pi}{2}$}
\psfrag{n}{$\frac{\pi}{2}$}
\psfrag{m}{$\pi$}
\psfrag{1}{$1$}
\psfrag{0}{$0$}
\psfrag{2}{$2$}
\psfrag{3}{$3$}
\psfrag{-1}{\hskip -0.02 cm$-1$}
\psfrag{-2}{\hskip -0.02 cm$-2$}
\psfrag{-3}{\hskip -0.02 cm$-3$}
\psfrag{k}{$k$}
\psfrag{E}{$E_{\rm tcb}(k)$}
\includegraphics[width=240pt]{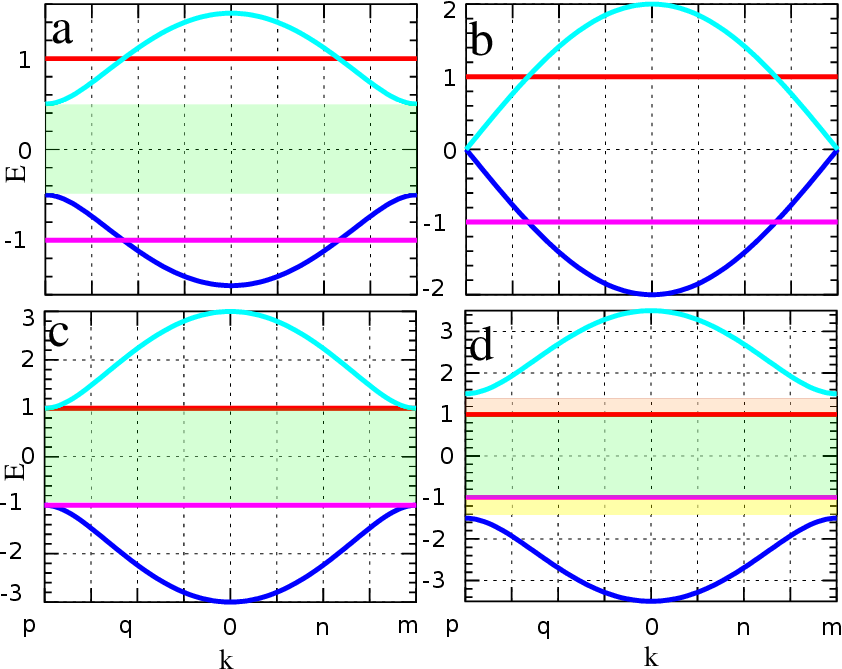}
\caption{Dispersion relations of $tcb$-pol for (a) $v=1/4$, 
  (b)  $v=1/2$,  (c)  $v=1$, and (d)  $v=5/4$, when $w=1$.
  Energy gaps in (a), (c) and (d) are shown
  by shades. Flat bands are shown in red and pink horizontal lines
  in every diagram.}
 \label{trans-cross-bridged-polyacene-dispersions}
 \end{figure}

The eigenvectors, 
$\phi_{\rm tcb}(k)$ are given by 
\[\phi_{\rm tcb}(k)=\left[\frac{E_{\rm tcb}-vx}{g^*}\;\;
  1\;\;x\;\;\frac{E_{\rm tcb}x-v}{g} \right]^{\rm T},\]
where
\[x=\frac{E_{\rm tcb}^2g-|g|^2g+v^2g^*}{vE_{\rm tcb}(g+g^*)}.\]
 Dispersion relations for
 four different parameters have been demonstrated in
 Fig \ref{trans-cross-bridged-polyacene-dispersions} when
 $v=1/4$ (a),  $v=1/2$  (b),  $v=1$  (c), and  $v=5/4$ (d), for $w=1$. 

\begin{figure}[h]
\psfrag{a}{(a)}
\psfrag{b}{(b)}
\psfrag{c}{(c)}
\psfrag{z}{\hskip 0.1 cm $Z_{2,3}$}
  \psfrag{I}{ \hskip -0.2 cm $I_{pr}$}
  \psfrag{00}{\hskip 0.05 cm $0.0$}
  \psfrag{0.25}{$0.25$}
  \psfrag{0.5}{\hskip -0.00 cm $0.5$}
 \psfrag{1.0}{\hskip -0. cm $1.0$} 
\psfrag{E}{ \hskip -0.5 cm Energies}
\psfrag{4}{$4$}
\psfrag{2}{$2$}
\psfrag{1}{$1$}
\psfrag{0}{$0$}
\psfrag{-1}{\hskip -0.14 cm$-1$}
\psfrag{-2}{\hskip -0.14 cm $-2$}
\psfrag{-4}{\hskip -0.14 cm $-4$}
\psfrag{v}{\hskip -0.15 cm $v/w$}
\psfrag{w}{$\nu$}
\hskip 0.16 cm
\includegraphics[width=240pt]{top-color-box-polyacene.eps}
\vskip 0.02 cm
\includegraphics[width=250pt]{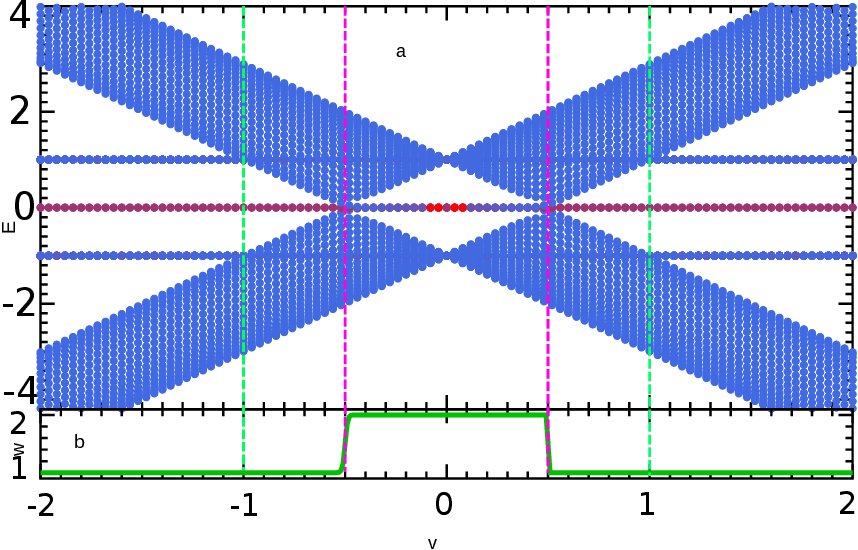}
\caption{Variation of energies for trans-cross-bridged polyacane open chain (a),
   winding number (b)
 have been plotted with respect to $v/w$. 
  Top colorbar indicates the variation of $I_{pr}$.
Vertical dashed lines are drawn at the phase transition points.}
 \label{trans-cross-bridged-polyacene-energies-obc}
\end{figure}

 The energy band diagram of the open system has been shown in
 Fig. \ref{trans-cross-bridged-polyacene-energies-obc} (a), for the
 parameter regime, $-2<v/w<+2$, along with the $I_{pr}$ of each state.
 The diagram indicates that
energy spectra satisfy the relation, $E_{\rm tcb}(-v)=E_{\rm tcb}(v)$. 
 The phase transition points have been indicated by the
 vertical dashed lines. For example, vertical lines (pink dashed)
 drawn at $v/w=\pm 1/2$, 
 correspond to the phase transition between the nontrivial phases
 characterized by $\nu=1$ and $\nu=2$. Band gap vanishes at these points.
 Another pair of vertical lines (green dashed) drawn at $v/w=\pm 1$, 
 where three distinct band gaps open up. The pair of flat bands
 becomes topologically nontrivial beyond that point as discussed later.
 These flat bands are shown by horizontal lines drawn at
 $E_{\rm tcb}=\pm w$. 
 Zero-energy edge modes are found to have larger values of 
 $I_{pr}$ indicating their higher degree of localization.
 
Under the unitary transformation $H'_{\rm tcb}(k)=U^{-1}H_{\rm tcb}(k)U$,
  $H'_{\rm tcb}(k)$ has been converted to  
 block off-diagonal form, as shown below. 
 \be
H'_{\rm tcb}(k)\!=\!\!\left(
\begin{array}{cc}0&h_{\rm tcb}(k)\\[0.4em]
  h_{\rm tcb}^\dag(k)&0
\end{array}\!\! \right)\!,\,{\rm with}\;
h_{\rm tcb}(k)\!=\!\left(
\begin{array}{cc}g^*(k)&v\\[0.4em]
  v&g^*(k)
\end{array}\!\! \right)\!.\nonumber
 \label{Htcbkprime}
 \ee
  Topological phases have been characterized by the 
 winding number as,
 \bea
 \nu&=&\frac{1}{2\pi i}\oint dk \;
    {\rm Tr}\left[ \frac{1}{h_{\rm tcb}(k)}\frac{dh_{\rm tcb}(k)}{dk}\right],\nonumber\\[0.4em]
    &=&  \frac{1}{2\pi i}\oint_{|z|=1} d\log{f_{\rm tcb}(z)}, \nonumber
\eea
where $f_{\rm tcb}(z)=w^2z^2+2vwz$, which has no pole.
The number of zero for $f_{\rm tcb}(z)$, within $|z|=1$, has been given as,
\be
\mathcal N_z=\left\{
\begin{array}{cc}1,& {\rm if}\; v>w/2,\nonumber\\[0.4em]
  2,&{\rm if}\; v<w/2.\nonumber
\end{array} \right.\nonumber\ee
Thus, the Cauchy principle argument leads to
\be \nu =\left\{\begin{array}{cc}1,& {\rm if}\; v>w/2,\nonumber\\[0.4em]
  2,&{\rm if}\; v<w/2.
\end{array} \right.
\ee
As $z=0$, is a root of the function $f_{\rm tcb}(z)$, 
it corresponds to the fact that 
 $\nu$ can never be zero in the entire parameter region. 
 %&=&\frac{1}{\pi}\int_0^{2\pi} dk \;\frac{v+we^{ik}}{2v+we^{ik}},\nonumber\\[0.4em]
% &=&\frac{3}{2}+\frac{1}{2}\left\{
%\begin{array}{cc}\frac{w^2-4v^2}{4v^2-w^2},& {\rm if}\; v>w/2,\nonumber\\[0.4em]
 % \frac{w^2-4v^2}{w^2-4v^2},&{\rm if}\; v<w/2,\nonumber\\[0.4em]
%\end{array} \right.,\nonumber\\[0.4em]
%&=&\left\{\begin{array}{cc}1,& {\rm if}\; v>w/2,\nonumber\\[0.4em]
%  2,&{\rm if}\; v<w/2.
%\end{array} \right.
% \eea
 Variation of winding number with respect to $v/w$ is shown in
 Fig. \ref{trans-cross-bridged-polyacene-energies-obc} (b), by
 green line, which indicates
 that the {\em tcb}-pol always remains topologically nontrivial.
 The system hosts a topological phase with
 $\nu=2$, when $-1/2< v/w < +1/2$, and another one with
 $\nu=1$, when $|v/w| > 1/2$. So, it undergoes
 a topological phase transition at the points $v/w=\pm 1/2$, when
 no band gap is noted as shown in Fig
 \ref{trans-cross-bridged-polyacene-dispersions} (b).
 Single band gap is
 observed in Fig
 \ref{trans-cross-bridged-polyacene-dispersions} (a) and (c), when the
 system exhibits distinct topological phase with  $\nu=2$, and  $\nu=1$,
 respectively. In these two topological regions band gap exists
 between the middle two bands around the Fermi energy, $E=0$, while
 the two top bands and two bottom bands cross each other separately
 yielding no gap upto $v/w=1$. Three distinct band gaps
 are found in this four-band model only when $|v/w|>1$,
 as shown in Fig
 \ref{trans-cross-bridged-polyacene-dispersions} (d).
 
% Variation of energies with $v/w$ for the
% open system has been shown in  Fig
% \ref{trans-cross-bridged-polyacene-energies-obc} (a) along with 
%the $I_{pr}$ of the cortresponding states.

%Existence of flat bands at $E_{\rm {tcb}}=\pm w$, 
%are shown here by blue spheres drawn over the parallel line
%at energies $E_{\rm {tcb}}=\pm 1$.
%Topological zero energy modes are shown by
%reddish spheres since their $I_{pr}$ is higher than
%non-zero energy modes. 
 
\begin{figure}[h]
  \psfrag{a}{(a)}
  \psfrag{b}{(b)}
   \psfrag{c}{(c)}
  \psfrag{d}{(d)}
   \psfrag{p}{\scriptsize {$|\psi|^2$}}  
 % \psfrag{p}{$|\psi|$}  
  \psfrag{Energy}{ \hskip -0.15 cm Energy}
  \psfrag{I}{ \hskip -0.2 cm $I_{pr}$}
  \psfrag{00}{\hskip 0.05 cm $0.0$}
  \psfrag{0.25}{$\hskip -0.05 cm 0.25$}
  \psfrag{0.00}{\hskip -0.05 cm $0.00$}
    \psfrag{0.50}{\hskip -0.1 cm $0.5$}  
\psfrag{0}{$0$}
\psfrag{20}{$20$}
\psfrag{40}{$40$}
\psfrag{60}{$60$}
\psfrag{80}{$80$}
\psfrag{0}{$0$}
\psfrag{1.0}{\hskip -0. cm $1.0$}
\psfrag{2.0}{\hskip -0. cm $2.0$}
\psfrag{-2.0}{\hskip -0.05 cm $-2.0$}
\psfrag{0.5}{\hskip -0. cm $0.5$}
\psfrag{0.0}{\hskip -0.0 cm $0.0$}
\psfrag{-1.4}{\hskip -0.05 cm$-1.4$}
\psfrag{-0.7}{\hskip -0.05 cm $-0.7$}
\psfrag{0.7}{\hskip -0.0 cm $0.7$}
\psfrag{1.4}{\hskip -0.0 cm $1.4$}
\psfrag{v}{\hskip -0.15 cm $v=0.2,\;w=1$}
\psfrag{v1}{\hskip -0.15 cm $v=1.2,\;w=1$}
\psfrag{site}{\hskip 0.2 cm site}
\hskip 0.22 cm
\includegraphics[width=230pt]{top-color-box-polyacene.eps}
\vskip -0.0 cm
\includegraphics[width=250pt]{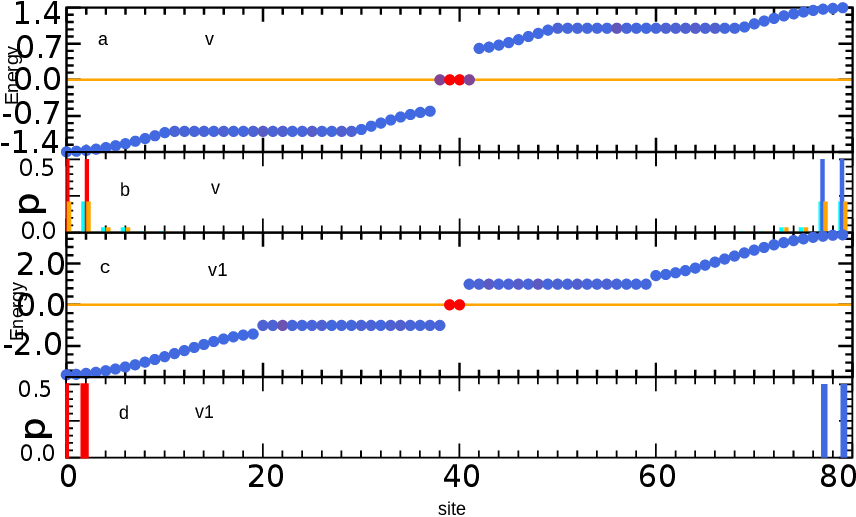}
\caption{Variation of energies in ascending order for $tcb$-pol open
  chain when $v=0.2$ (a), and
   $v=1.2$, when $w=1$ (c).
  Top colorbar indicates the variation of $I_{pr}$.
  Probability density, $|\psi_j|^2$ of edge states are
  shown in the histograms by different colors when $v/w=0.2$ (b), and
   $v/w=1.2$. (d)}
 \label{trans-cross-bridged-polyacene-edge-energies-set-I-II}
\end{figure}
In order to detect the zero-energy edge modes,
energies of open $tcb$-pol chain have been
plotted in ascending order at two different topological
points identified by  $v/w=0.2$, and $v/w=1.2$,
as shown in Fig \ref{trans-cross-bridged-polyacene-edge-energies-set-I-II}
(a) and (c), respectively. Existence of four (two) zero-energy states
supports the presence of topological phases with $\nu=2$ ($\nu=1$),
in accordance to the bulk-edge correspondence rule.
Existence of the four zero-energy modes can be explained
in the real space for $v = 0$, where atoms are connected only by
double bonds as shown in Fig. \ref{trans-polyacene-edge-modes} (b),
leaving four atoms free, 
which is identical to the $t$-pol in its topological regime. 
Probability densities of the zero-energy modes have been plotted in
Fig \ref{trans-cross-bridged-polyacene-edge-energies-set-I-II}
(b) and (d), where histograms of site-wise $|\psi|^2$ for 
four and two edge modes are drawn when $v/w=0.2$, and $v/w=1.2$, respectively.
 Nonetheless, the system exhibits nontrivial flat bands
 with energies $E_{\rm tcb}=\pm w$,
 (shown by blue spheres in Fig.
 \ref{trans-cross-bridged-polyacene-energies-obc} (a))
 with winding number, $\nu=1$, when $|v/w| > 1/2$ and
 with $\nu=2$, elsewhere. 

Comparing the three models, $t$-pol, $tb$-pol and $tcb$-pol, it appears that
although these systems preserve the same set of symmetries,
their topological character is different. $t$-pol is found to
exhibit two different phases with $\nu=0$, and $\nu=2$,
while $tb$-pol hosts three different phases with  $\nu=0$,
$\nu=1$, and $\nu=2$. On the other hand,  $tcb$-pol
exhibits two different phases $\nu=1$, and $\nu=2$. When $v=0$,
all the models exhibit a pair of doubly degenerate topological flat bands
with $\nu=2$. Four zero-energy edge modes are found to
appear on four extreme sites as shown in
Fig. \ref{trans-polyacene-edge-modes}, in this case.
Energy spectrum is found symmetric around $v=0$ for
$t$-pol and $tcb$-pol, but non-symmetric for $tb$-pol.
In the subsequent sections, topological properties of
four different models based on $cis$-polyacene will be
discussed. 
%%%%%%%%%%%%%%  Section V %%%%%%%%%%%%%%%%%%%%%%%%%%%%%%%%%%%%%%%%%%%%%%%%%
\section{THE {\em cis}-POLYACENE MODEL}
\label{cis}
\begin{figure}[h]
\psfrag{A}{A}
\psfrag{B}{B}
\psfrag{C}{C}
\psfrag{D}{D}
\psfrag{v}{$v$}
\psfrag{w}{$w$}
\psfrag{M}{M}
\includegraphics[width=230pt]{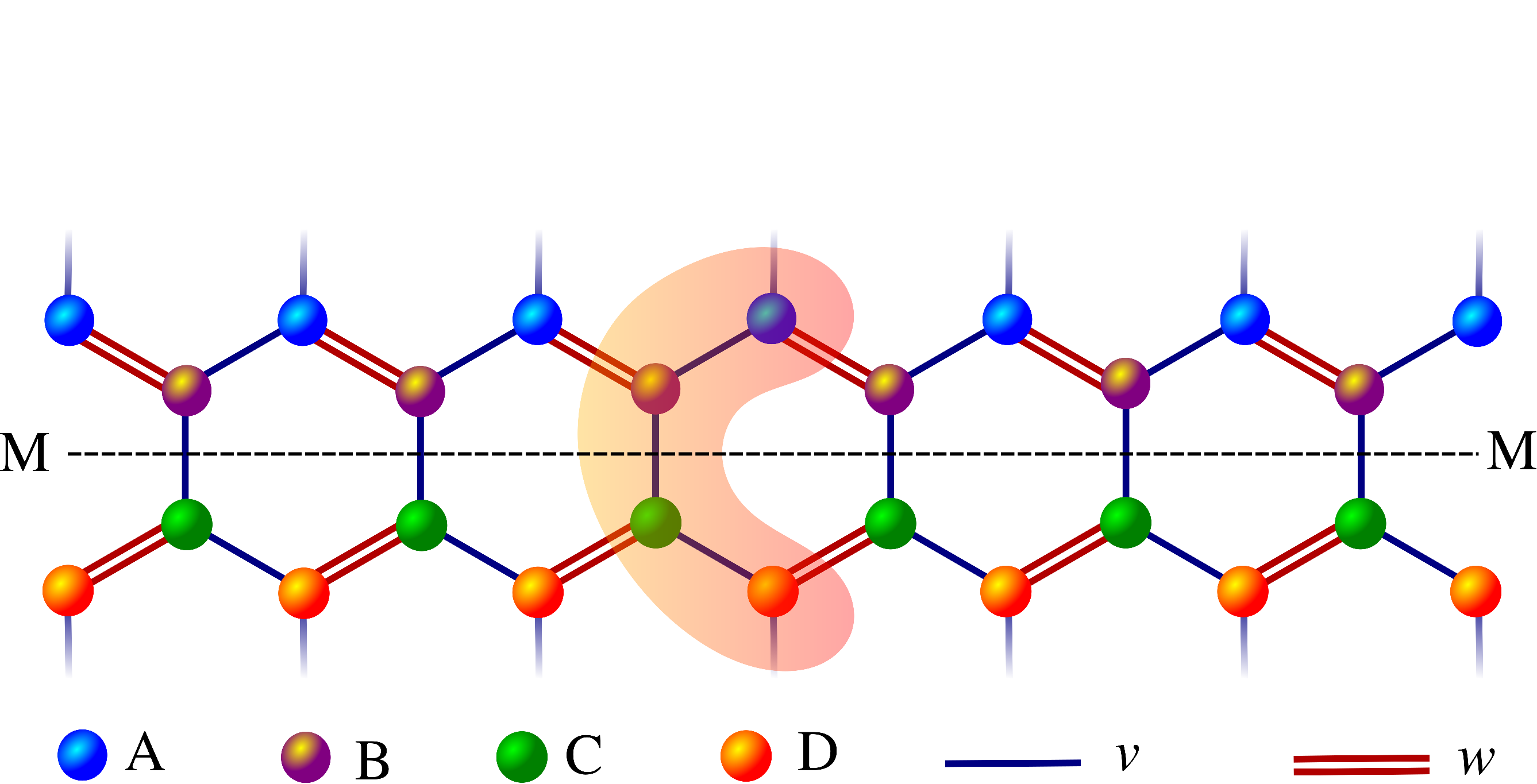}
\caption{Structure of {\em cis}-polyacene.
  Carbon atoms marked by A, B, C
  and D form a unit cell. One such unit cell is highlighted. 
Mirror plane is indicated by dashed horizontal line MM.}
 \label{cis-polyacene}
 \end{figure}

The structure of $c$-pol is shown in Fig. \ref{cis-polyacene},   
where the shape of unit cell is drawn in a different way.
Obviously, property of a system does not depend on
the choice of unit cell, however, adjacent unit cells are connected by
double bonds which is similar to that of $t$-pol. 
The TB Hamiltonian for the $c$-pol structure is
given by,
\bea H_{\rm c}&=&v\sum_{j=1}^N \left(a_j^\dag b_j+b_j^\dag c_j+c_j^\dag d_j \right)\nonumber \\
&\;\;+&w\sum_{j=1}^{N-1}
 \left(a_j^\dag b_{j+1}+d_j^\dag c_{j+1}\right)+h.c.,
 \label{cis-polyacene-ham}
 \eea
 where the symbols have their earlier meaning. 
 
 Hamiltonian in the $k$-space is written as
 \[H_{\rm c}=\sum_{k\in{\rm {BZ}}}\Psi_k^\dag H_{\rm c}(k)\Psi_k,\]
 where 
 \be H_{\rm c}(k)=\left(
\begin{array}{cccc}0&g^*(k)&0&0\\[0.4em]
  g(k) &0&v&0\\[0.4em]
  0&v&0&g(k)\\[0.4em]
  0&0&g^*(k) &0
\end{array} \right).
\label{Hck}  \ee
%where $g(k)=v+we^{-ik}$. 
% where $H_{\rm c}(k)=H_{\rm t}^\dag(k)$. 
Eigenvectors  of $H_{\rm c}(k)$ are given by, 
\[\phi_{\rm c}(k)=\left[\frac{vg^*}{E_{\rm c}^2-|g|^2}\;\;
  \frac{vE_{\rm c}}{E_{\rm c}^2-|g|^2}\;\;1\;\;\frac{g^*}{E_{\rm c}} \right]^{\rm T},\]
while eigenenergies are the same as that of $H_{\rm t}$. 
So, under PBC dispersion relations of both $t$-pol and
$c$-pol are the same. 
On the other hand, without PBC, eigenenergies are
different which will be explained later.
At the same time, estimated  value of 
band gap for $c$-pol by EHCO method is 0.002 eV,
which is much lower than that of $t$-pol \cite{Whangbo}. 

\begin{figure}[h]
  \psfrag{I}{ \hskip -0.2 cm $I_{pr}$}
  \psfrag{00}{\hskip 0.05 cm $0.0$}
  \psfrag{0.25}{$0.25$}
  \psfrag{0.5}{\hskip -0.00 cm $0.5$}
 \psfrag{1.0}{\hskip -0. cm $1.0$} 
\psfrag{Energy}{ \hskip -0.5 cm Energies}
\psfrag{4}{$4$}
\psfrag{2}{$2$}
\psfrag{1}{$1$}
\psfrag{0}{$0$}
\psfrag{-1}{\hskip -0.14 cm$-1$}
\psfrag{-2}{\hskip -0.14 cm $-2$}
\psfrag{-4}{\hskip -0.14 cm $-4$}
\psfrag{v}{\hskip -0.15 cm $v/w$}
\psfrag{w}{$w$}
\hskip 0.16 cm
\includegraphics[width=230pt]{top-color-box-polyacene.eps}
\vskip -0.1 cm
\includegraphics[width=250pt]{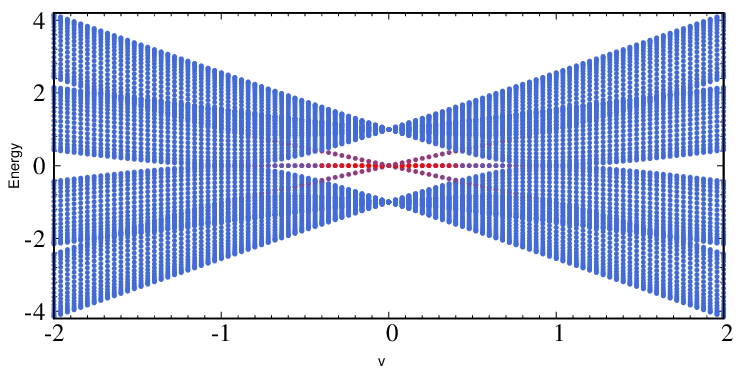}
\caption{Variation of energies with $v/w$ of {\em cis} polyacene model.
Top colorbar shows the variation of $I_{pr}$.}
 \label{cis-polyacene-energies}
\end{figure}

\begin{figure}[h]
  \psfrag{a}{(a)}
  \psfrag{b}{(b)}
   \psfrag{p}{\scriptsize {$|\psi|^2$}}  
%  \psfrag{p}{$|\psi|$}  
  \psfrag{Energy}{ \hskip -0.15 cm Energies}
  \psfrag{I}{ \hskip -0.2 cm $I_{pr}$}
  \psfrag{00}{\hskip 0.05 cm $0.0$}
  \psfrag{0.25}{$0.25$}
    \psfrag{0.50}{\hskip -0.00 cm $0.5$}  
\psfrag{0}{$0$}
\psfrag{20}{$20$}
\psfrag{40}{$40$}
\psfrag{60}{$60$}
\psfrag{80}{$80$}
\psfrag{0}{$0$}
\psfrag{1.0}{\hskip -0. cm $1.0$}
\psfrag{0.5}{\hskip -0. cm $0.5$}
\psfrag{0.0}{\hskip -0.0 cm $0.0$}
\psfrag{-1.4}{\hskip -0.05 cm$-1.4$}
\psfrag{-0.7}{\hskip -0.05 cm $-0.7$}
\psfrag{0.7}{\hskip -0.0 cm $0.7$}
\psfrag{1.4}{\hskip -0.0 cm $1.4$}
\psfrag{v}{\hskip -0.15 cm $v/w=0.2$}
\psfrag{site}{\hskip 0.2 cm site}
\hskip 0.22 cm
\includegraphics[width=230pt]{top-color-box-polyacene.eps}
\vskip -0.0 cm
\includegraphics[width=250pt]{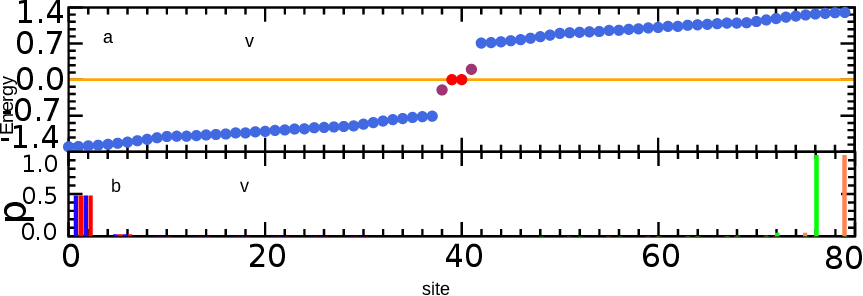}
\caption{Variation of energies  for {\em cis}-polyacene
  model is plotted in the ascending order when $v/w=0.2$ (a).
  It shows the presence of four edge
  states. Among them two are at the zero energy,
  one below the zero energy and one above the zero energy. 
  Top colorbar indicates the variation of $I_{pr}$.
  Probability amplitude of four edge state energies are
  shown in the histogram by different colors (b). }
 \label{cis-polyacene-edge-energies}
\end{figure}

This Hamiltonian preserves TRS, CS and PHS,
like the $t$-pol, but, it does not preserve the IS unlike
the $t$-pol, since
\[\mathit  \mathcal T_{xx}^{-1} H_{\rm t}( k)\mathit \mathcal T_{ xx}\ne H_{\rm t}(-k).\]
  Winding number in this case can be expressed as
 \be
 \nu=\frac{1}{2\pi i}\oint dk \, {\rm Tr}\left[ \frac{1}{h_{\rm c}(k)}\frac{dh_{\rm c}(k)}{dk}\right],\ee
 where,
 \[h_{\rm c}(k)\!=\!\left(
\begin{array}{cc}g^*(k)&0\\[0.4em]
  v&g(k)
\end{array} \right)\!.\]
However, after simplification, 
  \bea
  \nu &=&i\,\frac{vw}{\pi}\int_0^{2\pi} dk\; \frac{\sin{(k)}}{v^2+w^2+2vw\cos{(k)}},\nonumber\\[0.4em]
 &=&0. \label{winding-cis}
 \eea
%since  $\mathcal T_z^{-1} H_{\rm c}( k) \mathcal T_z\ne -H_{\rm c}(k)$, and 
% $\mathcal P^{-1} H_{\rm c}( k) \mathcal P\ne -H_{\rm c}( -k)$. 
%In addition, $\mathcal T_x^{-1} H_{\rm c}( k) \mathcal T_x\ne -H_{\rm c}(k)$,
%where  $\mathcal T_x=I \otimes \sigma_x$, and
%$\sigma_x$ is the $x$-component of Pauli matrix. 
%As a result, it does not conserve the  
%the inversion symmetry (IS). So, this system
%is bound to be topologically trivial, according to the
% tenfold classification scheme \cite{Ryu}.  
Although the system is non-topological, it exhibits coexistence of
 zero- and nonzero-energy
edge states in the region, $-1< v/w < +1$ as shown in
Fig. \ref{cis-polyacene-energies}. 
%, despite it lacks CS, PHS and IS.
In order to confirm the presence of edge states,
eigenenergies of the open chain with 80 sites
have been computed when $v=0.2$, and $w=1$. The 
energy spectrum of $c$-pol has been plotted in the
ascending order as shown in Fig. \ref{cis-polyacene-edge-energies} (a). 
It reveals that the system exhibits one pair of edge states each for
zero and nonzero energy in the region, $-1< v/w < +1$.
Property of those states has been further examined in terms of
$I_{pr}$, as well as by probability density per site.
$I_{pr}$ of those states are very high, which indicates
their strong localization. Distribution of $|\psi_j|^2$
confirms their localization near the both edges, which is 
shown in Fig. \ref{cis-polyacene-edge-energies} (b).
So, it further confirms that, eigenenergies of $c$-pol is almost similar
to that of $t$-pol, apart from the existence of
nonzero-energy edge states. Or, eventhough the energy dispersions under
PBC is identical for the $t$-pol and $c$-pol, feature of
edge modes under OBC is different.  

It has been mentioned before that according to
bulk-boundary correspondence rule %zero-energy 
zero- and nonzero-energy
edge states are the signature of topological states
belonging to the systems having CS  
and IS, respectively. This behaviour has been noted before in
SSH trimer model \cite{Alvarez,Du,Bomantara}.
However, the SSH trimer model exhibits a pair of
degenerate nonzero-energy edge modes in each case. 
Whereas, in this case, a pair of nondegenerate edge modes appear
with energy, $\pm v$, as shown in Fig. \ref{cis-polyacene-edge-energies} (a). 
%In this sense the system exhibits anomalous
%behaviour because of the fact that it simultaneously
%hosts zero- and nonzero-energy
%edge states which are not protected by %CS. 
%either CS or IS. 
%In other words it means although CS and IS are absent, their
%remnants still persist.
It is true that topological nontrivaility in 1D is always protected by
TRS, CS and IS, but on the other hand, presence of the TRS, CS or IS
does not ensure the nontriviality. 
%So, it seems that 
%the bulk-boundary correspondence rule has been broken in the other way,
%where edge states are present in the topologically trivial phase.
Although the zero- and nonzero-energy edge modes are protected by the
symmetries corresponding to BDI class, they are no more
topological, since both winding and Zak numbers are
simultaneously zero. 
The simultaneous presence of both zero- and
nonzero-energy edge states in a trivial system is very rare. 

On the other hand, Hamiltonian,
Eq \ref{cis-polyacene-ham}, possesses an additional symmetry 
in a sense that upon simultaneous interchanging of
the sublattice operators, $(a\leftrightarrow d)$,
and $(b\leftrightarrow c)$, the real-space 
Hamiltonian, $H_{\rm c}$ is found unchanged. 
After a closer look into the the geometrical structure in 
Fig \ref{cis-polyacene}, it turns out that this is actually a 
mirror symmetry about
the plane passing through the line MM. 
Mathematically this operation can be accomplished by the
operator, $\mathcal M_{(b\leftrightarrow c)}^{(a\leftrightarrow d)}$,
where  $H_{\rm c}$ satisfies the relation,
\be \left(\mathcal M_{(b\leftrightarrow c)}^{(a\leftrightarrow d)}\right)^{-1}
\!\!H_{\rm c}\,\mathcal M_{(b\leftrightarrow c)}^{(a\leftrightarrow d)}=H_{\rm c}.
\label{mirror}
\ee

In order to explain the existence of peculiar edge modes in the
topologically trivial system as 
shown in Fig. \ref{cis-polyacene-energies}, and in 
Fig. \ref{cis-polyacene-edge-energies} (a), 
Hamiltonian for $c$-pol has been rotated under a 
specific intracell operator transformations, which is
accomplished in the real space
in accordance to the MS of the Hamiltonian based on Eq. \ref{mirror}. 
The transformations in terms of new real space operators,
$\{\alpha^\pm_j,\beta^\pm_j\}$, within the $j$-th unit cell are
given by symmetric and antisymmetric combinations as, 
\be
\left\{\begin{array}{l}
\alpha^\pm_j=\dfrac{1}{\sqrt 2}\left(b_j\pm c_j \right),\\[0.5em]
\beta^\pm_j=\dfrac{1}{\sqrt 2}\left(a_j\pm d_j \right),
\end{array}\right.
\label{transformation}
\ee
which leads to
\bea H'_{\rm {c}}&=&\sum_{s=\pm} \left[H^{s}_{\rm {ssh}}+ sv\sum_{j=1}^N\alpha_j^{s\dag} \alpha_j^{s}\right],\;{\rm where}\nonumber \\[0.4em]
H^{s}_{\rm {ssh}}&=&\sum_{j=1}^N \left[v\,\alpha_j^{s\dag} \beta_j^{s}
+w\,\beta_j^{s\dag} \alpha_{j+1}^{s}+h.c.\right]\!.
 \label{cis-polyacene-ham-rotated}
 \eea
\begin{figure}[h]
\psfrag{A}{$\alpha^+$}
\psfrag{B}{$\beta^+$}
\psfrag{C}{$\alpha^-$}
\psfrag{D}{{$\beta^-$}}
\psfrag{v}{$v$}
\psfrag{w}{$w$}
\psfrag{a}{(a)}
\psfrag{b}{(b)}
\includegraphics[width=230pt]{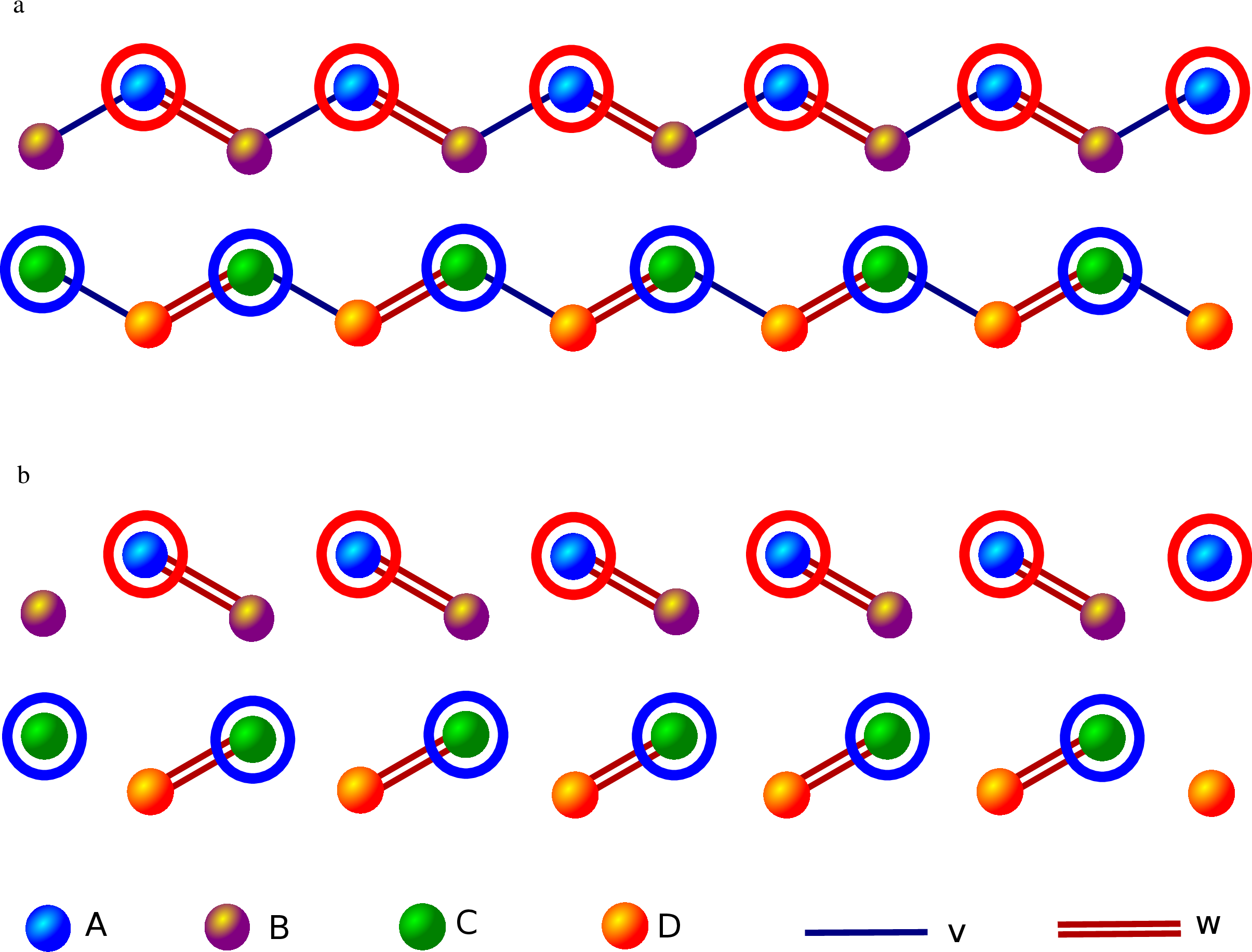}
\caption{Schematic view of the Hamiltonian for
  transformed {\em cis}-polyacene,
  $H'_{\rm {c}}$ (Eq. \ref{cis-polyacene-ham-rotated}),
  where two independent SSH chains 
have been shown separately by the sites ($\alpha^+$, $\beta^+$),
  and ($\alpha^-$, $\beta^-$) (a).
  The same structure when $v=0$ is shown in (b),
  with isolated dimers around the bond of strength $w$, 
  and four free sites with one at each four ends
  with different onsite potentials.}
 \label{cis-polyacene-rotated}
 \end{figure} 
It means the Hamiltonian ($H_{\rm {c}}$) in Eq. \ref{cis-polyacene-ham}
has been decoupled into a pair of identical SSH chains in the presence of 
non-uniform onsite potential with opposite signs, $\pm v$, for the
respective sublattices corresponding to operators $\alpha_j^{\pm}$.
It is pointed out earlier that polyacene chain can be viewed as
two coupled polyacetylene chains, but under this transformation
polyacene can be regarded as two decoupled polyacetylene chains
with carbon atoms of different kinds. 
Pictorial view of the transformed $c$-pol structure has been shown in
Fig. \ref{cis-polyacene-rotated} (a), where pair of isolated SSH chains 
have been constituted separately by the sites ($\alpha^+$, $\beta^+$),
and ($\alpha^-$, $\beta^-$), in the rotated space.
The onsite potential for $\alpha^+$
($\alpha^-$) is $+ v$ ($-v$), which are shown by red (blue) circles, while
that for other sites is zero. 
It is worth to mention that transformations
given in Eq. \ref{transformation}, can likewise be formulated 
in the $k$-space also. 

The transformed  Hamiltonians, 
$H^{\prime\pm}_{\rm {c}}=H^{\pm}_{\rm {ssh}}\pm v\sum_{j=1}^N \alpha_j^{\pm \dag} \alpha_j^{\pm}$, in the $k$-space 
can be written in terms of a $2\times 2$ matrix
for the dimer unit cell as:  
\be H^{\prime\pm}_{\rm {c}}=\sum_{k\in{\rm {BZ}}}\psi_k^{\pm \dag} H^{\pm}_{\rm {c}}(k)\psi^{\pm}_k.\ee
The state vector is $\psi^{\pm}_k=[\alpha^{\pm}_k\;\;\beta^{\pm}_k]^{\rm T}$, 
and
  \be H^{\pm}_{\rm {c}}(k)=\mathbf h^{\pm}\cdot {\boldsymbol \sigma},\label{SSH-k}
  \ee
  where 
  \[\left\{\begin{array}{l}h^{\pm}_0( k)=\pm v/2, \\[0.3em]
  h^{\pm}_1( k)=v+w\cos{( k)}, \\[0.3em]
      h^{\pm}_2( k)=w\sin{( k)}, \\[0.3em]
       h^{\pm}_3( k)=\pm v/2,
         \end{array}\right. \]
         and the $\sigma$ matrices are 
         \[\sigma_0 \!=\! \left(\begin{array}{cc}1 &\;0\\
    0&\;1\end{array}\! \right)\!,\;
         \sigma_1 \!=\! \left(\begin{array}{cc}0 &\;1\\
    1&\;0\end{array} \!\right)\!,\;
      \sigma_2\!=\! \left(\begin{array}{cr}0 &- i\\
        i&\,0\end{array}\! \right)\!,\;
      \sigma_3\!=\! \left(\begin{array}{cr}1 &0\\
        0&-1\end{array}\! \right)\!. \]
It shows that, the decoupled Hamiltonians, $H^{\pm}_{\rm {c}}(k)$,
clearly break the CS in the two-dimensional space, 
since $h^{\pm}_3\ne 0$.
This result is consistent with the earlier finding of winding
number as given in Eq. \ref{winding-cis}.
As a consequence, system exhibits
topological triviality, eventhough zero- and
nonzero-energy edge modes are present.
The existence of edge modes for the open chain can be explained
easily, by using the fact that 
$H^{\pm}_{\rm {ssh}}$ and $\pm v\sum_{j=1}^N \alpha_j^{\pm \dag} \alpha_j^{\pm}$,
in Eq. \ref{cis-polyacene-ham-rotated}
contribute zero-energy and nonzero-energy 
edge modes, respectively, as shown in
Fig. \ref{cis-polyacene-energies},
and in Fig. \ref{cis-polyacene-edge-energies} (a).
This feature can be explained with the help of the
Fig. \ref{cis-polyacene-rotated} (b), in which the schematic view of the
Hamiltonian, $H'_{\rm c}$ (Eq. \ref{cis-polyacene-ham-rotated})
has been depicted for the extreme limit, $v=0$.
It shows that the Hamiltonian at this point
is constituted by isolated dimers around the bond of strength $w$ along with
four free sites with one at each four ends of SSH chains,
among which two are without potential and other two
have nonzero potentials as shown by blue and red circles.
In the parameter region, $-1<v/w<1$, localized states are found to 
appear around the four ends of the chain as depicted in
Fig. \ref{cis-polyacene-energies}, and \ref{cis-polyacene-edge-energies}.
As a result, system exhibits two
zero-energy edge modes and two nonzero-edge modes with energy $\pm v$,
within the region, $-1<v/w<1$. 
%%%%%%%%%%%%%%  Section VI %%%%%%%%%%%%%%%%%%%%%%%%%%%%%%%%%%%%%%%%%%%%%%%%%
\section{THE {\em cis}-POLYACENE MODEL with bridging bonds}
\label{cis-bridged}
\begin{figure}[h]
\psfrag{A}{A}
\psfrag{B}{B}
\psfrag{C}{C}
\psfrag{D}{D}
\psfrag{v}{$v$}
\psfrag{w}{$w$}
\includegraphics[width=230pt]{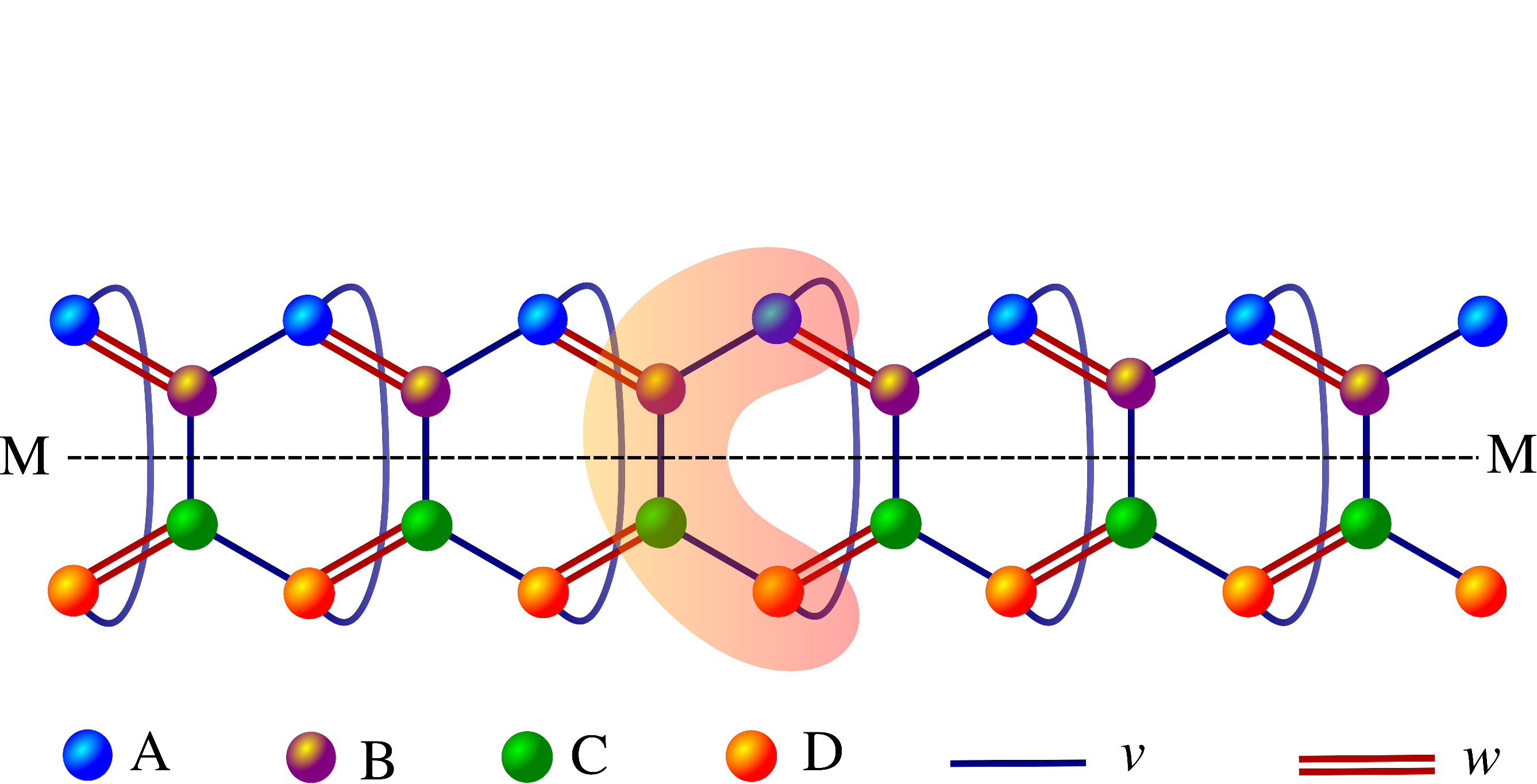}
\caption{Structure of $cb$-pol.  
  Carbon atoms marked by A, B, C
  and D form a unit cell. One such unit cell is highlighted.
Mirror plane is indicated by dashed line MM.}
 \label{achiral-cis-polyacene}
\end{figure}

In order to induce the nontriviality within $c$-pol, the primary structure
of $c$-pol has been modified by introducing an additional
intracell hopping path connecting A and D sublattice points
whose strength is the same to that of single bond ($v$). 
This structure is referred as the $cb$-pol, where
%because of the fact that 
this modification restores the inversion symmetry,
which was absent in the $c$-pol. 
%since 
%while keeping the chiral symmetry in abeyance. 
This modified structure has been shown in Fig \ref{achiral-cis-polyacene},
which preserves the same mirror symmetry of the original $c$-pol structure.
Additionally, this structure acquires the intracell rotational symmetry,
due to this extra hopping term. 

The Hamiltonian for this structure is given by
\be H_{\rm {cb}}=H_{\rm {c}} + v\sum_{j=1}^N \left(a_j^\dag d_j+h.c.\right),
 \label{achiral-cis-polyacene-ham}
 \ee
 which satisfies the relation,
\[\left(\mathcal M_{(b\leftrightarrow c)}^{(a\leftrightarrow d)}\right)^{-1}
\!\!H_{\rm cb}\,\mathcal M_{(b\leftrightarrow c)}^{(a\leftrightarrow d)}=H_{\rm cb}.\]
In $k$-space, it assumes the form
 \be
H_{\rm {cb}}(k)=\left(
\begin{array}{cccc}0&g^*(k)&0&v\\[0.4em]
  g(k) &0&v&0\\[0.4em]
  0&v&0&g(k)\\[0.4em]
  v&0&g^*(k) &0
\end{array} \right), 
 \label{Hcbk}
 \ee
 where $g(k)=v+we^{-ik}$. 
 However, in this case $H_{\rm {cb}}(k)$ preserves the
 IS in the $k$-space, since
 \[\mathcal T_{x}^{-1} H_{\rm cb}( k) \mathcal T_{ x}=H_{\rm cb}(-k),\]
 where $ \mathcal T_{x}=I\otimes \sigma_x$,
 which is different from $\mathcal T_{xx}$.
 It is worth to mention that $\mathcal T_{xx}$
 has been employed before in order to
 study the IS for $t$-pol, $tb$-pol and $tcb$-pol.
Obviously, operation of $ \mathcal T_{x}$ can be explained in the real space  
  by the operator, $\mathscr{ I}'=I'_{N/4}\otimes
 \mathcal T_{x}$.
 The inversion under $\mathscr{ I}'$ is accomplished by the
 transformations:
 \be
\left\{
\begin{array}{l}a_j \rightarrow b_{N+1-j},\\[0.4em]
b_j \rightarrow a_{N+1-j}, \\[0.4em]
 c_j \rightarrow d_{N+1-j}, \\[0.4em]
d_j \rightarrow c_{N+1-j}. 
\end{array} \right.
\label{transform2}
\ee
The transformation laid out in Eq. \ref{transform2} is
different from that in Eq. \ref{transform1}. Also this
transformation is different from
that of MS defined in Eq. \ref{mirror}. 

 Eigenenergies  of $H_{\rm cb}(k)$, are given by,
\be E_{\rm cb}(k)=\pm \sqrt{\frac{p_{\rm cb}\pm\sqrt{q_{\rm cb}}}{2}},
   \label{Eack}  \ee
  where
  $p_{\rm cb}= 4v^2+2w^2+4vw\cos{k}$, and
  $q_{\rm cb}=16v^4+16v^2w^2+32v^3w\cos{k}$,  
while the corresponding eigenvectors are 
\[\phi_{\rm cb}(k)=\left[\beta\;\;
  \frac{g\beta+v}{E_{\rm cb}}\;\;1\;\;\frac{E^2_{\rm cb}-vg\beta-v^2}{E_{\rm cb}g} \right]^{\rm T},\]
where
\[\beta=\frac{v^3-v(E^2_{\rm cb}+|g|^2)}{g^*g^2-g(E^2_{\rm cb}+v^2)}.\]

\begin{figure}[h]
\psfrag{a}{(a)}
\psfrag{b}{(b)}
\psfrag{c}{(c)}
\psfrag{d}{(d)}
\psfrag{p}{$-\pi$}
\psfrag{q}{$-\frac{\pi}{2}$}
\psfrag{n}{$\frac{\pi}{2}$}
\psfrag{m}{$\pi$}
\psfrag{1}{$1$}
\psfrag{0}{$0$}
\psfrag{-1}{\hskip -0.00 cm$-1$}
\psfrag{k}{$k$}
\psfrag{E}{$E(k)$}
\includegraphics[width=240pt]{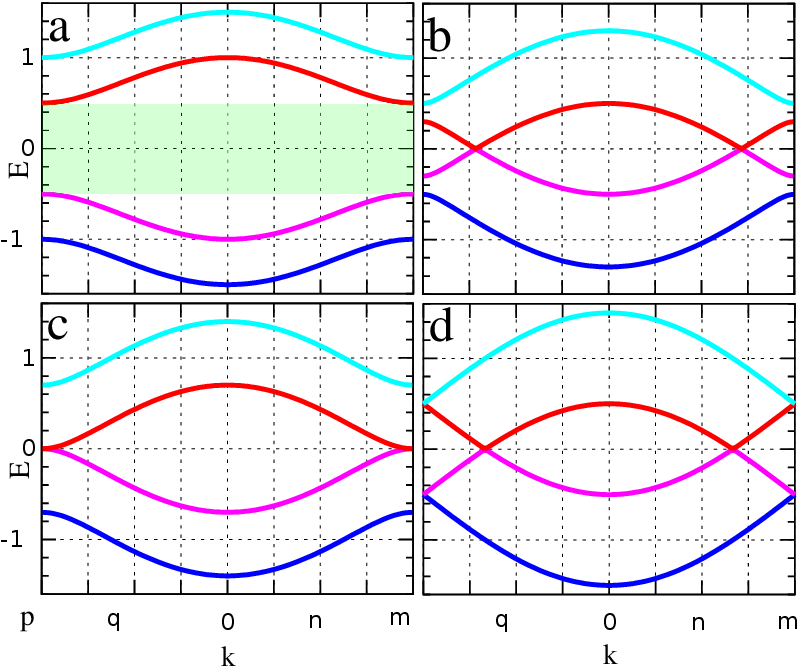}
\caption{Dispersion relations for (a) $v=1/4$, $w=1$; 
   (b)  $v=0.4$, $w=1/2$;  (c)  $v=0.35$, $w=0.7$; and (d)  $v=1/2$, $w=1/2$; 
  of $cb$-pol model. Energy gap in (a) is
shown by green shade.}
 \label{achiral-cis-polyacene-dispersions}
\end{figure}

The system exhibits a pair of doubly degenerate flat bands, 
$E_{\rm cb}=\pm w$, when $v=0$, which is similar to the earlier models.
In contrast, when $w=0$, it yields three flat bands with doubly
degenerate $E_{\rm cb}=0$, and nondegenerate $E_{\rm cb}=\pm 2v$. 
Dispersion relations are shown in
Fig \ref{achiral-cis-polyacene-dispersions} (a), (b), (c) and
(d) for four different sets of parameters. 
The system exhibits true band gap
when $v=1/4$, $w=1$, as depicted in (a)
with a green shade. In this case, four energy bands are separated. 
However in (c) the middle bands touch at the zone boundary,
for $v=0.35$, $w=0.7$, or when $v/w=1/2$. With the increase of
$v/w$, middle bands cross each other twice
as shown in (b) for $v=0.4$, $w=1/2$.
The top and bottom bands are
still separated from the middle bands as shown in (b) and (c),
even though the true band gap is missing in the band structure.
%This phase is known as the semi-metallic phase \cite{Sil3}. 
Finally, these pseudo gaps are found to disappear at $v=1/2$, $w=1/2$,
or when $v/w=1$, as shown in Fig \ref{achiral-cis-polyacene-dispersions} (d).

\begin{figure}[h]
\psfrag{a}{(a)}
\psfrag{b}{(b)}
\psfrag{c}{(c)}
\psfrag{d}{(d)}
\psfrag{e}{(e)}
\psfrag{z1}{\hskip 0.1 cm $Z_1$}
\psfrag{z2}{\hskip 0.1 cm $Z_2$}
\psfrag{z3}{\hskip 0.1 cm $Z_3$}
\psfrag{z4}{\hskip 0.1 cm$Z_4$}
  \psfrag{I}{ \hskip -0.2 cm $I_{pr}$}
  \psfrag{00}{\hskip 0.05 cm $0.0$}
  \psfrag{0.25}{$0.25$}
  \psfrag{0.5}{\hskip -0.00 cm $0.5$}
 \psfrag{1.0}{\hskip -0. cm $1.0$} 
\psfrag{Energy}{ \hskip -0.5 cm Energies}
\psfrag{4}{$4$}
\psfrag{2}{$2$}
\psfrag{1}{$1$}
\psfrag{0}{$0$}
\psfrag{-1}{\hskip -0.14 cm$-1$}
\psfrag{-2}{\hskip -0.14 cm $-2$}
\psfrag{-4}{\hskip -0.14 cm $-4$}
\psfrag{v}{\hskip -0.15 cm $v/w$}
\psfrag{w}{$w$}
\hskip 0.16 cm
\includegraphics[width=240pt]{top-color-box-polyacene.eps}
\vskip 0.02 cm
\includegraphics[width=250pt]{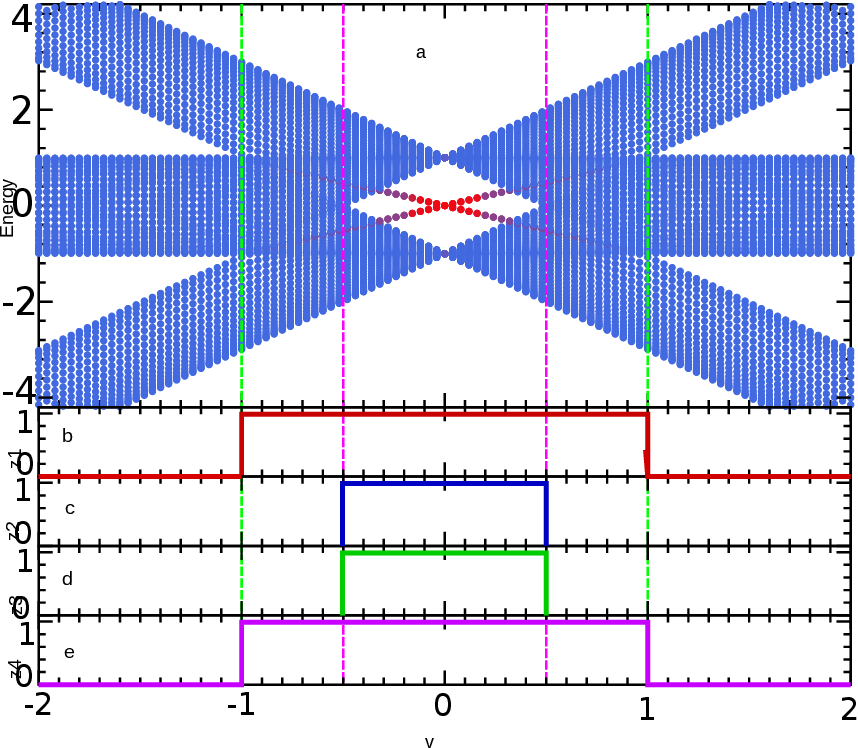}
\caption{Variation of energies for $cb$-pol open chain (a),
  and Zak numbers of four bands,
  $Z_1$ (top) (b),  $Z_2$ (middle upper) (c),
   $Z_3$ (middle lower) (d) and  $Z_4$ (bottom) (e)
  with respect to $v/w$. 
  Top colorbar indicates the variation of $I_{pr}$.
   Vertical dashed lines are drawn at the points $v/w=\pm 1/2$, and
  $\pm 1$.}
 \label{achiral-cis-polyacene-energies-obc}
\end{figure}

In order to investigate the feature of edge states,
eigenenergies of the open system
have been plotted with respect to $v/w$ as shown in
Fig. \ref{achiral-cis-polyacene-energies-obc} (a),
as long as $|v/w|<2$. For this purpose,
Hamiltonian matrix of the real space Hamiltonian 
(Eq. \ref{achiral-cis-polyacene-ham}), has been
spanned for $N=80$ sites. The resulting matrix has been diagonalized
in order to obtain the eigenenergies and 
eigenstates $\psi$, where $\psi_j$
is the probability amplitude of $\psi$ on the site $j$. 
It is observed that nonzero-energy edge states exist 
within  $-1< v/w < +1$, while they vanish beyond $|v/w|>1$. 
The localization of eigenvectors is further examined by
estimating the value of $I_{pr}$. 
Colorbar indicates that value of 
$I_{pr}$ for the nonzero-energy edge states which 
is much higher than that for the other energy states. 
It confirms the fact that nonzero-energy states are also highly localized,
like the zero-energy edge states.
\begin{figure}[h]
  \psfrag{a}{(a)}
  \psfrag{b}{(b)}
   \psfrag{p}{\scriptsize {$|\psi|^2$}}  
  \psfrag{Energy}{ \hskip -0.15 cm Energy}
  \psfrag{I}{ \hskip -0.2 cm $I_{pr}$}
  \psfrag{00}{\hskip 0.05 cm $0.0$}
  \psfrag{0.25}{$\hskip -0.05 cm 0.25$}
  \psfrag{0.00}{\hskip -0.05 cm $0.00$}
    \psfrag{0.50}{\hskip -0.1 cm $0.5$}  
\psfrag{0}{$0$}
\psfrag{20}{$20$}
\psfrag{40}{$40$}
\psfrag{60}{$60$}
\psfrag{80}{$80$}
\psfrag{0}{$0$}
\psfrag{1.0}{\hskip -0. cm $1.0$}
\psfrag{0.5}{\hskip -0. cm $0.5$}
\psfrag{0.0}{\hskip -0.0 cm $0.0$}
\psfrag{-1.4}{\hskip -0.05 cm$-1.4$}
\psfrag{-0.7}{\hskip -0.05 cm $-0.7$}
\psfrag{0.7}{\hskip -0.0 cm $0.7$}
\psfrag{1.4}{\hskip -0.0 cm $1.4$}
\psfrag{v}{\hskip -0.15 cm $v=0.2,\;w=1$}
\psfrag{site}{\hskip 0.2 cm site}
\hskip 0.22 cm
\includegraphics[width=230pt]{top-color-box-polyacene.eps}
\vskip -0.0 cm
\includegraphics[width=250pt]{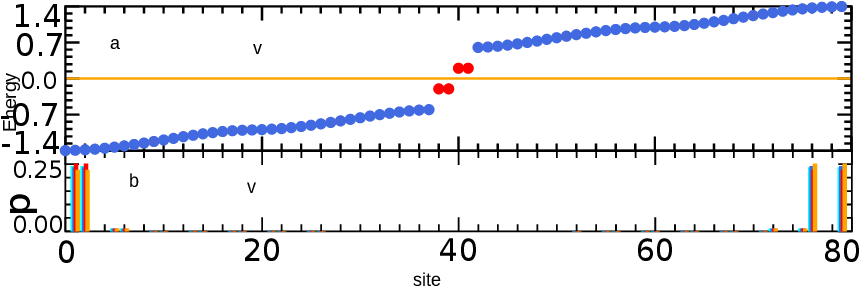}
\caption{Variation of energies in ascending order for $cb$-pol
  open chain when $v/w=0.2$ (a).
  It shows the presence of four edge
  states within the band gap. Among them two are below the zero
  energy and two are above the zero energy. 
  Top colorbar indicates the variation of $I_{pr}$.
  Probability density, $|\psi_j|^2$ of four edge states are
  shown in the histograms by different colors (b). }
 \label{achiral-cis-polyacene-edge-energies-set-1}
\end{figure}

\begin{figure}[h]
  \psfrag{a}{(a)}
  \psfrag{b}{(b)}
   \psfrag{p}{\scriptsize {$|\psi|^2$}}  
 % \psfrag{p}{$|\psi|$}  
  \psfrag{Energy}{ \hskip -0.15 cm Energy}
  \psfrag{I}{ \hskip -0.2 cm $I_{pr}$}
  \psfrag{00}{\hskip 0.05 cm $0.0$}
  \psfrag{0.1}{$\hskip -0.05 cm 0.1$}
  \psfrag{0.0}{\hskip -0.05 cm $0.0$}
    \psfrag{0.50}{\hskip -0.1 cm $0.5$}  
\psfrag{0}{$0$}
\psfrag{20}{$20$}
\psfrag{40}{$40$}
\psfrag{60}{$60$}
\psfrag{80}{$80$}
\psfrag{0}{$0$}
\psfrag{1.0}{\hskip -0. cm $1.0$}
\psfrag{0.5}{\hskip -0. cm $0.5$}
\psfrag{0.0}{\hskip -0.0 cm $0.0$}
\psfrag{-1.4}{\hskip -0.05 cm$-1.4$}
\psfrag{-0.7}{\hskip -0.05 cm $-0.7$}
\psfrag{0.7}{\hskip -0.0 cm $0.7$}
\psfrag{1.4}{\hskip -0.0 cm $1.4$}
\psfrag{v}{\hskip -0.15 cm $v=0.4,\;w=0.5$}
\psfrag{site}{\hskip 0.2 cm site}
\hskip 0.22 cm
\includegraphics[width=230pt]{top-color-box-polyacene.eps}
\vskip -0.0 cm
\includegraphics[width=250pt]{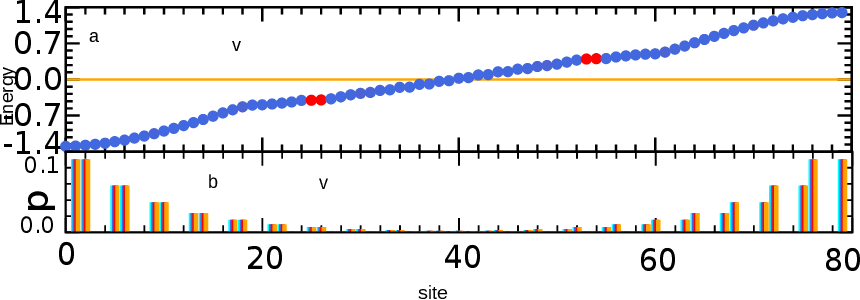}
\caption{Variation of energies in ascending order for $cb$-pol
  open chain when $v/w=0.8$ (a). It shows the existence of four edge
  states. Among them two are below the zero energy
  and two are above the zero energy. 
  Top colorbar indicates the variation of $I_{pr}$.
  Probability density, $|\psi_j|^2$ of four edge states are
  shown in the histograms by different colors (b). }
 \label{achiral-cis-polyacene-edge-energies-set-2}
\end{figure}

In order to determine the feature of localized states, 
energy spectrum of the $cb$-pol open chain 
of 80 sites has been plotted in
Fig \ref{achiral-cis-polyacene-edge-energies-set-1} (a), 
when $v/w=0.2$. 
System hosts two pairs of nonzero-energy edge states with
higher values of $I_{pr}$. Among them one pair lies above 
the zero energy while another pair lies below, 
and value of these energies are $\pm v$. 
But both the pairs lie within the true band gap in this case.
Histograms of site-wise probability density, $|\psi_j|^2$,
for all the four edge states have been shown
in Fig \ref{achiral-cis-polyacene-edge-energies-set-1} (b), which
indicates that they are highly localized towards the edges of both
open sides. 

Similarly, energy spectrum of the system has been demonstrated in
Fig \ref{achiral-cis-polyacene-edge-energies-set-2} (a)
for $v/w=0.8$. 
Two pairs of nonzero-energy edge states with
higher values of $I_{pr}$ have been observed where 
each pair lies above and below the zero energy,
which is similar to the previous case,
but the value of energies are different. 
Probability density, $|\psi_j|^2$ of four edge states are
shown in the histograms by different colors in
Fig \ref{achiral-cis-polyacene-edge-energies-set-2} (b).

In order to find the winding number for this model, 
Hamiltonian, $H_{\rm cb}(k)$ in Eq. \ref{Hcbk},
has been block off-diagonalized
under unitary transformation or, $H'_{\rm cb}(k)=U^{-1}H_{\rm cb}(k)U$,
where $H'_{\rm cb}(k)$ looks like
 \be
H'_{\rm cb}(k)\!=\!\!\left(
\begin{array}{cc}0&h_{\rm cb}(k)\\[0.4em]
  h_{\rm cb}^\dag(k)&0
\end{array}\!\! \right)\!,\,{\rm with}\;
h_{\rm cb}(k)\!=\!\left(
\begin{array}{ll}g^*(k)&v\\[0.4em]
  v&g(k)
\end{array}\!\! \right)\!.\nonumber
 \label{Htcbkprime}
 \ee
  Winding number can be evaluated as,
 \bea
 \nu&=&\frac{1}{2\pi i}\oint dk \;
    {\rm Tr}\left[ \frac{1}{h_{\rm cb}(k)}\frac{dh_{\rm cb}(k)}{dk}\right],\nonumber\\[0.4em]
    &=& -\frac{2}{2\pi i}\int_{0}^{2\pi} dk \;\frac{vw\sin{(k)}}{w(w+2v\cos{(k)})}, \nonumber\\[0.4em]
    &=&0.
    \label{winding0}
\eea
The winding number again vanishes in this case like $c$-pol. 
However, topological character of this
model has been examined further in terms of
the Zak number,
(Zak phase per $\pi$) which is defined by \cite{Zak}
\be
Z_n=\frac{i}{\pi}\int_{-\pi}^\pi dk
\langle \phi^{(n)}_{\rm tcb}(k)|\partial_k| \phi^{(n)}_{\rm tcb}(k)\rangle.
\label{Zaknumber}
\ee
$Z_n$ basically is the Pancharatnam-Berry phase evaluated 
for systems of one spatial dimension,
where $n$ indicates the band index of the state
$\phi_{\rm tcb}(k)$ for four different bands, $n=1,2,3,4$. 

The variation of $Z_n$ has been shown in Fig
\ref{achiral-cis-polyacene-energies-obc}
(b), (c), (d) and (e) for $n=1,2,3$, and 4, respectively.
It has been noted that topological character of top ($n=1$)
and bottom ($n=4$) bands are identical, which is also true for
the middle two bands ($n=2,3$). The difference between the two sets
is governed by the dissimilar distribution of Zak number as follows:
\be Z_n (n=1,4)=\left\{\begin{array}{cc}1,&\;-1<\frac{v}{w}<1,
\\[0.2em]
0,&\; {\rm otherwise},\;
\end{array}\right. \;\ee
while, 
\be Z_n (n=2,3)=\left\{\begin{array}{cc}1,&\;-\frac{1}{2}<\frac{v}{w}<\frac{1}{2},
\\[0.2em]
{\rm undefined},&\; {\rm otherwise}.\;
\end{array}\right. \;\ee
However, this peculiar distribution of Zak number
does not represent the true topological character of
the system. The standard nontrivial topological phase 
in this chiral model is defined by 
either $\nu=1$, or 
$Z_n=1$, $n=1,2,3,4$, where Zak number for all bands has been
specified uniquely. This conventional character is found within the 
limited region, $-1/2<v/w<+1/2$,  as shown by vertical pink dashed lines
in Fig. \ref{achiral-cis-polyacene-energies-obc}. 
The Zak number $Z_n$, for $n=2,3$, in the intermediate regions, 
$\frac{1}{2}<\left|\frac{v}{w}\right|<1$, 
are noted undefined, although nonzero-energy
edge modes exist in the entire region, $-1<v/w<+1$,
as shown in Fig. \ref{achiral-cis-polyacene-energies-obc} (a). It will be
shown later that this anomally is simply corresponds to the band overlap
between the two middle bands.

In order to explain this feature,  
Hamiltonian for $cb$-pol has been rotated under the
same operator transformations as introduced before (Eq. \ref{transformation}), 
which leads to
\be H'_{\rm {cb}}=\sum_{s=\pm}\left[ H^{s}_{\rm {ssh}}+sv\sum_{j=1}^N \left(\alpha_j^{s\dag} \alpha_j^{s}+
\beta_j^{s\dag} \beta_j^{s}\right)\right].
 \label{cis-bridged-polyacene-ham-rotated}
\ee
The transformed Hamiltonian, $H'_{\rm {cb}}$ is found to decompose into
two independent SSH chains in the presence of 
uniform onsite potentials of opposite signs, $\pm v$
for $H^{\pm}_{\rm {ssh}}$.
\begin{figure}[h]
\psfrag{A}{$\alpha^+$}
\psfrag{B}{$\beta^+$}
\psfrag{C}{$\alpha^-$}
\psfrag{D}{{$\beta^-$}}
\psfrag{v}{$v$}
\psfrag{w}{$w$}
\psfrag{a}{(a)}
\psfrag{b}{(b)}
\includegraphics[width=230pt]{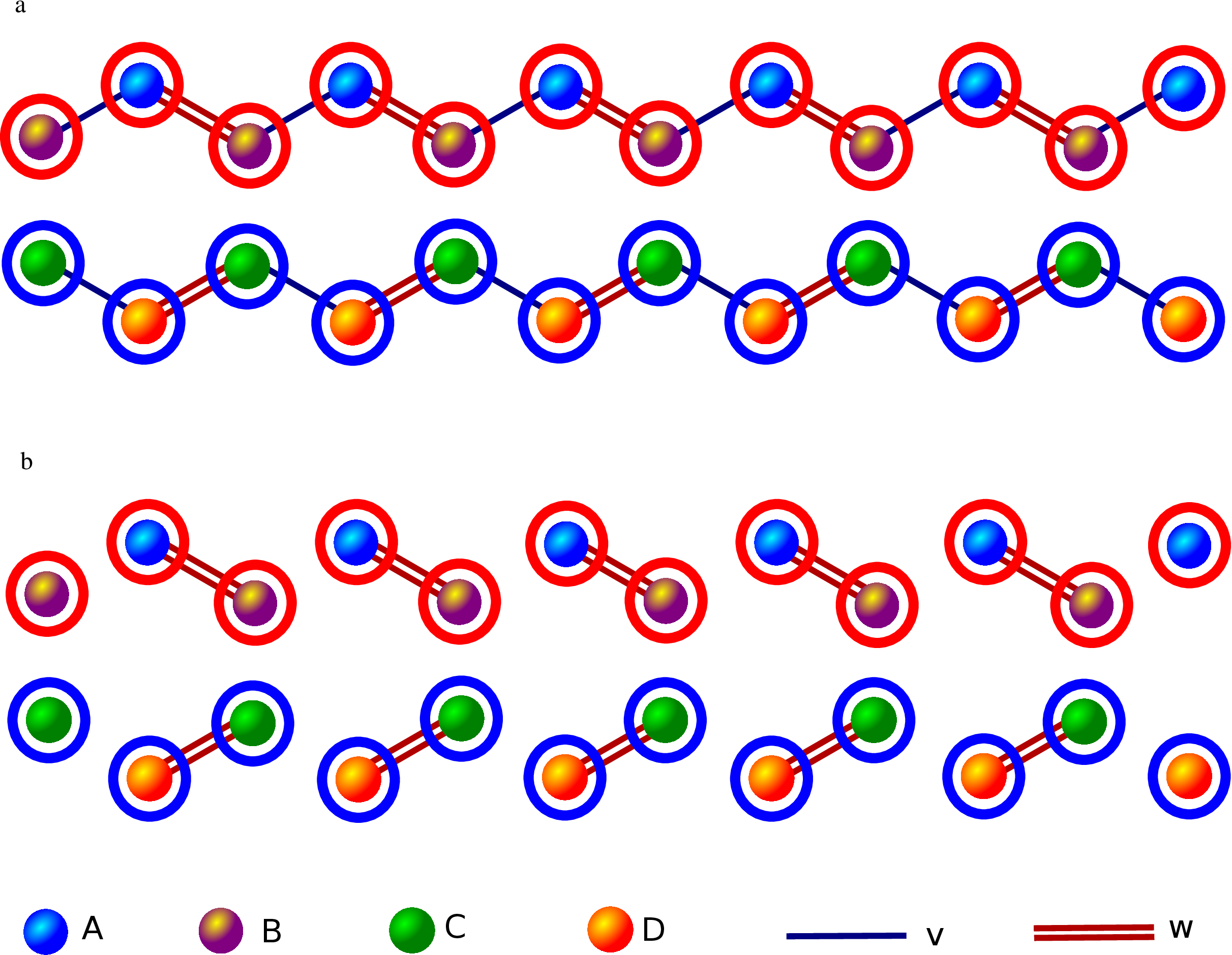}
\caption{Schematic view of the rotated Hamiltonian for
   {\em cis}-bridged-polyacene,
  $H'_{\rm {cb}}$ (Eq. \ref{cis-bridged-polyacene-ham-rotated}),
  where two independent SSH chains 
have been shown separately by the sites ($\alpha^+$, $\beta^+$),
  and ($\alpha^-$, $\beta^-$) (a).
  The same structure when $v=0$ is shown in (b),
  with isolated dimers around the bond of strength $w$, 
  and four free sites with one at each four ends
  with different onsite potentials shown by red and blue circles.}
 \label{cis-bridged-polyacene-rotated}
\end{figure}
Pictorial view of $H'_{\rm {cb}}$ has been
demonstrated in Fig. \ref{cis-bridged-polyacene-rotated} (a),
and for $v=0$ in (b), where two separated SSH are shown with different
onsite potentials by red and blue circles. 

The transformed  Hamiltonians, $H'_{\rm {cb}}$
(Eq. \ref{cis-bridged-polyacene-ham-rotated}) 
in the $k$-space can be written as:  
\be H'_{\rm {cb}}=\sum_{s=\pm}\sum_{k\in{\rm {BZ}}}\psi_k^{s \dag} H^{s}_{\rm {cb}}(k)\psi^{s}_k,\ee
and
  \be H^{s}_{\rm {cb}}(k)=\mathbf h^{s}\cdot {\boldsymbol \sigma},\label{SSH-k}
  \ee
  where 
  \[\left\{\begin{array}{l}h^{s}_0( k)=s v, \\[0.3em]
  h^{s}_1( k)=v+w\cos{( k)}, \\[0.3em]
  h^{s}_2( k)=w\sin{( k)}, \\[0.3em]
  h^{s}_3( k)=0.
              \end{array}\right. \]
  It shows that, as $h^{s}_3=0$,
  the decoupled Hamiltonians, $H^{s}_{\rm {cb}}(k)$,
separately preserve CS for $s=\pm$,
and thus the presence of uniform onsite potential does not break
the symmetry of the standard SSH model.

Obviously, energy spectra for $H_{\rm {cb}}$ and $H'_{\rm {cb}}$
will be identical. Therefore, it can be concluded at this moment that 
topological feature of $H_{\rm {cb}}$ will be identical to that
of the standard SSH model, which is given by,
\be \nu=\left\{\begin{array}{lc}1,&\, {\rm when}\; -1<v/w<+1, \\[0.3em]
 0,&{\rm elsewhere.}
\end{array}\right. \ee
although, the direct estimation of that using Eq. \ref{winding0}
yields zero. 

Zak number, the topological invariant as defined by Eq \ref{Zaknumber},
is a definite property for individual bands of a system, which means 
different bands may admit different values at a time.
In contrast, for the chiral system different bands admit the unique value
of $Z_n$ in its topological phase. For example, in the
standard SSH model, $Z_n$ has the same value for the two bands.
For the $t$-pol, estimated value of Zak number for each band
is $Z_n=2$, when $v<w$, which is equal to the winding number. 
It means Zak number and winding number yield the same value,
since they are equivalent to each other in chiral symmetric systems. 
However, when the band gap vanishes or bands merge, the eigenvalues become
degenerate at some values of Bloch wave vector, $k$,
and so the eigenvectors are no longer
uniquely defined, leading to the Zak phase undefined in such cases. 
This is the reason behind the situation occurring 
in the intermediate regions, 
$\frac{1}{2}<\left|\frac{v}{w}\right|<1$, 
where Zak number $Z_n$, for $n=2,3$, becomes undefined
for the accidental degeneracy appearing due to the
relevant band overlap. This is also true for
$tb$-pol and $tcb$-pol, where winding number is
defined uniquely, but Zak number of every band
is not always defined due to band merging. 

As a result, each of the SSH chain
will exhibit the nontrivial topology within 
the region $-1<v/w<+1$, with the Zak number $Z_n=1$,
alongwith triviality beyond that region with $Z_n=0$.
At the same time, pair of topological edge modes will
appear at the energy $+v$ ($-v$) for the
$H^{+}_{\rm {ssh}}$ ($H^{-}_{\rm {ssh}}$) chain,
replacing the conventional zero-energy modes.
This happens due to the fact that the presence of
uniform onsite potential shifts the overall
energy spectrum by the value of potential itself,
without distorting the feature of spectrum with no 
potential. This means energy spectra for $H^{+}_{\rm {ssh}}$
and $H^{-}_{\rm {ssh}}$ are identical as long as onsite
potential is zero for obvious reason. In contrast,
due to the presence of onsite potential $+v$ ($-v$)
for $H^{+}_{\rm {ssh}}$ ($H^{-}_{\rm {ssh}}$) chain, its
spectra shifts upward (downward) by the amount $|v|$.
As a result, edge modes appear at the energy $+v$ ($-v$) for the
$H^{+}_{\rm {ssh}}$ ($H^{-}_{\rm {ssh}}$) chain,
instead of the conventional zero-energy modes.
This phenomena have been demonstrated in Fig. 
\ref{achiral-cis-polyacene-edge-energies-set-1} (a),
where edge modes are found to appear at $\pm 0.2$,
and in Fig. \ref{achiral-cis-polyacene-edge-energies-set-2} (a),
where edge modes appear at $\pm 0.4$. In the later case,
it has been drawn for $v/w=0.8$, which lies within the regions, 
$\frac{1}{2}<\left|\frac{v}{w}\right|<1$, where band overlapping occurs. 
As a result, edge modes appear within the bulk modes for $v/w=0.8$,
in contrast to the case, $v/w=0.2$. 

On closer inspection to 
Fig. \ref{achiral-cis-polyacene-energies-obc} (a),
it reveals that edge modes cross each other at zero-energy
when $v=0$, because at this point all the four edge modes
appear at zero-energy. Since the variation of energy spectra
have been drawn with respect to $v$, edge modes for 
$H^{+}_{\rm {ssh}}$ ($H^{-}_{\rm {ssh}}$) move upward (downward),
with increasing $v/w$. However in each case,
edge modes survive within the topological region,
$-1<v/w<+1$, as shown by vertical green dashed lines
in Fig. \ref{achiral-cis-polyacene-energies-obc}. 

Dispersion relations in Fig. \ref{achiral-cis-polyacene-dispersions}
can be explained in terms of the disperions for $H^{+}_{\rm {ssh}}$
and $H^{-}_{\rm {ssh}}$. In the absence of onsite potential,
pair of dispersions for both the SSH chains are degenerate, since 
$H^{+}_{\rm {ssh}}$ and $H^{-}_{\rm {ssh}}$ are identical.
In contrast, in the presence of uniform onsite potentials of opposite signs,
pair of bands for $H^{+}_{\rm {ssh}}$ shifts upward, and that for 
 $H^{-}_{\rm {ssh}}$ shifts downward leaving a gap of $2|v|$. 
For example in Fig. \ref{achiral-cis-polyacene-dispersions},
top and lower-middle bands correspond to dispersions for
$H^{+}_{\rm {ssh}}$, while upper-middle and bottom bands
correspond to those for $H^{-}_{\rm {ssh}}$. As a result,
vertical shift is exactly 0.5,
in Fig. \ref{achiral-cis-polyacene-dispersions} (a), since $v=1/4$. 
Energy bands for all the diagrams in
Fig. \ref{achiral-cis-polyacene-dispersions}
can be explained in this way, as well as their vertical shifts.

\begin{figure}[h]
\psfrag{a}{(a)}
\psfrag{b}{(b)}
\psfrag{c}{(c)}
\psfrag{d}{(d)}
\psfrag{e}{(e)}
\psfrag{z1}{\hskip 0.0 cm$Z_1$}
\psfrag{z2}{\hskip 0.0 cm$Z_2$}
\psfrag{z3}{\hskip 0.0 cm$Z_3$}
\psfrag{z4}{\hskip 0.0 cm$Z_4$}
  \psfrag{I}{ \hskip -0.2 cm $I_{pr}$}
  \psfrag{00}{\hskip 0.05 cm $0.0$}
  \psfrag{0.25}{$0.25$}
  \psfrag{0.5}{\hskip -0.00 cm $0.5$}
 \psfrag{1.0}{\hskip -0. cm $1.0$} 
\psfrag{Energy}{ \hskip -0.5 cm Energies}
\psfrag{4}{$4$}
\psfrag{2}{$2$}
\psfrag{1}{$1$}
\psfrag{0}{$0$}
\psfrag{-1}{\hskip -0.14 cm$-1$}
\psfrag{-2}{\hskip -0.14 cm $-2$}
\psfrag{-4}{\hskip -0.14 cm $-4$}
\psfrag{w}{$w/v$}
\hskip 0.16 cm
\includegraphics[width=240pt]{top-color-box-polyacene.eps}
\vskip 0.02 cm
\includegraphics[width=250pt]{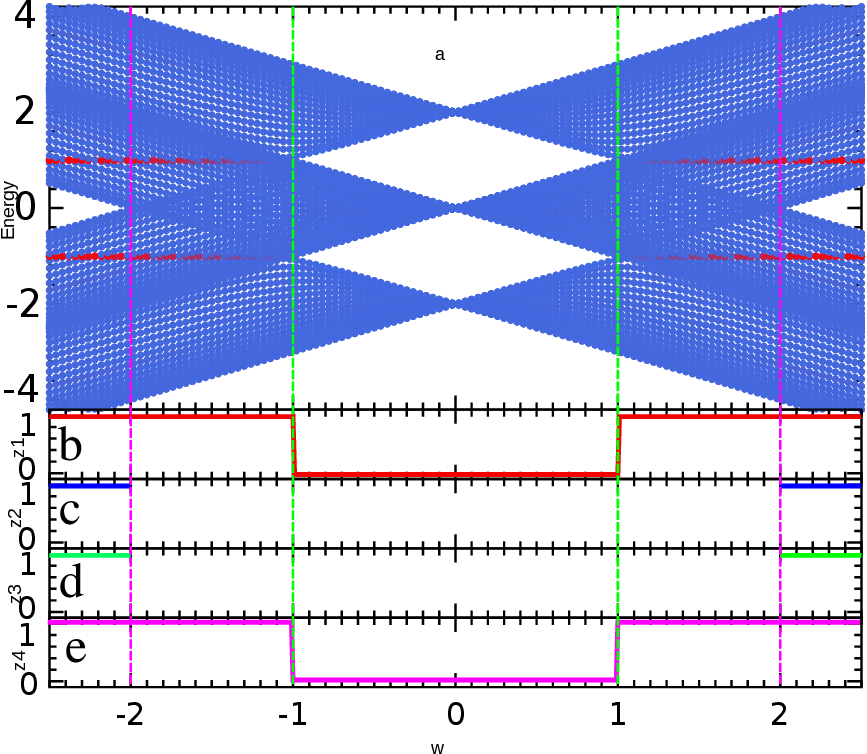}
\caption{Variation of energies for $cb$-pol open chain (a),
  and Zak numbers of four bands,
  $Z_1$ (top) (b),  $Z_2$ (middle upper) (c),
   $Z_3$ (middle lower) (d) and  $Z_4$ (bottom) (e)
  with respect to $w/v$. 
  Top colorbar indicates the variation of $I_{pr}$.
  Vertical dashed lines are drawn at the points $v/w=\pm 1/2$, and
  $\pm 1$.}
 \label{cis-bridged-polyacene-energies-obc-varying-w}
\end{figure}

For better understanding on the topological feature of $cb$-pol, 
eigenenergies of the open system
have been plotted with respect to $w/v$ as shown in
Fig. \ref{cis-bridged-polyacene-energies-obc-varying-w} (a),
as long as $|w/v|\le 2.5$.
The Fig. \ref{cis-bridged-polyacene-energies-obc-varying-w}
plotted with respect to $w/v$ 
can be treated as complementary of
the Fig \ref{achiral-cis-polyacene-energies-obc},
which has been plotted with respect to $v/w$. 
It is mentioned above that 
when $w=0$, the system yields three flat bands with doubly
degenerate state at $E_{\rm cb}=0$, and
nondegenerate state at $E_{\rm cb}=\pm 2v$.
This feature is clear in Fig.
\ref{cis-bridged-polyacene-energies-obc-varying-w} (a),
when $w=0$. In this diagram topological triviality
exists in the region, $-1<w/v<+1$, which is enclosed with the
green dashed lines. Nonzero-energy topological edge modes exist
outside this region as evident form this diagram.
However, Zak number $Z_n$, for $n=2,3$, in the intermediate regions, 
$1<\left|\frac{w}{v}\right|<2$, 
are noted undefined, although nonzero-energy
edge modes exist. This feature is noted before in the
complementary diagram in Fig \ref{achiral-cis-polyacene-energies-obc} (a). 
This anomaly is attributed to the overlap of
the two middle bands. So, topological phase is noted
in the region, $-1<v/w<1$,
or $1<\left|w/v\right|<\infty$, with $Z_n=1$,
which is identical to the standard SSH model.
Therefore, four-band energy diagrams in 
Fig. \ref{achiral-cis-polyacene-energies-obc} (a) and in 
Fig. \ref{cis-bridged-polyacene-energies-obc-varying-w} (a),
can be regarded as the superposition of two-band energy diagrams of
two uncoupled SSH chains. 

As previously mentioned, topological phase of a multiband achiral
system can be assigned by different values of $Z_n$ for its different bands.
For example, different topological phases of an achiral SSH
trimer model have been defined
by the different values for ($Z_1,Z_2,Z_3$), where 
$Z_n$ denotes the Zak number of $n$-th band \cite{Alvarez,Du,Bomantara}.
These models are found to possess the IS instead of the CS.
Remarkably, this feature is similar to the
Chern insulating state for 2D systems \cite{Sil2,Moumita1,Sil3}, where
the bands can admit different Chern numbers
including nonzero-energy edge states in accordance to the 
bulk-boundary correspondence rule \cite{Hatsugai,Sil4}. 
On the other hand,
winding number is the property of the entire system possessing
chiral symmetry because of the fact that
for a multiband system, an unique value of $\nu$ is assigned
in the nontrivial phase \cite{Rakesh2,Rakesh3,Rittwik}.  

\begin{figure}[h]
\includegraphics[width=230pt]{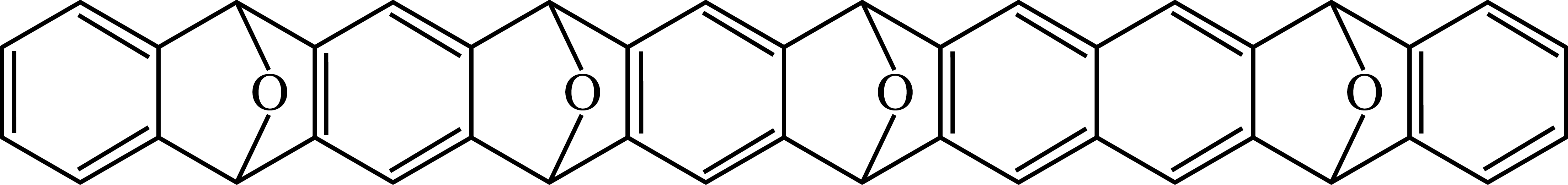}
\caption{Structure of decacene precursors}
\label{decacene}
\end{figure}

Material realization of the structure of $cb$-pol is
challenging because of the infeasibility of direct bond
between the para-positioned carbon atoms in the benzene ring.
However, a bridge comprising of some atom or
group is possible between the intracell para positioned carbon atoms,
and that could explain the existence of 
the additional hopping terms. 
Existence of bond connecting A and D sublattice
through oxygen atom has been observed in decacene precursors,
which is shown in Fig \ref{decacene} \cite{Kruger}.
Similar type of bonding bridged by carbonyl group is
observed in another polyacene precursors \cite{Jancarik}.
In this sense, presence of the extra hopping terms
considered in $cb$-pol is not exceptional. 
However, this type of bond is found to exist neither in every benzene ring
nor in a regular interval in these polymers \cite{Kruger,Jancarik}.
%%%%%%%%%%%%%%  Section VII %%%%%%%%%%%%%%%%%%%%%%%%%%%%%%%%%%%%%%%%%%%%%%%%%
\section{THE {\em cis}-POLYACENE MODEL with cross-bridging bonds}
\label{cis-cross-bridged}
\begin{figure}[h]
\psfrag{A}{A}
\psfrag{B}{B}
\psfrag{C}{C}
\psfrag{D}{D}
\psfrag{v}{$v$}
\psfrag{w}{$w$}
\includegraphics[width=230pt]{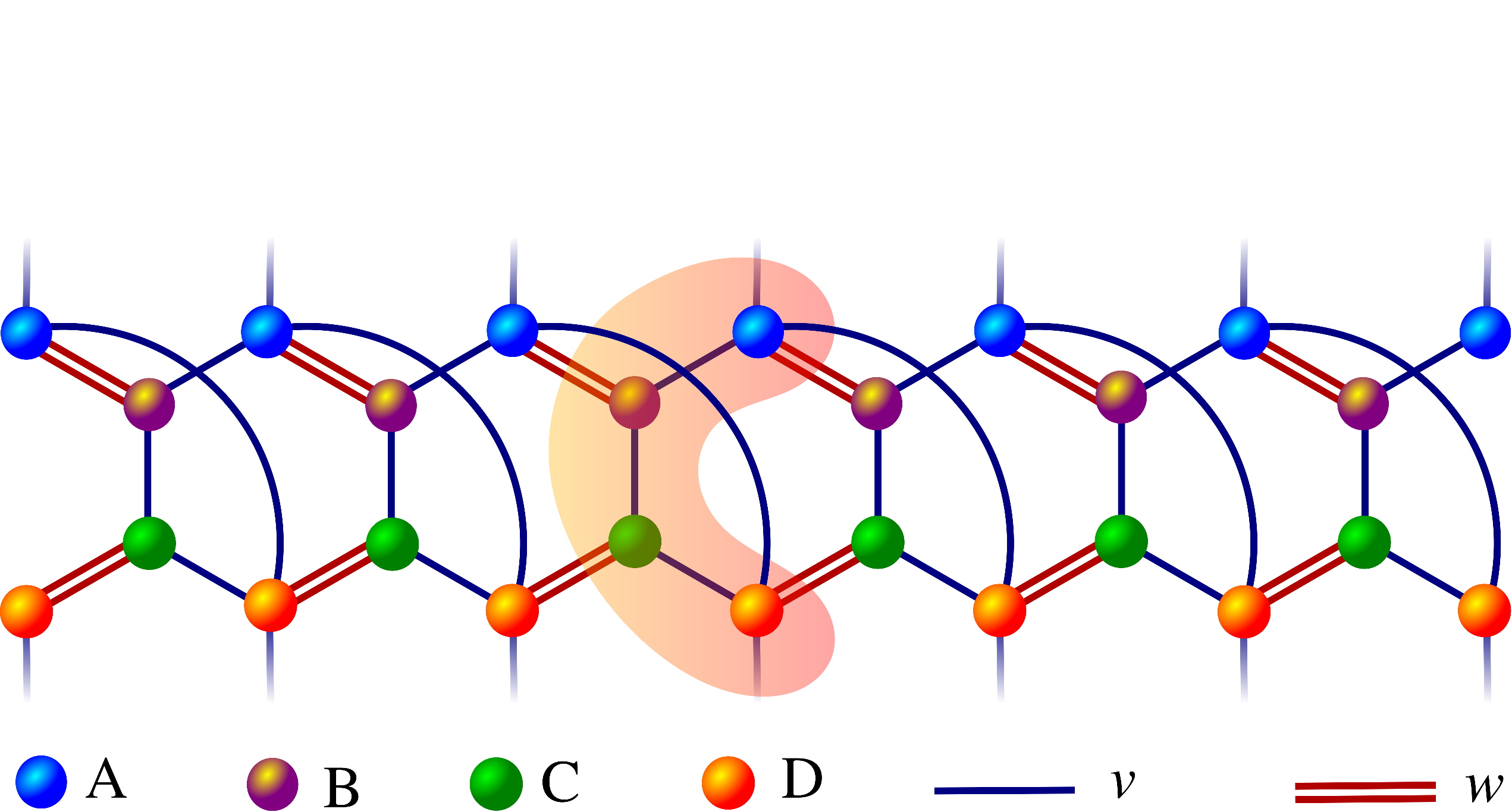}
\caption{Structure of $ccb$-pol.  
  Carbon atoms marked by A, B, C
  and D form a unit cell. One such unit cell is highlighted.}
 \label{cis-cross-bridged-polyacene}
\end{figure}

In order to induce the nontriviality within $c$-pol,
in a different way, the primary structure
of $c$-pol has been modified by introducing an additional
intercell hopping path connecting A and D sublattice points 
of adjacent cells whose strength is the same to that of single bond ($v$).
Topological character of this model has been studied
in terms of winding number. 
This structure is referred as the cis cross-bridged polyacene
or $ccb$-pol, where
this variant breaks the mirror symmetry which is present in $c$-pol and
$cb$-pol. 
%because of the fact that 
%However, this modification restores the inversion symmetry in the $k$-space,
%which was absent in the $cb$-pol???? 
%since 
%while keeping the chiral symmetry in abeyance. 
This modified structure has been shown in Fig
\ref{cis-cross-bridged-polyacene}. 

\begin{figure}[h]
\psfrag{a}{(a)}
\psfrag{b}{(b)}
\psfrag{c}{(c)}
\psfrag{d}{(d)}
\psfrag{p}{$-\pi$}
\psfrag{q}{$-\frac{\pi}{2}$}
\psfrag{n}{$\frac{\pi}{2}$}
\psfrag{m}{$\pi$}
\psfrag{1}{$1$}
\psfrag{0}{$0$}
\psfrag{2}{$2$}
\psfrag{3}{$3$}
\psfrag{-1}{\hskip -0.02 cm$-1$}
\psfrag{-2}{\hskip -0.02 cm$-2$}
\psfrag{-3}{\hskip -0.02 cm$-3$}
\psfrag{k}{$k$}
\psfrag{E}{$E_{\rm ccb}(k)$}
\includegraphics[width=240pt]{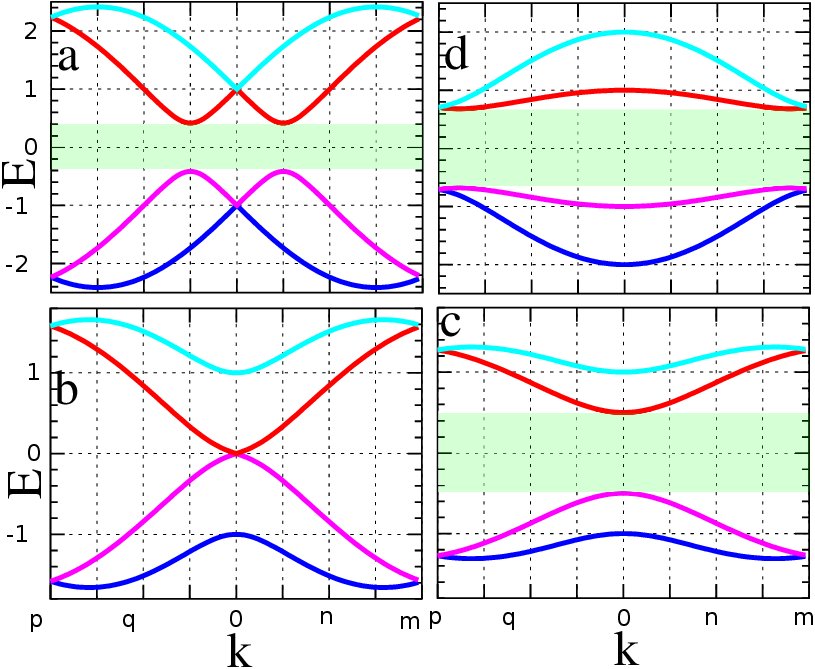}
\caption{Dispersion relations of $ccb$-pol for (a) $v=-1$, 
  (b)  $v=-1/2$,  (c)  $v=-1/4$, and (d)  $v=1/2$, when $w=1$.
  Energy gaps in (a), (c) and (d) are shown
  by shades.}
 \label{cis-cross-bridged-polyacene-dispersions}
 \end{figure}

The Hamiltonian for this structure is given by
\be H_{\rm {ccb}}=H_{\rm {c}} + v\sum_{j=1}^N \left(a_j^\dag d_{j+1}+h.c.\right),
 \label{cis-cross-bridged-polyacene-ham}
 \ee
and in the $k$-space it is written as
 \[H_{\rm ccb}=\sum_{k\in{\rm {BZ}}}\Psi_k^\dag H_{\rm ccb}(k)\Psi_k,\]
 where 
 \be H_{\rm ccb}(k)=\left(
\begin{array}{cccc}0&g^*(k)&0&ve^{ik}\\[0.4em]
  g(k) &0&v&0\\[0.4em]
  0&v&0&g(k)\\[0.4em]
  ve^{-ik}&0&g^*(k) &0
\end{array} \right).
\label{Hccbk}  \ee
%where $g(k)=v+we^{-ik}$.
The system preserves the
TRS, CS, and PHS, but no IS.  
The dispersion relations are given by
\[E_{\rm ccb}(k)=\pm\sqrt{v^2+f^2\pm 2vf\cos{(k/2)}},\]
with $f^2=v^2+w^2+2vw\cos{k}$.
Band diagrams have been shown in
Fig \ref{cis-cross-bridged-polyacene-dispersions}
(a), (b), (c) and (d) for $v=-1$,  $v=-1/2$,
$v=-1/4$, and $v=1/2$, respectively, when $w=1$.
 Two upper and two lower bands always touch
 each other separately at the Brillouin zone boundaries,
 $k=\pm \pi$, for any values of the
 parameters, $v$ and $w$. This feature is evident from all the
 diagrams, (a), (b), (c) and (d) in
 Fig \ref{cis-cross-bridged-polyacene-dispersions},
 which is similar to  $tb$-pol. 
 Pair of doubly degenerate flat bands
 appear at $E_{\rm ccb}=\pm w$, when $v=0$. 
Eigenvectors  of $H_{\rm ccb}(k)$ are given by, 
\[\phi_{\rm ccb}(k)=\left[\gamma\;\;\frac{g\gamma+v}{E_{\rm ccb}}\;\;1\;\;
  \frac{E^2_{\rm ccb}-vg\gamma-v^2}{gE_{\rm ccb}} \right]^{\rm T},\]
where $\gamma=\dfrac{v^3e^{ik}-v(E^2_{\rm ccb}e^{ik}+|g|^2)}
{g^*g^2-g(E^2_{\rm ccb}+v^2e^{ik})}$.

\begin{figure}[h]
\psfrag{a}{(a)}
\psfrag{b}{(b)}
\psfrag{z}{\hskip 0.1 cm $Z_{2,3}$}
  \psfrag{I}{ \hskip -0.2 cm $I_{pr}$}
  \psfrag{00}{\hskip 0.05 cm $0.0$}
  \psfrag{0.25}{$0.25$}
  \psfrag{0.5}{\hskip -0.00 cm $0.5$}
 \psfrag{1.0}{\hskip -0. cm $1.0$} 
\psfrag{E}{ \hskip -0.5 cm Energies}
\psfrag{4}{$4$}
\psfrag{2}{$2$}
\psfrag{1}{$1$}
\psfrag{0}{$0$}
\psfrag{-1}{\hskip -0.14 cm$-1$}
\psfrag{-2}{\hskip -0.14 cm $-2$}
\psfrag{-4}{\hskip -0.14 cm $-4$}
\psfrag{v}{\hskip -0.15 cm $v/w$}
\psfrag{w}{$\nu$}
\hskip 0.16 cm
\includegraphics[width=240pt]{top-color-box-polyacene.eps}
\vskip 0.02 cm
\includegraphics[width=250pt]{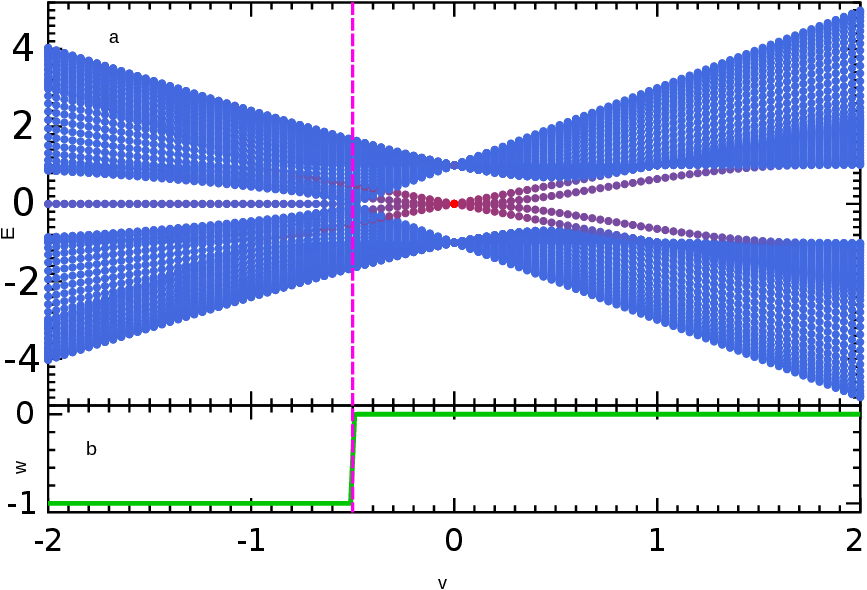}
\caption{Variation of energies for cis cross-bridged polyacane open chain (a), and 
   winding number (b) with respect to $v/w$. 
  Top colorbar indicates the variation of $I_{pr}$.
Vertical dashed line is drawn at the phase transition point.}
 \label{cis-cross-bridged-polyacene-energies-obc}
\end{figure}

Under the unitary transformation $H'_{\rm ccb}(k)=U^{-1}H_{\rm ccb}(k)U$,
  $H'_{\rm ccb}(k)$ has assumed 
 block off-diagonal form, as shown below. 
 \be
H'_{\rm ccb}(k)\!=\!\!\left(
\begin{array}{cc}0&h_{\rm ccb}(k)\\[0.4em]
  h_{\rm ccb}^\dag(k)&0
\end{array}\!\! \right)\!,\,{\rm with}\;
h_{\rm ccb}(k)\!=\!\left(
\begin{array}{ll}g^*(k)&ve^{-ik}\\[0.4em]
  v&g(k)
\end{array}\!\! \right)\!.\nonumber
 \label{Htcbkprime}
 \ee
  Winding number can be evaluated as,
 \bea
 \nu&=&\frac{1}{2\pi i}\oint dk \;
    {\rm Tr}\left[ \frac{1}{h_{\rm ccb}(k)}\frac{dh_{\rm ccb}(k)}{dk}\right],\nonumber\\[0.4em]
    &=&  \frac{1}{2\pi i}\oint_{|z|=1} d\log{f_{\rm ccb}(z)}-1, \nonumber
\eea
where $f_{\rm ccb}(z)=vw +(v^2+w^2)z+(vw-v^2)z^2$, which has no pole.
The number of zero for $f_{\rm ccb}(z)$, within $|z|=1$, has been given as,
\be
\mathcal N_z=\left\{
\begin{array}{cc}0,& {\rm if}\; v<-w/2,\nonumber\\[0.4em]
  1,&{\rm if}\; v>-w/2.\nonumber
\end{array} \right.\nonumber\ee
As a result, 
\be \nu =\left\{\begin{array}{cc}-1,& {\rm if}\; v<-w/2,\nonumber\\[0.4em]
  0,&{\rm if}\; v>-w/2.
\end{array} \right.
\ee
So, this model exhibits a single nontrivial
topological phase with $\nu=-1$, when $v/w<-1/2$. 
Variation of winding number with $v/w$ is shown in Fig
\ref{cis-cross-bridged-polyacene-energies-obc} (b),
where vertical dashed line indicates the phase transition point. 

\begin{figure}[h]
  \psfrag{a}{(a)}
  \psfrag{b}{(b)}
   \psfrag{p}{\scriptsize {$|\psi|^2$}}  
   \psfrag{Energy}{ \hskip -0.15 cm Energy}
  \psfrag{I}{ \hskip -0.2 cm $I_{pr}$}
  \psfrag{00}{\hskip 0.05 cm $0.0$}
  \psfrag{0.25}{$\hskip -0.05 cm 0.25$}
  \psfrag{0.00}{\hskip -0.05 cm $0.00$}
    \psfrag{0.50}{\hskip -0.1 cm $0.5$}  
\psfrag{0}{$0$}
\psfrag{20}{$20$}
\psfrag{40}{$40$}
\psfrag{60}{$60$}
\psfrag{80}{$80$}
\psfrag{0}{$0$}
\psfrag{1.0}{\hskip -0. cm $1.0$}
\psfrag{0.5}{\hskip -0. cm $0.5$}
\psfrag{0.0}{\hskip -0.0 cm $0.0$}
\psfrag{-1}{\hskip -0.05 cm$-1$}
\psfrag{-2}{\hskip -0.05 cm$-2$}
\psfrag{-0.7}{\hskip -0.05 cm $-0.7$}
\psfrag{2}{\hskip -0.0 cm $2$}
\psfrag{1}{\hskip -0.0 cm $1$}
\psfrag{v}{\hskip -0.15 cm $v=-1,\;w=1$}
\psfrag{site}{\hskip 0.2 cm site}
\hskip 0.22 cm
\includegraphics[width=230pt]{top-color-box-polyacene.eps}
\vskip -0.0 cm
\includegraphics[width=250pt]{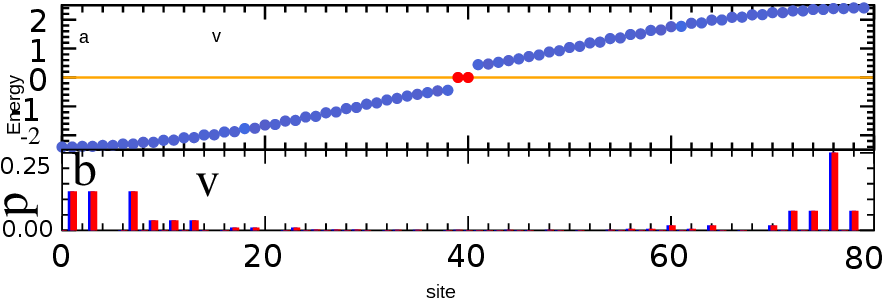}
\caption{Variation of energies in ascending order for $ccb$-pol
  open chain when $v/w=-1$ (a).
  It shows the presence of two zero-energy 
  modes within the band gap.  
  Top colorbar indicates the variation of $I_{pr}$.
  Probability density, $|\psi_j|^2$ of the edge modes are
  shown in the histograms by different colors (b). }
 \label{cis-cross-bridged-polyacene-edge-energies-set-1}
\end{figure}

Variation of energy spectrum with $v/w$ for the open chain
of 80 sites is shown in
Fig \ref{cis-cross-bridged-polyacene-energies-obc} (a), along with $I_{pr}$
according to top color bar. A single pair of zero-energy modes
are found in the topological region, $v/w<-1/2$,
in addition to two pairs of zero-energy modes at the point $v/w=0$.
This diagram shows that energy spectra satisfy the
relation, $E(-v) \ne E(v)$, like $tb$-pol,
but unlike $c$-pol and $cb$-pol. 
%Zak phase would not be defined in this model since
%no band gap exists between upper and lower two bands separately.
In order to examine the property of zero-energy modes,
energy spectrum in ascending order along with the $I_{pr}$, 
for the system in the topological regime
has been shown in Fig
\ref{cis-cross-bridged-polyacene-edge-energies-set-1} (a),
when $v/w=-1$. Higher values of $I_{pr}$
account their strong localization. 
Probability amplitudes of the zero-energy modes have been plotted in
the histograms of Fig \ref{cis-cross-bridged-polyacene-edge-energies-set-1} (b),
which confirm that these states represent the true zero-energy edge modes. 
%%%%%%%%%%%%%%  Section VIII %%%%%%%%%%%%%%%%%%%%%%%%%%%%%%%%%%%%%%%%%%%%%%%%%
\section{THE NONTRIVIAL {\em cis}-POLYACENE MODEL}
\label{cis-nontrivial}
\begin{figure}[h]
\psfrag{A}{A}
\psfrag{B}{B}
\psfrag{C}{C}
\psfrag{D}{D}
\psfrag{u}{$u$}
\psfrag{v}{$v$}
\psfrag{w}{$w$}
\includegraphics[width=280pt]{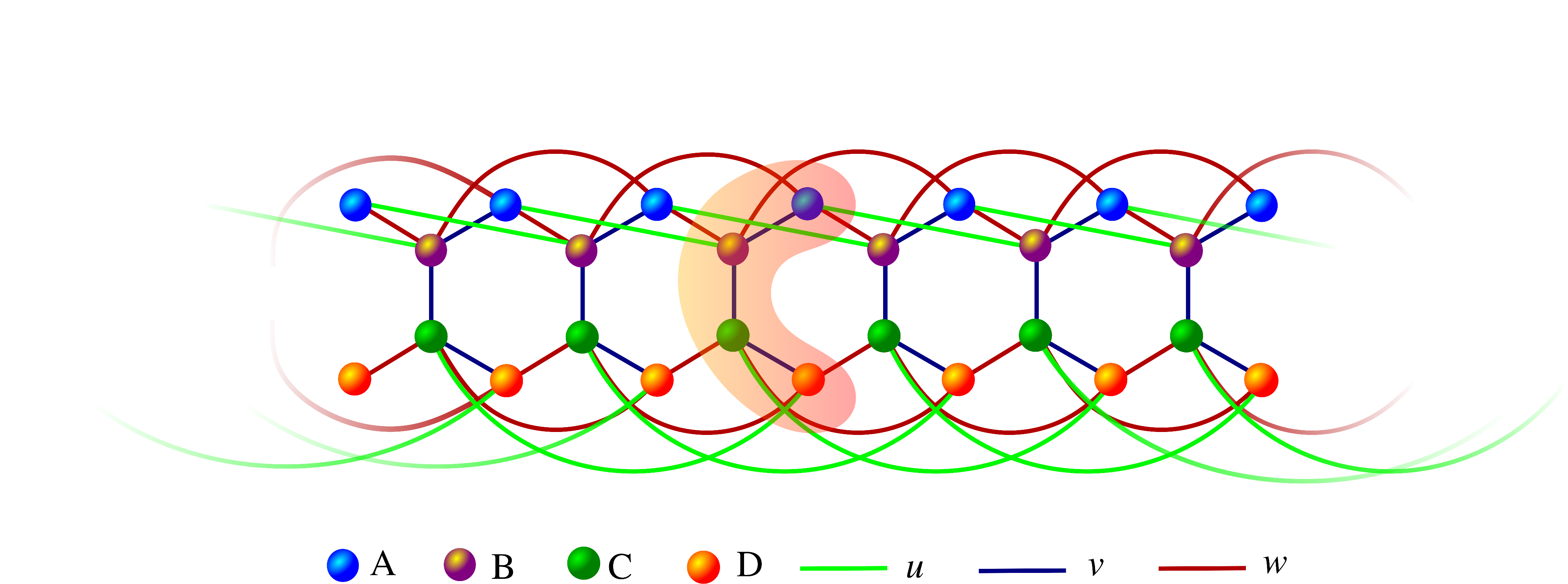}
\caption{Structure of  $cn$-pol. 
Atoms marked by A, B, C
and D form a unit cell. One such unit cell is highlighted. 
Two extra bonds of strength $w$ exit between NN unit cells
and two new bonds of strength $u$ act between NNN cells.}
\label{chiral-cis-polyacene-FN}
\end{figure}

Topological nature of six different polyacene models have
been discussed in the previous sections. 
Nonetheless, another modified form of $cis$-polyacene
structure has been introduced in this section which possesses
TRS, CS, PHS and IS 
like the $t$-pol, but at the cost of natural four-valency
configuration of carbon atom.
So, physical realizations of this model are unavailable.
This model is found to exhibit topological phases of higher winding numbers 
than the previous models. 
Since the topological phase of this model can be
studied in terms of several winding numbers, it is referred as nontrivial
$c$-pol or $cn$-pol. 
Structure of the $cn$-pol model is shown in Fig
\ref{chiral-cis-polyacene-FN}. Four new intercell hopping terms 
have been added in the $c$-pol structure,
where two bonds are formed sharing A and B atoms between 
NN cells, while the remaining two terms connect
C and D atoms between NNN cells. 
Amplitude of the new NN hopping terms is $w$, while 
that of NNN hopping is $u$. So, the resulting model is now
specified by three parameters, $v$, $w$ and $u$. 
%Two additional FN hopping terms  
%have been introduced within $c$-pol structure whose amplitude is different from
%that of the NN terms. At the same time two new NN hopping terms are added.
It is important to note that
this minimal modification is found necessary in order to induce
non-triviality in the $c$-pol structure in terms of higher winding numbers.
At the same time, mirror symmetry of $c$-pol
is lost along with the rotational symmetry within the unit cell. 
Unit cell of this modified structure
as usual accommodates four atoms, where three of them
marked by A, B and D have four bonds whereas 
the fourth atom C has five bonds, which looks artificial.
%No physical realization behind this structure is found so far. 

The TB Hamiltonian for the $cn$-pol model is given by
\bea H_{\rm {cn}}=H_{\rm {c}}&+&w\sum_{j=1}^{N-1} \left(b_j^\dag a_{j+1}
+c_j^\dag d_{j+1}+h.c.\right)\nonumber \\
&+& u\sum_{j=1}^{N-2} \left(a_j^\dag b_{j+2}+c_j^\dag d_{j+2}+h.c.\right).
 \label{chiral-cis-polyacene-ham}
 \eea
 %where $u$ be the amplutude of NNN hopping terms, while $v$ and $w$
 %are that for the NN hopping as defined earlier. 
 In $k$-space it is written as
 \[H_{\rm cn}=\sum_{k\in{\rm {BZ}}}\Psi_k^\dag H_{\rm cn}(k)\Psi_k,\]
 where 
 \be H_{\rm cn}(k)=\left(
\begin{array}{cccc}0&g'(k)&0&0\\[0.4em]
  g'^*(k) &0&v&0\\[0.4em]
  0&v&0&g'(k)\\[0.4em]
  0&0&g'^*(k) &0
\end{array} \right),
 \ee
 with $g'(k)=v+2w\cos{(k)}+ue^{2ik}$. $H_{\rm cn}(k)$
 looks identical to $H_{\rm t}(k)$, if $g'$ is replaced by $g$. 
  The eigenenergies of $H_{\rm cn}(k)$ are given by
  \be E_{\rm cn}(k)=\pm \sqrt{\frac{p_{\rm cn}\pm\sqrt{q_{\rm cn}}}{2}},
   \label{Hcck}  \ee
  where
  $p_{\rm cn}=3v^2+8vw\cos{(k)}+4vu\cos{(2k)}+8wu\cos{(k)}\cos{(2k)}
  +8w^2\cos^2{(k)}+2u^2$, and
  $q_{\rm cn}=5v^4+16v^3w\cos{(k)}+8v^3u\cos{(2k)}+16v^2wu\cos{(k)}\cos{(2k)}
   +16v^2w^2\cos^2{(k)}+4v^2u^2$. 
The corresponding eigenvectors are 
\[\phi_{\rm cn}(k)=\left[\frac{vg'}{E_{\rm cn}^2-|g'|^2}\;\;
  \frac{vE_{\rm cn}}{E_{\rm cn}^2-|g'|^2}\;\;1\;\;\frac{g'^*}{E_{\rm cn}}
  \right]^{\rm T}.\]
$H_{\rm cn}(k)$ preserves the TRS, CS, PHS and IS
like the $t$-pol, $tb$-pol, and $tcb$-pol. 
%\[\left\{\begin{array}{l}
%\mathcal K^{-1} H_{\rm cn}( k) \mathcal K=H_{\rm cn}(-k),\\ [0.4em]
% \mathcal T_z^{-1} H_{\rm cn}( k) \mathcal T_z=-H_{\rm cn}(k),\\ [0.4em]
% \mathcal P^{-1} H_{\rm cn}( k) \mathcal P=-H_{\rm cn}( -k).\end{array}\right.
% \]
% It means this modified $c$-pol structure regains all the
% symmetries of $t$-pol sturucture as discussed before. 
 
\begin{figure}[h]
\psfrag{a}{(a)}
\psfrag{b}{(b)}
\psfrag{win}{$\nu$}
  \psfrag{I}{ \hskip -0.2 cm $I_{pr}$}
  \psfrag{00}{\hskip 0.05 cm $0.0$}
  \psfrag{0.25}{$0.25$}
  \psfrag{0.5}{\hskip -0.00 cm $0.5$}
     \psfrag{-0.5}{\hskip -0.00 cm $-0.5$}
 \psfrag{1.0}{\hskip -0. cm $1.0$} 
 \psfrag{Energy}{ \hskip -0.25 cm Energy}
 \psfrag{4}{$4$}
\psfrag{3}{$3$}
\psfrag{2}{$2$}
\psfrag{1}{$1$}
\psfrag{0}{$0$}
\psfrag{-1}{\hskip -0.06 cm$-1$}
\psfrag{-2}{\hskip -0.06 cm $-2$}
\psfrag{-3}{\hskip -0.06 cm $-3$}
\psfrag{v}{\hskip -0.15 cm $v/w$}
\psfrag{w}{$w$}
\hskip 0.16 cm
\includegraphics[width=240pt]{top-color-box-polyacene.eps}
\vskip 0.02 cm
\includegraphics[width=250pt]{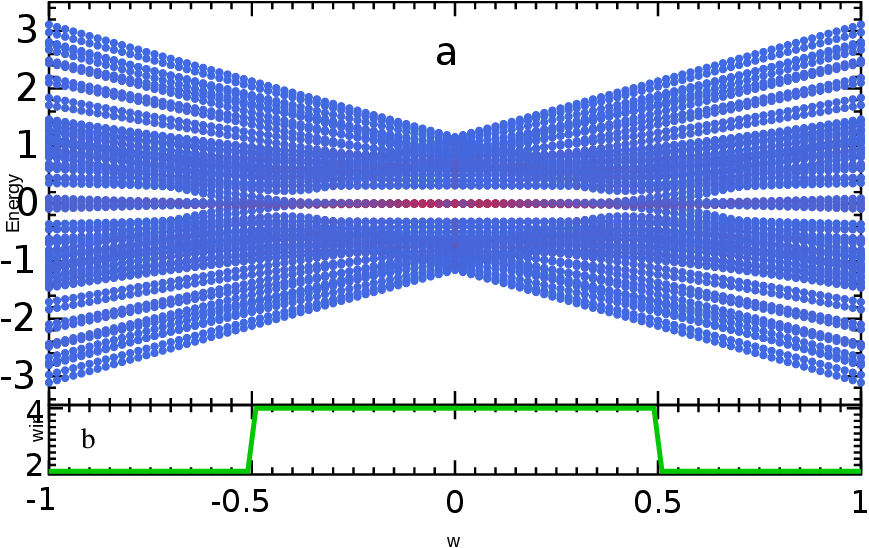}
\caption{Variation of energy (a) and winding number (b) 
  with $w$ of $cn$-pol model when $u=0.7$ and $v=0.3$.
  Top colorbar indicates the variation of $I_{pr}$.}
 \label{chiral-cis-polyacene-FN-energies-winding-number-set-1}
\end{figure}

\begin{figure}[h]
\psfrag{a}{(a)}
\psfrag{b}{(b)}
\psfrag{win}{$\nu$}
  \psfrag{I}{ \hskip -0.2 cm $I_{pr}$}
  \psfrag{00}{\hskip 0.05 cm $0.0$}
  \psfrag{0.25}{$0.25$}
  \psfrag{0.5}{\hskip -0.00 cm $0.5$}
   \psfrag{-0.5}{\hskip -0.00 cm $-0.5$}
 \psfrag{1.0}{\hskip -0. cm $1.0$} 
\psfrag{Energy}{ \hskip -0.4 cm Energy}
\psfrag{3}{$3$}
\psfrag{2}{$2$}
\psfrag{1}{$1$}
\psfrag{0}{$0$}
\psfrag{-1}{\hskip -0. cm$-1$}
\psfrag{-2}{\hskip -0. cm $-2$}
\psfrag{-3}{\hskip -0. cm $-3$}
\psfrag{v}{\hskip -0.15 cm $v/w$}
\psfrag{w}{$w$}
\hskip 0.16 cm
\includegraphics[width=240pt]{top-color-box-polyacene.eps}
\vskip 0.02 cm
\includegraphics[width=250pt]{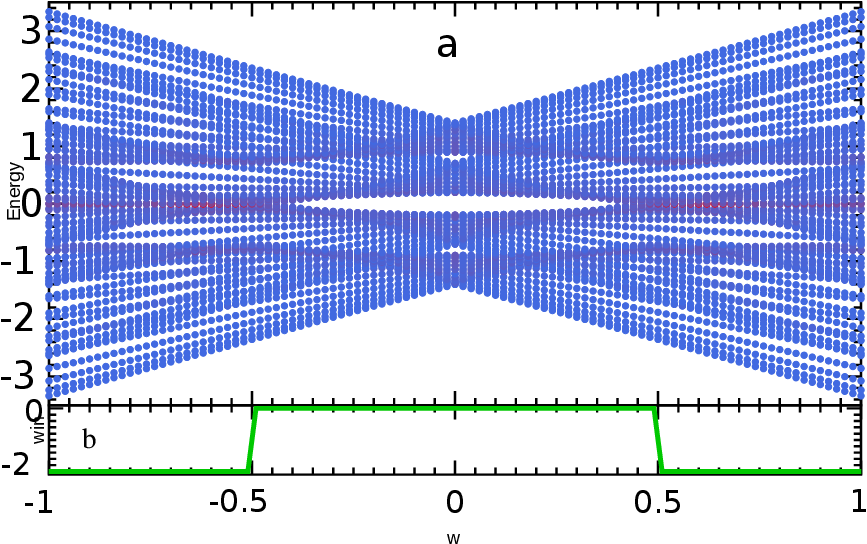}
\caption{Variation of energy (a) and winding number (b) 
  with $w$ of  $cn$-pol model when $u=0.3$ and $v=0.7$.
  Top colorbar indicates the variation of $I_{pr}$.}
 \label{chiral-cis-polyacene-FN-energies-winding-number-set-2}
\end{figure}

Under the unitary transformation,
$H'_{\rm cn}(k)=U^{-1}H_{\rm cn}(k)U$, 
the new Hamiltonian admits 
$2 \times 2$ block off-diagonal form as shown below,
which is similar to the $t$-pol model for obvious reason.  
 \be
H'_{\rm cn}(k)\!=\!\left(\!
\begin{array}{cc}0&h_{\rm cn}(k)\\[0.4em]
  h_{\rm cn}^\dag(k)&0
\end{array}\! \right)\!,\;{\rm with}\;\;
h_{\rm cn}(k)\!=\!\left(\!
\begin{array}{cc}g'^*(k)&0\\[0.4em]
  v&g'^*(k)
\end{array} \!\right)\!.
 \label{Hcckprime}
 \ee
 Winding number can be evaluated numerically, where 
 \be
 \nu=\frac{1}{2\pi i}\oint dk \; {\rm Tr}\left[ \frac{1}{h_{\rm cn}(k)}\frac{dh_{\rm cn}(k)}{dk}\right].\ee
 Variation of $\nu$ in two different parameter regimes have been
 shown in Fig \ref{chiral-cis-polyacene-FN-energies-winding-number-set-1} (b), 
 and Fig \ref{chiral-cis-polyacene-FN-energies-winding-number-set-2} (b). 
% Therefore, by invoking four extra hopping terms
% the chiral symmetry within modified $cis$-polyacene
% is brought back, and as a result, winding number is set to play the
% role of topological invariant
% for this model again.
 
\begin{figure}[h]
\psfrag{a}{\hskip -0.07 cm (a)}
\psfrag{b}{\hskip -0.07 cm (b)}
\psfrag{c}{\hskip -0.07 cm (c)}
\psfrag{d}{\hskip -0.07 cm (d)}
\psfrag{e}{\hskip -0.07 cm (e)}
\psfrag{f}{\hskip -0.02 cm (f)}
\psfrag{p}{$-\pi$}
\psfrag{q}{$-\frac{\pi}{2}$}
\psfrag{n}{$\frac{\pi}{2}$}
\psfrag{m}{$\pi$}
\psfrag{2}{$2$}
\psfrag{1}{$1$}
\psfrag{0}{$0$}
\psfrag{-1}{\hskip -0.05 cm$-1$}
\psfrag{-2}{\hskip -0.05 cm$-2$}
\psfrag{k}{$k$}
\psfrag{E}{$E(k)$}
\includegraphics[width=240pt]{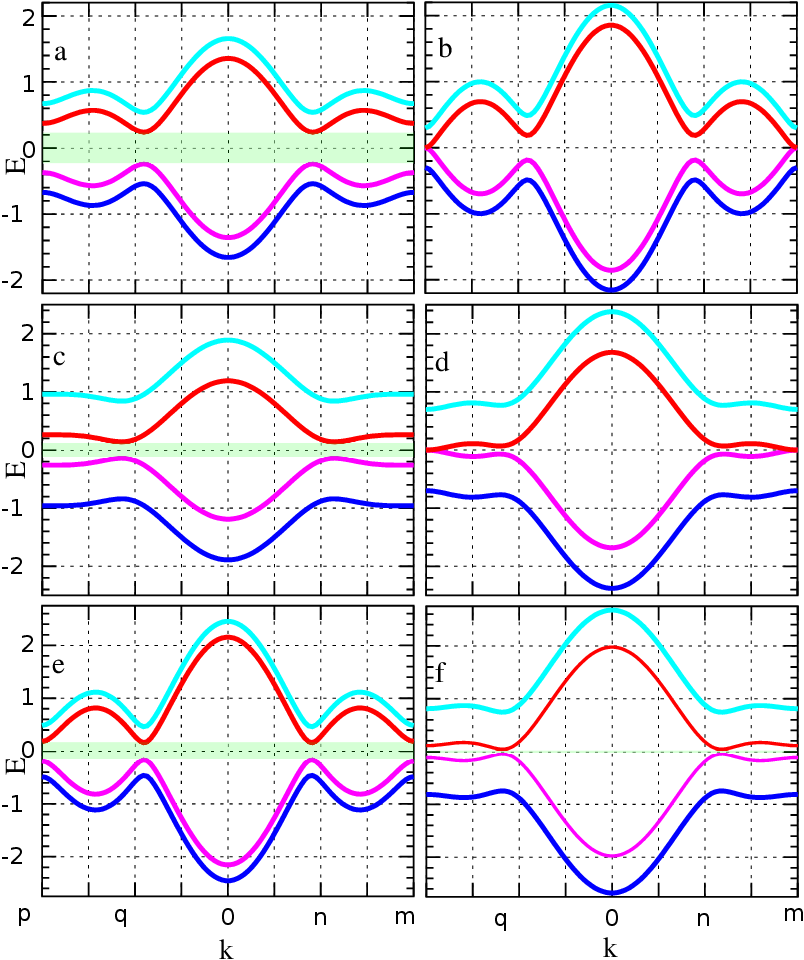}
\caption{Dispersion relations for (a) $u=0.7$, $v=0.3$, $w=1/4$; 
  (b) $u=0.7$,  $v=0.3$, $w=1/2$;  (c) $u=0.3$, $v=0.7$, $w=1/4$;
  (d) $u=0.3$, $v=0.7$, $w=1/2$; (e) $u=0.7$, $v=0.3$, $w=0.65$; and
  (f) $u=0.3$, $v=0.7$, $w=0.65$; 
  of chiral {\em cis}-polyacene model.
Energy gap in (a), (c), (e)  and (f) is shown by green shade.}
 \label{chiral-cis-polyacene-dispersions}
\end{figure}

The band structure of the $cn$-pol model has been shown
in Fig \ref{chiral-cis-polyacene-dispersions}, by plotting the
Eq \ref{Hcck} for six different sets of values for $v$, $w$ and $u$.
Four diagrams, Fig \ref{chiral-cis-polyacene-dispersions} (a), (c),
(e) and (f), among the total six indicate the presence of band gap
of the system. The remaining two diagrams, 
Fig \ref{chiral-cis-polyacene-dispersions} (b) and (d)
exhibit no band gap in the system. These two are drawn at the
phase transition points while the former four diagrams
are drawn for either topologically trivial or nontrivial phases,
which will be discussed later. 

\begin{figure}[h]
  \psfrag{a}{(a)}
  \psfrag{b}{(b)}
   \psfrag{p}{\scriptsize {$|\psi|^2$}}  
 % \psfrag{p}{$|\psi|$}  
  \psfrag{Energy}{ \hskip -0.15 cm Energies}
  \psfrag{I}{ \hskip -0.2 cm $I_{pr}$}
  \psfrag{00}{\hskip 0.05 cm $0.0$}
  \psfrag{0.25}{$\hskip -0.05 cm 0.25$}
  \psfrag{0.00}{\hskip -0.05 cm $0.00$}
    \psfrag{0.50}{\hskip -0.1 cm $0.5$}  
\psfrag{0}{$0$}
\psfrag{20}{$20$}
\psfrag{40}{$40$}
\psfrag{60}{$60$}
\psfrag{80}{$80$}
\psfrag{0}{$0$}
\psfrag{1.0}{\hskip -0. cm $1.0$}
\psfrag{0.5}{\hskip -0. cm $0.5$}
\psfrag{0.0}{\hskip -0.0 cm $0.0$}
\psfrag{-1.4}{\hskip -0.05 cm$-1.4$}
\psfrag{-0.7}{\hskip -0.05 cm $-0.7$}
\psfrag{0.7}{\hskip -0.0 cm $0.7$}
\psfrag{1.4}{\hskip -0.0 cm $1.4$}
\psfrag{v}{\hskip -0.15 cm $u\!=\!0.7,\,v\!=\!0.3,\,w\!=\!0.1$}
\psfrag{site}{\hskip 0.2 cm site}
\hskip 0.22 cm
\includegraphics[width=230pt]{top-color-box-polyacene.eps}
\vskip -0.0 cm
\includegraphics[width=250pt]{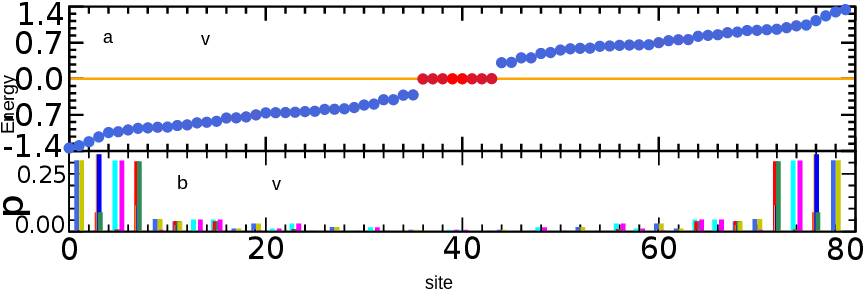}
\caption{Variation of energies of ascending order for  $cn$-pol
  model with $N=80$, when $u=0.7$, $v=0.3$ and $w=0.1$ (a).
  It shows the presence of eight zero-energy edge
  states. Top colorbar indicates the variation of $I_{pr}$.
  Probability density of eight edge states are
  shown in the histograms by different colors (b). }
 \label{chiral-cis-polyacene-FN-edge-energies-set-1}
\end{figure}
In order to investigate the topological properties of the $cn$-pol
model, variation of energy of the open chain with 80 sites
with respect to $w$ have been shown in
Fig \ref{chiral-cis-polyacene-FN-energies-winding-number-set-1} (a) and 
Fig \ref{chiral-cis-polyacene-FN-energies-winding-number-set-2} (a)
for two different sets, ($u=0.7,\,v=0.3$), and
($u=0.3,\,v=0.7$), respectively. Variation of winding number 
for the respective sets has been shown in
Fig \ref{chiral-cis-polyacene-FN-energies-winding-number-set-1} (b) and 
Fig \ref{chiral-cis-polyacene-FN-energies-winding-number-set-2} (b).
Diagrams are found symmetric about $v=0$, since
$E(-v)= E(v)$.   
The diagrams reveal that the system hosts three nontrivial
phases with $\nu=-2,2,4$. 
Specifically,
Fig \ref{chiral-cis-polyacene-FN-energies-winding-number-set-1} (b)
demonstrates the existence of topological phase with $\nu=4$,
in the region, $-0.5<w<0.5$, and that with $\nu=2$,
in the regions, $w>0.5$, and $w<-0.5$, when $u=0.7,\,v=0.3$.
It means that the system undergoes a transition at the points,
$w=\pm 0.5$, between the two nontrivial topological phases, 
when $u=0.7,\,v=0.3$, as shown in
Fig \ref{chiral-cis-polyacene-FN-energies-winding-number-set-1} (a).
Band gap is missing at these two points as revealed again in
Fig \ref{chiral-cis-polyacene-FN-energies-winding-number-set-1} (a),
while finite band gap is found in the two adjacent topological regions. 
In order to confirm the existence of the phase, $\nu=4$,
energy spectrum of the open system for $u=0.7,\,v=0.3,w=0.1$,
has been plotted in Fig \ref{chiral-cis-polyacene-FN-edge-energies-set-1} (a), 
following the ascending order when $N=80$. 
It shows the presence of eight zero-energy edge modes 
which is consistent to the phase with $\nu=4$, according to the
bulk boundary correspondence rule.
Higher values of $I_{pr}$ for the zero-energy modes ensure the
strong localization of those. 
Probability density per site,
$|\psi_j|^2$, of the eight edge states are
shown in the histograms by different colors as displayed in 
Fig \ref{chiral-cis-polyacene-FN-edge-energies-set-1} (b).
It confirms that all the zero-energy modes are localized near the
two opposite edges of the chain.  

\begin{figure}[h]
  \psfrag{a}{(a)}
  \psfrag{b}{(b)}
  \psfrag{p}{\scriptsize {$|\psi|^2$}}  
  \psfrag{Energy}{ \hskip -0.15 cm Energy}
  \psfrag{I}{ \hskip -0.2 cm $I_{pr}$}
  \psfrag{00}{\hskip 0.05 cm $0.0$}
  \psfrag{0.25}{$\hskip -0.05 cm 0.25$}
  \psfrag{0.00}{\hskip -0.05 cm $0.00$}
  \psfrag{0.50}{\hskip -0.1 cm $0.5$}
  \psfrag{1.0}{\hskip -0. cm $1.0$}
\psfrag{0.5}{\hskip -0. cm $0.5$}
\psfrag{0}{$0$}
\psfrag{20}{$20$}
\psfrag{40}{$40$}
\psfrag{60}{$60$}
\psfrag{80}{$80$}
\psfrag{0}{$0$}
\psfrag{1}{\hskip -0. cm $1$}
\psfrag{2}{\hskip -0. cm $2$}
\psfrag{0.0}{\hskip -0.0 cm $0.0$}
\psfrag{-2}{\hskip -0.05 cm$-2$}
\psfrag{-1}{\hskip -0.05 cm $-1$}
\psfrag{0.14}{\hskip -0.03 cm $0.14$}
\psfrag{1.4}{\hskip -0.0 cm $1.4$}
\psfrag{v}{\hskip -0.15 cm $u\!=\!0.3,\,v\!=\!0.7,\,w\!=\!3/4$}
\psfrag{site}{\hskip 0.2 cm site}
\hskip 0.22 cm
\includegraphics[width=230pt]{top-color-box-polyacene.eps}
\vskip -0.0 cm
\includegraphics[width=250pt]{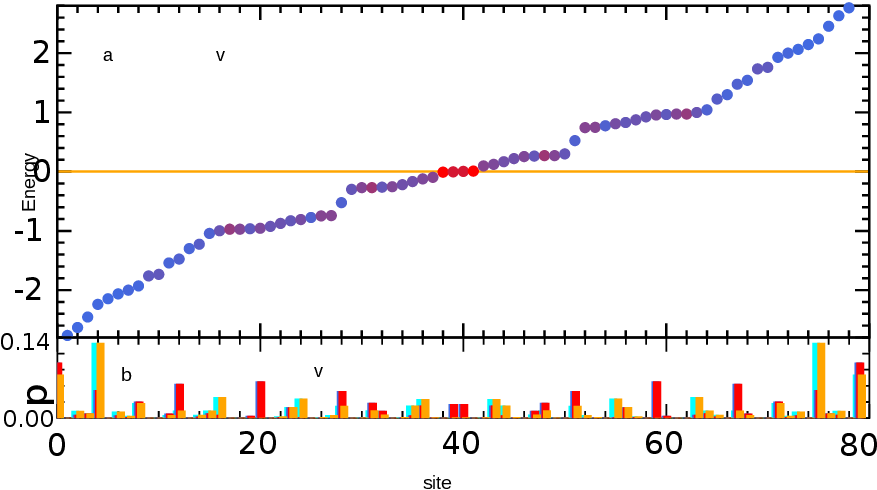}
\caption{Variation of energies in ascending order for $cn$-pol
  model with $N=80$, when $u=0.3$, $v=0.7$ and $w=3/4$ (a).
  It shows the existence of four zero-energy edge
  states. 
  Top colorbar indicates the variation of $I_{pr}$.
  Probability density of four edge state are
  shown in the histograms by different colors (b). }
 \label{chiral-cis-polyacene-FN-edge-energies-set-2}
\end{figure}
On the other hand,
Fig \ref{chiral-cis-polyacene-FN-energies-winding-number-set-2} (b) 
demonstrates that the system hosts another topological phase with $\nu=-2$,
for the regions, $w>0.5$, and $w<-0.5$, when $u=0.3$ and $v=0.7$.
The system remains trivial in the region, $-0.5<w<0.5$, as a result, it
undergoes phase transitions at the points $w=\pm 0.5$,
when $u=0.3$ and $v=0.7$. Energy spectrum of the open
system with respect to $w$ has been
plotted in Fig \ref{chiral-cis-polyacene-FN-energies-winding-number-set-2}
(a), which shows no zero-energy modes in the topologically trivial
region as expected from the bulk boundary correspondence rule. 
In contrast, zero-energy modes with higher degrees of
localization have been noted
in the regions, $w>0.5$, and $w<-0.5$. As usual localization
has been estimated by the values of $I_{pr}$. 
In order to verify the existence of zero-energy edge modes corresponding to
the nontrivial phase, $\nu=-2$, energy spectrum of the system for 
$u=0.3$, $v=0.7$ and $w=0.75$ has been shown in Fig
\ref{chiral-cis-polyacene-FN-edge-energies-set-2} (a).
Values of $I_{pr}$ for the respective states are shown according to the
colorbar. It indicates the strong localization for the
zero-energy modes.
Band gap vanishes at the transition points $w=\pm 0.5$, as noted
in Fig \ref{chiral-cis-polyacene-FN-energies-winding-number-set-2} (a). 
Finite band gap is found in the trivial region, while
very narrow band gap is found in the nontrivial regions for $\nu=-2$.

Now the properties of band diagrams shown in Fig
\ref{chiral-cis-polyacene-dispersions} will be explained
in terms of the topology of the system. Figs
\ref{chiral-cis-polyacene-dispersions} (a), (b) and (e)
have been drawn in accordance to the
topological phases shown in Fig
\ref{chiral-cis-polyacene-FN-energies-winding-number-set-1},
where (a) and (e) correspond to the topological phases 
with $\nu=4$, and $\nu=2$, respectively,
while (b) corresponds to their intermediate phase transition.
True band gap is observed for both of the topological
phases, while it is missing at the transition point.
Similarly, Figs
\ref{chiral-cis-polyacene-dispersions} (c), (d) and (f)
have been drawn in agreement with the
topological phases shown in Fig
\ref{chiral-cis-polyacene-FN-energies-winding-number-set-2}, where
(c) and (f) correspond to the trivial ($\nu=0$) and topological
($\nu=-2$) phases, while (d) corresponds to their
intermediate phase transition. Large band gap is found in the
trivial phase whereas very narrow band gap is
noticed in the topological phase.

%The primary concern behind the $cn$-pol
%structure is its material realization, even though,
%\begin{figure}[h]
%\psfrag{u}{ \hskip -.4cm $u/w$}
%\psfrag{v}{\hskip -.1cm $v/w$}
% \psfrag{nu}{\hskip 0.03cm \large $\nu$}
% \psfrag{nu0}{\hskip -.3cm \color{blue} \large $\nu=0$}
% \psfrag{nu2}{\hskip -.3 cm \color{white} \large $\nu=2$}
% \psfrag{nu-2}{\hskip -.5cm \color{white} \large $\nu=-2$}
%  \psfrag{nu4}{\hskip -.3 cm \color{white} \large $\nu\!=\!4$}
%\psfrag{1.0}{\hskip -0. cm $1.0$}
%\psfrag{0.5}{\hskip -0. cm $0.5$}
%\psfrag{0.0}{\hskip -0.0 cm $0.0$}
%\psfrag{2.0}{\hskip -0. cm $2.0$}
%\psfrag{1.5}{\hskip -0. cm $1.5$}
%\psfrag{ 0}{$0$}
%\psfrag{ 2}{$2$}
%\psfrag{ 4}{$4$}
%\psfrag{ 3}{$3$}
%\psfrag{ 1}{$1$}
%\psfrag{-1}{$-1$}
%\psfrag{-2}{$-2$}
%\includegraphics[width=230pt]{Topological-phase-polyacene-chiral-trans-FN-varyi%ng-u-and-v.eps}
%\caption{Topological phase diagram in $u$-$v$ plane.}
%\label{polyacene-chiral-FN}
%\end{figure}

Four different models on $cis$-polyacene structure have been
formulated where three modified structures, say, $cb$-pol, $ccb$-pol and
$cn$-pol are found to exhibit topological non-triviality
in terms of winding number. Although they all possess
TRS, CS and PHS but they are different in terms of preserving IS and MS. 
$ccb$-pol breaks IS and MS, while 
$cb$-pol preserves all of them. The primary structure, $c$-pol
does not preserve IS while MS is broken in $cn$-pol. 
As a result their topological character are totally different. 
Energy diagrams for $c$-pol, $cb$-pol and $cn$-pol
are symmetric around $v=0$, while $ccb$-pol is non-symmetric, 
but it is similar to $tb$-pol.
$c$-pol and $cb$-pol have been mapped onto a pair of
uncoupled SSH models in the presence of staggered and
uniform onsite potentials, respectively,
under a specific intracell operator transformation. 
As a result, $c$-pol turns into topologically trivial,
while $cb$-pol is found topologically identical to the SSH model. 
Nontrivial topology is found in $ccb$-pol, when $v<-w/2$, as
$\nu=-1$ in this region. On the other hand, three different
nontrivial phases with $\nu=-2,2,4$ have been identified in $cn$-pol.
Topological feature of all the seven models are different.
Phase transition points of these models can be identified
in terms of fidelity susceptibility \cite{Sirker,Surajit}. 

Since $c$-pol, $cb$-pol and $cn$-pol have been mapped onto a pair of
SSH model, the obvious question is whether the remaining models can be 
expressed in terms of interacting SSH models like, two-leg SSH ladders
or else. A rigorous analysis on the other models deny this hope, as
they are found not connected to SSH chains by any means.
In addition, these models are totally different from the
existing ladder models those are investigated
before\cite{Liu,Li,Mao,Vishveshwara}. 

\begin{table}[htbp]
  \centering
\begin{tabular}{|c|c|c|c|c|c|c|c|} 
\hline  &&&&&\\[-1.2em]\;Symmetry\;&$t$-pol&$tb$-pol&$tcb$-pol&$c$-pol &$cb$-pol&$ccb$-pol&$cn$-pol
\\[0.2em] \hline
&&&&&\\[-1.1em] TRS& \CheckmarkBold  & \CheckmarkBold &  \CheckmarkBold &  \CheckmarkBold &\CheckmarkBold&  \CheckmarkBold &\CheckmarkBold \\[0.1em] \hline
&&&&&\\[-1.1em] CS  & \CheckmarkBold & \CheckmarkBold &   \CheckmarkBold & \CheckmarkBold& \CheckmarkBold &  \CheckmarkBold &\CheckmarkBold \\[0.1em] \hline
&&&&&\\[-1.1em] PHS &\CheckmarkBold & \CheckmarkBold & \CheckmarkBold & \CheckmarkBold& \CheckmarkBold&  \CheckmarkBold &\CheckmarkBold  \\[0.1em] \hline
&&&&&\\[-1.1em] IS${}^*$ & \CheckmarkBold & \CheckmarkBold & \CheckmarkBold & \XSolidBold& \CheckmarkBold &  \XSolidBold &\CheckmarkBold  \\[0.1em] \hline
&&&&&\\[-1.1em] MS & \XSolidBold & \XSolidBold & \XSolidBold & \CheckmarkBold& \CheckmarkBold&\XSolidBold&\XSolidBold \\[0.1em] \hline
\end{tabular}
\caption{Table for symmetries of all the models, where
  \CheckmarkBold (\XSolidBold)
  means preserved (broken). ${}^*$Transformation used for IS in
  $t$-pol, $tb$-pol, $tcb$-pol and
$cn$-pol is different from that used in $cb$-pol as explained before.}
\label{Table}
\end{table}
Symmetries of all the models have been presented in the Table \ref{Table}.
It reveals that $cb$-pol is the only model which possesses every symmetry,
and $ccb$-pol breaks two symmeteries, while rest are found to break
one symmetry. However, all the models preserve
TRS, CS and PHS. 
%%%%%%%%%%%%%%%%% Discussion %%%%%%%%%%%%%%%%%%%%%%%%%%%%
\section{Discussion}
\label{Discussion}
In this investigation, TB model has been formulated on
seven different polyacene structures in order to
study their topological character. Among them
two are based on the $cis$- and $trans$-polyacene 
structures, the most common geometric isomers of polyacene.
Although both the models yield the same band structure,  
$c$-pol owns an extra MS due to its 
particular arrangement of single and double bonds, which is missing in $t$-pol.
On the otherhand, $t$-pol possesses IS which is absent in $c$-pol. 
Otherwise, in $k$-space, both the structures possess the
same set of symmetries as shown in the Table \ref{Table}.
%$t$-pol possesses CS and PHS,
%which are absent in $c$-pol. This difference
However, they lead to opposite scenario in their topological 
behavior in a sense that, where $t$-pol is found nontrivial,  
$c$-pol turns out trivial. More precisely, 
$t$-pol possesses topological phase with $\nu=2$, in
the region $-1<v/w<+1$. So, the system exhibits
zero-energy edge modes in the same region
owing to the bulk-boundary correspondence
rule valid for the model with CS. Interestingly, the 
$c$-pol model exhibits both the zero- and nonzero-energy 
edge modes in the same region $-1<v/w<+1$,
although the system is topologically trivial
albeit possesses all the relevant symmetries.
In order to understand this feature, $c$-pol
has been examined in a rotated space, where it turns out that
$c$-pol is essentially composed of two decoupled
SSH chains with staggered fields. As a result, it exhibits
triviality. 

Two additional variants of the $t$-pol, referred as
$tb$-pol and $tcb$-pol have been constructed out of 
$t$-pol with an additional hopping term of strength $v$,
exists between A and D sites of the NN and same unit cell,
respectively. Although they possess the same set of symmetries, 
topological character of these models are different. 

In order to imprint topological signature in the 
trivial $c$-pol model, three different modified variants of 
this structure have been introduced, where extra hopping terms
are taken into consideration.
In the first attempt, a single intracell hopping term 
between A and D sites has been added, where the resulting model ($cb$-pol)
inherits all the symmetries, CS, TRS, PHS, IS and MS.
Under the same rotation which is employed before
for $c$-pol, $cb$-pol is found to decouple into
two independent SSH chains with uniform fields of different signs.
As a result, topological feature of $cb$-pol is identical to
that of SSH chain, which means nontrivial topology with $\nu=1$, 
within $-1<v/w<+1$.
In the second attempt, effect of an additional intercell 
hopping term between A and D sites
on $c$-pol ($ccb$-pol) has been studied 
which exhibits topological phase with $\nu=-1$ in the region
$v<-w/2$. However, only two types of bond strengths, $v$ and $w$
have been considered so far for all the previous models. 
In the third attempt, modified $c$-pol structure
has been constructed ($cn$-pol) by adding two pairs of intercell 
hopping terms, among which one pair exists 
between NN cells of strength $w$, 
and another pair acts between NNN cells of strength $u$, 
(Fig. \ref{chiral-cis-polyacene-FN}).  
The resulting model possesses all the symmetries except the MS.
%At the same time, mirror and intracell rotational symmetry are
%lost in the real space.  This model has been termed as $cn$-pol,
$cn$-pol exhibits three different nontrivial phases
with winding numbers, $\nu=-2,2,4$.
All the modes introduced in this study belongs to BDI class,
and have different topological characters. 

The primary motivation behind the study is the search of
topological phase in organic polymer called polyacene which has been 
synthesized very recently.
So, the models are very close to real materials,
like the SSH model which is based on polyacetylene. 
Further polyacene is regarded as the 1D analogue of
2D graphene whose electrical and thermal conductances
are extremely high. Thus, in the topological regime
these conductivities are expected to increase manifold
due to their ability to conduct electricity along their boundaries
without backscattering. 

Besides the common geometric isomers,
$cis$- and $trans$-polyacene, five modified structures
have been considered in this study, whose %Although
material realization %of the $cn$-pol structure
is a big challenge. However, 
the experimental realization of their topological phases
can be made possible on various hybrid material platforms
by duplicating their physical structures.
For example, topological phase of SSH model
has been demonstrated in the following hybrid systems:
(i) Stub lattice of coupled waveguides \cite{Vicencio},
(ii) mechanical oscillator \cite{Merlo},
(iii) acoustic waveguides, \cite{Chen}
(iv) optical lattices composed of ultracold atoms\cite{Atala}, 
(v) photonic systems of helical and cylindrical waveguides \cite{Rechtsman},
%(v) vacancy lattice of a chlorine monolayer on a Cu(100) surface,
(vi) topolectrical circuits \cite{Thomale}, and
(vii) water wave channel \cite{Pagneux}.
Topological phase of SSH trimer model
has been realized upon fabricating its structure by 
In atoms on the Si(553)-Au surface\cite{Alvarez,Kwapinski}.
Topological phase of Kitaev's 1D $p$-wave superconducting model
has been realized recently in topolectrical circuits \cite{Kitaev,Ezawa}.
The covalent organic frameworks blending synthetic chemistry
and nanotechnology have proven the potential route in producing
1D and 2D organic topological compounds of desired structure \cite{Wang}. 
These scientific achievements suggest that realization of
topological phases found in the polyacene models
can be accomplished on these potential hybrid platforms
as well.

  \section{ACKNOWLEDGMENTS}
  RKM acknowledges the DST/INSPIRE Fellowship/2019/IF190085.
  \section{Data availability statement}
  All data that support the findings of this study are
  included within the article.
   \section{Conflict of interest}
  Authors declare that they have no conflict of interest.
  

\begin{thebibliography}{99}
\bibitem{Qi}Qi, X. L. and  Zhang, S. C., Reviews of Modern Physics,
    {\bf 83}(4), 1057-1110 (2011).    
  \bibitem{SSH1} W. Su, J. Schrieffer and A. J. Heeger, 
Phys. Rev. Lett. {\bf 42}, 1698 (1979).
    \bibitem{SSH2} W. -P. Su, J. Schrieffer and A. Heeger, 
      Phys. Rev. B {\bf 22}, 2099 (1980).
%https://link.aps.org/doi/10.1103/PhysRevB.22.2099
\bibitem{SSH3}  Heeger A. J., Kivelson S., Schrieffer J. R. and  Su W. -P., 
  Rev. Mod. Phys. {\bf 60}, 781 (1988).
  \bibitem{Rakesh1}R K Malakar and A K Ghosh, J . Phys. Condens. Matter, {\bf  35}, 335401 (2023).
    \bibitem{Vicencio}  G Caceres-Aravena, B Real, D Guzman-Silva, A Amo,
  L E F Foa Torres, and R A Vicencio, Phys. Rev. Research {\bf 4}, 013185
  (2022)
\bibitem{Browaeys} S de Leseleuc, V Lienhard, P Scholl, D Barredo, 
  S Weber, N Lang, H P B\"uchler, T Lahaye, A Browaeys,
  Science {\bf 365}, 775-780 (2019)
%\bibitem{Zhu}F Mei, D-W Zhang, S-L Zhu, Annals of Physics {\bf 358}
%   58-82 (2015)
\bibitem{Chen} Z-G Chen, L Wang, G Zhang and G Ma, Phys. Rev. Applied {\bf 14}, 024023 (2020)
\bibitem{Merlo}L Thatcher, P Fairfield, L Merlo-Ramirez and J M Merlo,
  Phys. Scr. {\bf 97} 035702 (2022)
 \bibitem{Pagneux} A Anglart, P Obrepalski, A Maurel, P Petitjeans 
and V Pagneux, Phys. Rev. B {\bf 111}, 224311 (2025). 
 \bibitem{Atala} M. Atala, M. Aidelsburger, J. T. Barreiro, D. Abanin,
T. Kitagawa, E. Demler, and I. Bloch, Nature Physics {\bf 9},
795 (2013).
\bibitem{Xie} D Xie, W Gou, T Xiao, B Gadway and B Yan,
  Npj Quantum Inf.   {\bf 5} 55 (2019).
\bibitem{Thomale}R Thomale et al, Commun. Phys., {\bf 1}, 39 (2018).
\bibitem{Alvarez}  V M M Alvarez and M D Coutinho-Filho,
  Phys. Rev. A {\bf 99} 013833 (2019)
\bibitem{Du} T Du, Y-X Li, H-L Lu and H Zhang,
  New J. Phys. {\bf 26}  023044 (2024).
\bibitem{Bomantara}M Ghuneim and R W Bomantara,
  J. Phys.: Condens. Matter {\bf 36}, 495402 (2024).
\bibitem{Rechtsman} Rechtsman M C, Zeuner J M, Plotnik Y, Lumer Y, Podolsky D,
  Dreisow F, Nolte S, Segev M and Szameit A, Nature {\bf 496}, 196 (2013)
\bibitem{Clar-John}E. Clar and F. John,
  Ber. Dtsch. Chem. Ges. B {\bf 63}, 2967 (1930).
\bibitem{Bettinger} H F Bettinger, Pure Appl. Chem., {\bf 82},
  905-915 (2010)
\bibitem{Chin} M. Mosaferi, M. Aktas, D. Romanin, and A. W. Chin, 
  Phys. Rev. Mat. {\bf 9} 044204 (2025).
\bibitem{Romanin}M. Aktas, M. Mosaferi, J. Saint-Martin, A. W. Chin, and D. Romanin, 
Phys. Rev. B {\bf 112} 045130 (2025).
\bibitem{Kitao} T Kitao, T Miura, R Nakayama, Y Tsutsui, 
  Y Chan, H Hayashi, H Yamada, S Seki, T Hitosugi and T Uemura,
  Nat. Synth. {\bf 2} 848-854 (2023)
\bibitem{Melo}A L S da Rosa and C P de Melo,
  Phys. Rev. B {\bf 38}, 5430 (1988).
\bibitem{Whangbo} M. H. Whanghbo, R. Hoffmann, and R. B. Woodward, Proc.
R. Soc. Lond. A {\bf 366} 23-46 (1979).
\bibitem{Tanaka} K Tanaka, K Ohzeki, S Nankai, T Yamabe and H Shirakawa,
  J. Phys. Chem. Solids. {\bf 44} 1069-1075 (1983).
 \bibitem{Zak} J. Zak, Phys. Rev. Lett., {\bf 62} 2747, (1989).
\bibitem{Ryu}Ryu S, Schnyder A, Furusaki A and Ludwig A, 
  New J. Phys. {\bf 12} 065010 (2010)
\bibitem{Liu} J S Liu, Y Z Han and C S Liu,  Chin. Phys. B {\bf 28}, 100304 (2019)
  \bibitem{Sil2} Sil A. and Ghosh A. K., J. Phys.: Condens. Matter {\bf 32}, 025601 (2019).
  \bibitem{Moumita1} M Deb and A K Ghosh, J. Phys.: Condens. Matter {\bf 31},
  345601 (2019).
 \bibitem{Sil3} Sil A. and Ghosh A. K., J. Phys.: Condens. Matter {\bf 36}, 125401 (2023).  
   \bibitem{Sil4} Sil A. and Ghosh A. K., J. Phys.: Condens. Matter {\bf 32}, 025601 (2020).
\bibitem{Kruger} J Kruger et al, Angew. Chem., {\bf 129}, 12107-12110 (2017)
\bibitem{Jancarik} A Jancarik, J Holec, Y Nagata, M Samal and A Gourdon
  Nature Commun. {\bf 13}, 223 (2022).
\bibitem{Kwapinski} M Jalochowski, M Krawiec and T Kwapinski,
  ACS Nano, {\bf 18}, 12861-12869 (2024).
   \bibitem{Kitaev} A Kitaev, Phys. Usp. {\bf 44}, 131 (2001)
   \bibitem{Ezawa} M Ezawa et al, Commun. Phys., {\bf 6}, 279 (2023).
\bibitem{Wang} Z.F. Wang, Z Liu and F Liu, Nature Commun. {\bf 4},
  1471 (2013).
\bibitem{Hatsugai}Y. Hatsugai, Phys. Rev. Lett. {\bf 71}, 3697 (1993).
\bibitem{Rakesh2}R K Malakar and A K Ghosh, 
  Phys. Scr. {\bf  99}, 035944 (2024).
\bibitem{Rakesh3}R K Malakar and A K Ghosh, J . Phys. Condens. Matter,
  {\bf  36}, 325401 (2024).
\bibitem{Rittwik} R Chatterjee and  A K Ghosh, Phys. Scr. {\bf 100}, 105945 (2025).
\bibitem{Sirker} J Sirker, M Maiti, N P Konstantinidis and
  N Sedlmayr, J. Stat. Mech.  P10032 (2014).
\bibitem{Surajit} S Mandal and A K Ghosh, arXiv:2608.22875 (2026)
\bibitem{Li} C. Li, S. Lin, G. Zhang, and Z. Song, Phys. Rev. B {\bf 96} 125418 (2017)
\bibitem{Mao} Y-y Mao {\em et al}, Phys. Rev. Res. {\bf 8} L022006 (2026)
\bibitem{Vishveshwara}K Padavic, S S Hegde, W DeGottardi, and S Vishveshwara, Phys. Rev. B {\bf 98} 024205 (2018)  
%\bibitem{Moumita3} M Deb and A K Ghosh, J. Phys.: Condens. Matter {\bf 32},   365601 (2020).
%\bibitem{Moumita4} M Deb and A K Ghosh, Eur. Phys. J. B {\bf 93}, 145 (2020).
%\bibitem{Suzuki}M Suzuki, Phys. Lett. A 34, 94 (1971); Prog. Theor. Phys.
%  46, 1337 (1971)
\end{thebibliography}
\end{document}